\documentclass[aos,preprint]{imsart}

\RequirePackage[OT1]{fontenc}
\RequirePackage{amsthm,amsmath}
\RequirePackage[numbers]{natbib}
\RequirePackage[colorlinks,citecolor=blue,urlcolor=blue]{hyperref}

\usepackage{latexsym, epsfig, amssymb, amsmath, graphicx, mathrsfs}
\usepackage{natbib}
\usepackage[OT1]{fontenc}
\usepackage{multirow}

\usepackage{color}
\usepackage{dsfont}
\usepackage{booktabs}
\usepackage{url}
\usepackage{float}

\startlocaldefs
\theoremstyle{plain}

\newtheorem{condition}{Condition}
\newtheorem{lemma}{Lemma}
\newtheorem{theorem}{Theorem}
\renewcommand{\theequation}{
\arabic{equation}%
}

\makeatother

\newcommand{\corr}{\mathrm{corr}}
\newcommand{\ba}{{\bf a}}
\newcommand{\bs}{{\bf s}}
\newcommand{\bt}{{\bf t}}
\newcommand{\bu}{{\bf u}}
\newcommand{\bv}{{\bf v}}
\newcommand{\bw}{{\bf w}}
\newcommand{\bx}{{\bf x}}
\newcommand{\by}{{\bf y}}
\newcommand{\bz}{{\bf z}}

\newcommand{\bB}{\mbox{\bf B}}
\newcommand{\bM}{\mbox{\bf M}}
\newcommand{\bW}{\mbox{\bf W}}
\newcommand{\bX}{\mbox{\bf X}}
\newcommand{\bY}{\mbox{\bf Y}}
\newcommand{\bZ}{\mbox{\bf Z}}

\newcommand{\balpha}{\mbox{\boldmath${\alpha}$}}
\newcommand{\bDel}{\mbox{\boldmath $\Delta$}}
\newcommand{\bSig}{\mbox{\boldmath $\Sigma$}}

\newcommand{\dcov}{\mbox{dcov}}
\newcommand{\dcorr}{\mbox{dcorr}}

\newcommand{\bzero}{\mbox{\bf 0}}
\newcommand{\tr}{\mathrm{tr}}
\def\t{^T}

\endlocaldefs

\begin{document}

\begin{frontmatter}
\title{Interaction pursuit in high-dimensional multi-response regression via distance correlation\thanksref{T1}}
\runtitle{IPDC}
\thankstext{T1}{This work was supported by NSF CAREER Awards DMS-0955316 and DMS-1150318 and a grant from the Simons Foundation. Yinfei Kong and Daoji Li contributed equally to this work. The authors would like to thank the Co-Editor, Associate Editor, and referees for their valuable comments that have helped improve the paper significantly. Part of this work was completed while the last two authors visited the Departments of Statistics at University of California, Berkeley and Stanford University. They sincerely thank both departments for their hospitality.}

\begin{aug}
\author{\fnms{Yinfei}~\snm{Kong}\thanksref{m1}\ead[label=e1]{yinfeiko@usc.edu}},
\author{\fnms{Daoji}~\snm{Li}\thanksref{m2}\ead[label=e2]{daoji.li@ucf.edu}},
\author{\fnms{Yingying}~\snm{Fan}\thanksref{m3}\ead[label=e3]{fanyingy@marshall.usc.edu}}
\and
\author{\fnms{Jinchi}~\snm{Lv}\thanksref{m3}\ead[label=e4]{jinchilv@marshall.usc.edu}
}

\runauthor{Y. Kong, D. Li, Y. Fan and J. Lv}

\affiliation{California State University at Fullerton\thanksmark{m1},  University of Central Florida\thanksmark{m2} and University of Southern California\thanksmark{m3}}

\address{Department of Information Systems and Decision Sciences\\
Mihaylo College of Business and Economics\\
California State University at Fullerton\\
Fullerton, CA 92831\\
USA\\
\printead{e1}}

\address{Department of Statistics\\
University of Central Florida\\
Orlando, FL 32816-2370\\
USA\\
\printead{e2}}

\address{Data Sciences and Operations Department\\
Marshall School of Business\\
University of Southern California\\
Los Angeles, CA 90089\\
USA\\
\printead{e3}\\
\phantom{E-mail:\ }\printead*{e4}}

\end{aug}

%
%
%
%
%

\begin{abstract}
Feature interactions can contribute to a large proportion of variation in many prediction models. In the era of big data, the coexistence of high dimensionality in both responses and covariates poses unprecedented challenges in identifying important interactions. In this paper, we suggest a two-stage interaction identification method, called the interaction pursuit via distance correlation (IPDC), in the setting of high-dimensional multi-response interaction models that exploits feature screening applied to transformed variables with distance correlation followed by feature selection. Such a procedure is computationally efficient, generally applicable beyond the heredity assumption, and effective even when the number of responses diverges with the sample size. Under mild regularity conditions, we show that this method enjoys nice theoretical properties including the sure screening property, support union recovery, and oracle inequalities in prediction and estimation for both interactions and main effects. The advantages of our method are supported by several simulation studies and real data analysis.
\end{abstract}

\begin{keyword}[class=MSC]
\kwd[Primary ]{62H12}
\kwd{62J02}
\kwd[; secondary ]{62F07}
\kwd{62F12.}
\end{keyword}

\begin{keyword}
\kwd{Interaction pursuit}
\kwd{distance correlation}
\kwd{square transformation}
\kwd{multi-response regression}
\kwd{high dimensionality}
\kwd{sparsity.}
\end{keyword}

\end{frontmatter}

\section{Introduction}\label{sec: Intro}

Recent years have seen a surge of interests on interaction identification in the high-dimensional setting by many researchers. 
For instance, 
\citet{hall2014selecting}
proposed a 
recursive approach to 
{identify} 
important interactions among covariates, where all $p$ covariates are first ranked by the generalized correlation and then only the top $p^{1/2}$ ones are retained to construct pairwise interactions of order $O(p)$ for further screening and selection of both interactions and main effects. A forward selection based screening procedure was introduced in \cite{hao2014interaction} for identifying interactions in a greedy fashion under the heredity assumption.  
Such an assumption in the strong sense requires that an interaction between two covariates {should} be included in the model only if both main effects are important, while the weak version relaxes such a constraint to the presence of at least one main effect being important. Regularization methods have also been used for interaction selection under the heredity assumption. See, for example, \cite{yuan2009structured}, \cite{choi2010variable}, and \cite{bien2013lasso}. Under the inverse modeling framework, \cite{jiang2014sliced} proposed a new method, called the sliced inverse regression for interaction detection (SIRI), which can detect pairwise interactions among
covariates without the heredity assumption. 
The theoretical development in \cite{jiang2014sliced} relies primarily on the joint normality assumption on the covariates. The innovated interaction screening procedure was introduced in \cite{fan2015IIS} for high-dimensional nonlinear classification with no heredity assumption.


Although the aforementioned methods can perform well in many scenarios, they may have two potential limitations. First, those approaches assume mainly interaction models with a single response, while the coexistence of multiple responses becomes increasingly common in the big data era. Second, those developments 
are usually built upon the strong or weak heredity assumption, or the normality assumption, which may not be satisfied in certain real applications. 

To enable broader applications in practice, in this paper we consider the following high-dimensional multi-response interaction model 
   \begin{eqnarray}\label{eq: model}
     \by = \balpha + \bB_{\bx}\t\bx + \bB_{\bz}\t\bz + \bw,
   \end{eqnarray}
where $\by=(Y_1, \cdots, Y_q)^T$ is a $q$-dimensional vector of responses, 
$\bx=(X_1, \cdots,\\ X_p)^T$ is a $p$-dimensional vector of covariates, $\bz$ is a 
$p(p-1)/2$-dimensional vector of all pairwise interactions between covariates $X_j$'s, $\balpha=(\alpha_1, \cdots, \alpha_q)^T$ is a $q$-dimensional vector of intercepts, $\bB_{\bx}\in \mathbb{ R}^{p\times q}$ and $ \bB_{\bz}\in \mathbb{R}^{[p(p-1)/2]\times q}$ are regression coefficient matrices for the main effects and interactions, respectively, and $\bw=(W_1, \cdots, W_q)^T$ is a $q$-dimensional vector of random errors with mean zero and being independent of $\bx$.  Each response in this model is allowed to have its own regression coefficients, and to simplify the presentation, the covariate vector $\bx$ is assumed to be centered with mean zero. 
Commonly encountered is the setting of high dimensionality in both responses and covariates, where the numbers of responses and covariates, $q$ and $p$, can diverge with the sample size. It is of practical importance to consider sparse models in which the rows of the coefficient matrices $\bB_{\bx}$ and $\bB_{\bz}$ are sparse with only a 
fraction of nonzeros. We aim at identifying the important interactions and main effects, which have nonzero regression coefficients, that contribute to the responses.
%

Interaction identification in the multi-response interaction model \eqref{eq: model} with large $p$ and $q$ 
{is} intrinsically challenging. The difficulties include the high dimensionality in responses, the high computational cost caused by the existence of a large number of interactions among covariates, and the technical challenges associated with the complex model structure. The idea of variable screening can speed up the computation. Yet, under model setting \eqref{eq: model} most existing variable screening methods based on the marginal correlation may no longer work. To appreciate this, let us consider a specific case of model \eqref{eq: model} with only one response
   \begin{eqnarray}\label{eq: simple-model0}
     Y = \alpha + \sum_{j=1}^p\beta_jX_j+\sum_{k=1}^{p-1}\sum_{\ell=k+1}^p\gamma_{k\ell}X_kX_{\ell} + W,
   \end{eqnarray}
where all the notation is the same as therein with $\bB_{\bx} = (\beta_j)_{1 \leq j \leq p}$ and $\bB_{\bz} = (\gamma_{k\ell})_{1\leq k \leq p -1,\, k+1 \leq \ell \leq p}$. For simplicity, assume that the covariates $X_1,\cdots, X_p$ are independent of each other. Then under the above model setting (\ref{eq: simple-model0}), it is easy to see that
\begin{eqnarray}\label{eq: conditional-mean}
E(Y|X_j) = \alpha + \beta_j X_j.
  \end{eqnarray}
This representation shows that if some covariate $X_j$ is an unimportant main effect with $\beta_j=0$, then the conditional mean of $Y$ given $X_j$ is free of $X_j$, regardless of whether $X_j$ contributes to interactions or not.
When such a covariate $X_j$ indeed appears in an important interaction, variable screening methods based on the marginal correlations of $Y$ and $X_k$'s are not capable of detecting $X_j$ if the heredity assumption fails to hold. As a consequence, there is an important need for new proposals 
of interaction screening. When the covariates are correlated, the conditional mean \eqref{eq: conditional-mean} may depend on $X_j$ indirectly through correlations with other covariates when $\beta_j=0$. Such a relationship can, however, be still weak if the correlations between $X_j$ and other covariates are weak.


To address the aforementioned challenges, we suggest a new two-stage approach to interaction identification, named the interaction pursuit via distance correlation (IPDC), exploiting the idea of interaction screening and selection.
In the screening step, we first transform the responses and covariates and then perform variable screening based on the transformed 
responses and covariates. Such a transformation enhances the dependence of responses on covariates that contribute to important interactions or main effects. The novelty of our interaction screening method is that it aims at recovering variables that contribute to important interactions instead of finding these interactions directly, which reduces the computational cost substantially from a factor of $O(p^2)$ to $O(p)$. To take advantage of the correlation structure among multiple responses, we build our marginal utility function using the distance correlation proposed in \cite{szekely:Rizzo:Bakirov:2007}.
After the screening step, we conduct interaction selection by constructing pairwise interactions with the retained variables from the first step, and applying the group regularization method to further select important interactions and main effects for the multi-response model in the reduced feature space.

The main contributions of this paper are twofold.  First, the suggested IPDC method provides a computationally efficient approach to interaction screening and selection in ultra-high dimensional interaction models. Such a procedure accommodates the model setting with a diverging number of responses, and is generally applicable without the requirement of the heredity assumption. Second, our procedure is theoretically justified to be capable of retaining all covariates that contribute to important interactions or main effects with asymptotic probability one, the so-called sure screening property \citep{Fan:Lv:2008, Lv2013}, in the screening step. In the selection step, it is also shown to enjoy nice sampling properties for both interactions and main effects such as the support union recovery and oracle inequalities in prediction and estimation. In particular, there are two key messages that are delivered in this paper: a separate screening step for interactions can significantly enhance the screening performance if one aims at finding important interactions, and screening interaction variables can be more effective and efficient than screening interactions directly due to the noise accumulation. The former message is elaborated more with a numerical example presented in Section \ref{newsimu.dcsis2} of the Supplementary Material.

The rest of the paper is organized as follows. Section \ref{sec: screening} introduces the new interaction screening approach and studies its theoretical properties. We illustrate the advantages of the proposed procedure using several simulation studies in Section \ref{sec: simulation} and a real data example in Section \ref{sec: RealData}. Section \ref{sec: discussion} discusses some possible extensions of our method. The proofs of main results are relegated to the Appendix. Additional proofs of main results and technical details as well as additional numerical studies are provided in the Supplementary Material.

\section{A new interaction screening approach}\label{sec: screening}

\subsection{Motivation of the new method} \label{Sec2.1}
To facilitate the presentation, we call $X_k X_{\ell}$ an important interaction if the corresponding row of $\bB_{\bz}$ is nonzero, and $X_k$ an active interaction variable if there exists some 
{$1\leq \ell\neq k\leq p$} such that $X_kX_{\ell}$ is an important interaction. Denote by $\mathcal{I}$ the set of all important interactions. Similarly, $X_j$ is referred to as an important main effect if its associated row of $\bB_{\bx}$ is nonzero. It is of crucial importance to identify both the set $\mathcal{A}$ of all active interaction variables and the set $\mathcal{M}$ of all important main effects.

Before presenting our main ideas, let us revisit the specific example \eqref{eq: simple-model0} discussed in the Introduction. A phenomenon mentioned there is that variable screening methods using the marginal correlations between the response and covariates can fail to detect active interaction variables that have no main effects. We now consider the square transformation for the response. 
Some standard calculations (see Section \ref{AppD} of the Supplementary Material) yield
\begin{align*}
E& (Y^2|X_j)  = \big[\beta_j^2 + \sum\nolimits_{k=1}^{j-1}\gamma_{kj}^2E(X_k^2) + \sum\nolimits_{\ell=j+1}^p\gamma_{j\ell}^2E(X_{\ell}^2)\big]X_j^2 \\
& \quad\quad+2\big[\beta_j\alpha+\sum\nolimits_{k=1}^{j-1}\beta_k\gamma_{kj}E(X_k^2) +\sum\nolimits_{\ell=j+1}^{p}\beta_{\ell}\gamma_{j\ell}E(X_{\ell}^2)\big]X_j+C_j,
\end{align*}
where $C_j$ is some constant that is free of $X_j$. This shows that the conditional mean $E(Y^2|X_j)$ is linear in $X_j^2$ as long as $X_j$ is an active interaction variable, that is, $\gamma_{kj}$ or $\gamma_{j\ell} \neq 0$ for some $k$ or $\ell$, regardless of whether it is also an important main effect or not. In fact, we can see from the above representation that the coefficient of $X_j^2$ reflects the cumulative contribution of covariate $X_j$ to response $Y$ as both an interaction variable and a main effect.

Motivated by the above example, we consider the approach of screening interaction variables via some marginal utility function for the transformed variables $Y^2$ and $X_j^2$, with the square transformation applied to both the response and covariates. Such an idea has been exploited in \cite{fan2014ip} for interaction screening in the setting of single-response interaction models. To rank the relative importance of features, they calculated the Pearson correlations between $Y^2$ and $X_j^2$. This idea is, however, no longer applicable when there are multiple responses, since the Pearson correlation is not well defined for the pair of $q$-vector $\by$ of responses with $q > 1$ and covariate $X_j$. A naive strategy is to screen the interaction variables for each response $Y_k$ with $1 \leq k \leq q$ using the approach of \cite{fan2014ip}. Such a naive procedure can suffer from several potential drawbacks. First, it may be inefficient and can result in undesirable results since the correlation structure among the responses $Y_1,\cdots, Y_q$ is completely ignored. Second, when $q$ is large it may retain too many interaction variables in total, which can in turn cause difficulty in model interpretation and high computational cost when further selecting active interaction variables.

To address the afore-discussed issues, we propose to construct the marginal utility function exploiting the distance correlation introduced in \cite{szekely:Rizzo:Bakirov:2007}. More specifically, we identify the set of all active interaction variables $\mathcal{A}$ by ranking the distance correlations 
between the squared covariates $X_j^2$ and the squared response vector $\by\circ\by$, where $\circ$ denotes the Hadamard (componentwise) product of two vectors.
The distance correlation
\begin{equation*}
      \dcorr(\bu, \bv)=\frac{\dcov(\bu, \bv)}{\sqrt{\dcov(\bu, \bu)\,\dcov(\bv, \bv)}}
   \end{equation*}
is well defined for any two random vectors $\bu\in\mathbb{R}^{d_u}$ and $\bv\in\mathbb{R}^{d_v}$ of arbitrary mixed dimensions, where the distance covariance between $\bu$ and $\bv$ is given by
   \begin{equation*}
            \dcov^2(\bu, \bv)
        =\frac{1}{c_{d_u}c_{d_v}}\int_{\mathbb{R}^{d_u+d_v}}
          \frac{|\varphi_{\bu, \bv}(\bs, \bt)-\varphi_{\bu}(\bs)\varphi_{\bv}(\bt)|^2}{\|\bs\|^{d_u+1}\|\bt\|^{d_v+1}} d\bs d\bt.
   \end{equation*}
Here $c_{m}=\pi^{(m+1)/2}/\Gamma\{(m+1)/2\}$ is the half area of the
unit sphere $S^{m} \subset \mathbb{R}^{m+1}$, $\varphi_{\bu, \bv}(\bs, \bt)$, $\varphi_{\bu}(\bs)$, and $\varphi_{\bv}(\bt)$ are the characteristic functions of $(\bu, \bv)$, $\bu$, and $\bv$, respectively, and $\|\cdot\|$ denotes the Euclidean norm. Compared to the Pearson correlation, it also has the advantage that the distance correlation of two random vectors is zero if and only if they are independent. Moreover, the distance correlation of two univariate Gaussian random variables is a strictly increasing function of the absolute value of the Pearson correlation between them. See \cite{szekely:Rizzo:Bakirov:2007} for more properties and discussions of the distance correlation, and \cite{HuoSzekely2015} for a fast algorithm for computing the distance correlation.


\begin{figure}[ht]
\vspace{-0.1in}
\centering
\includegraphics[scale=0.31]{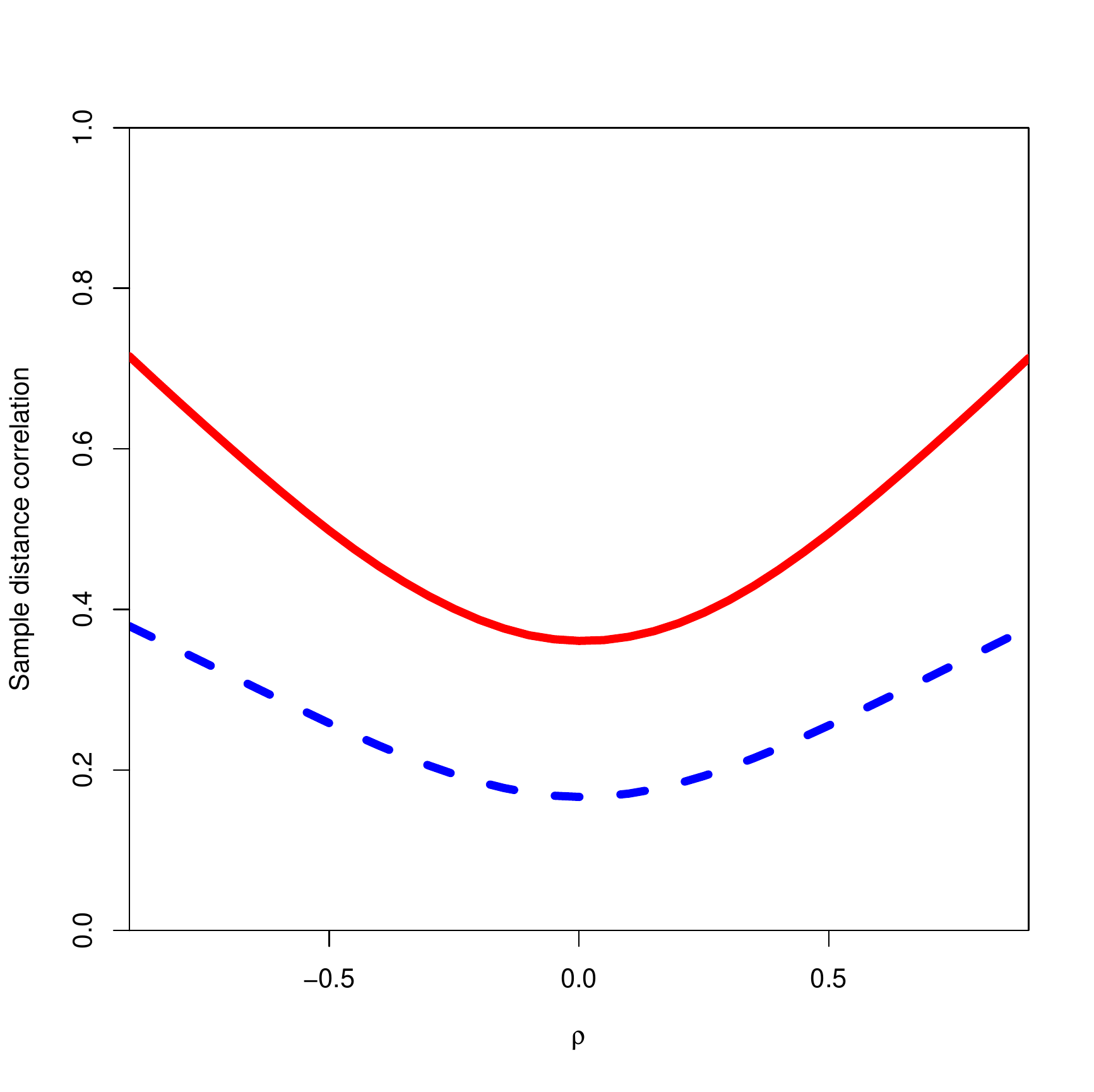}%
\includegraphics[scale=0.31]{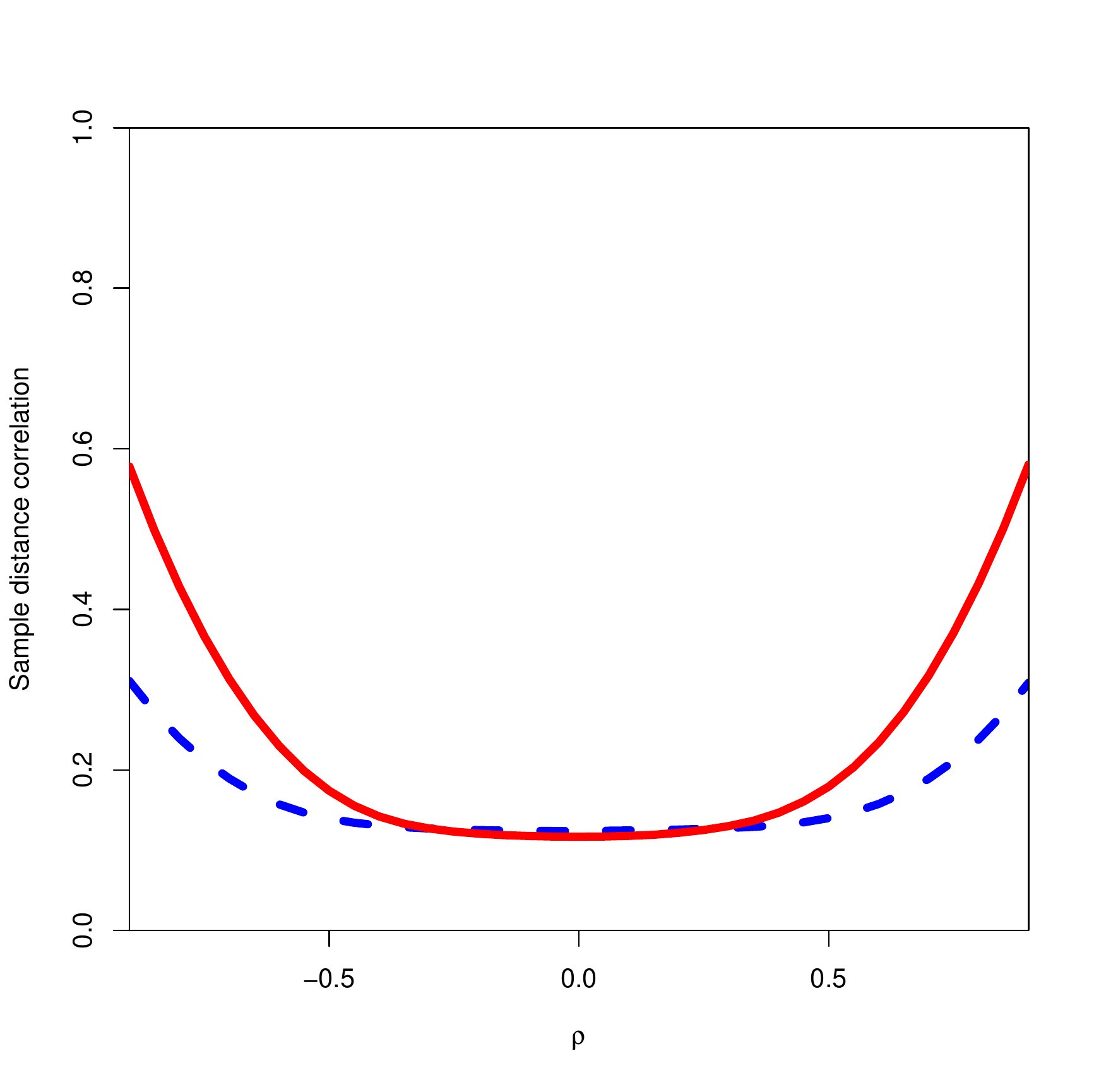} \\%
\vspace{-0.2in}
\includegraphics[scale=0.31]{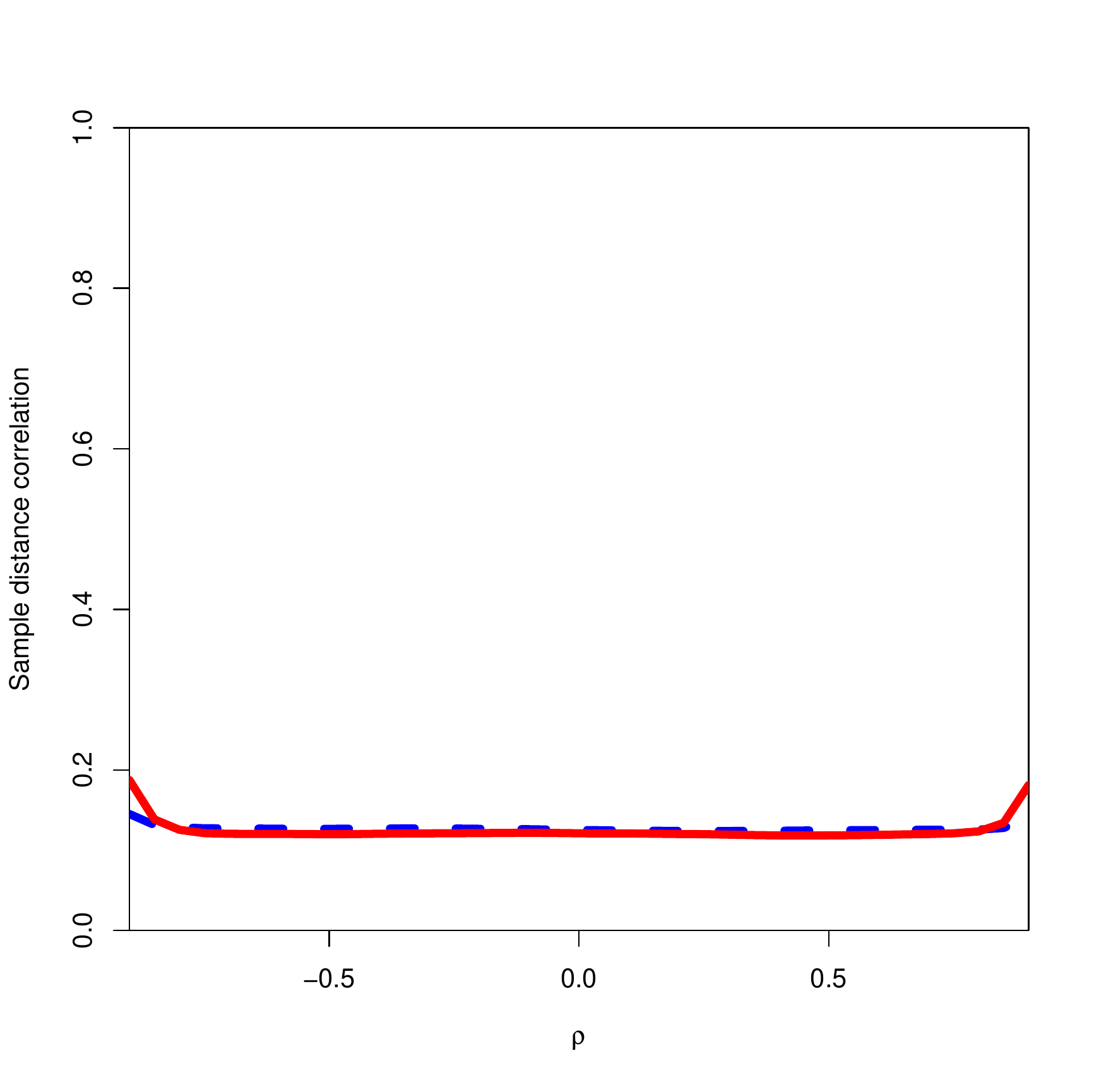}%
\includegraphics[scale=0.31]{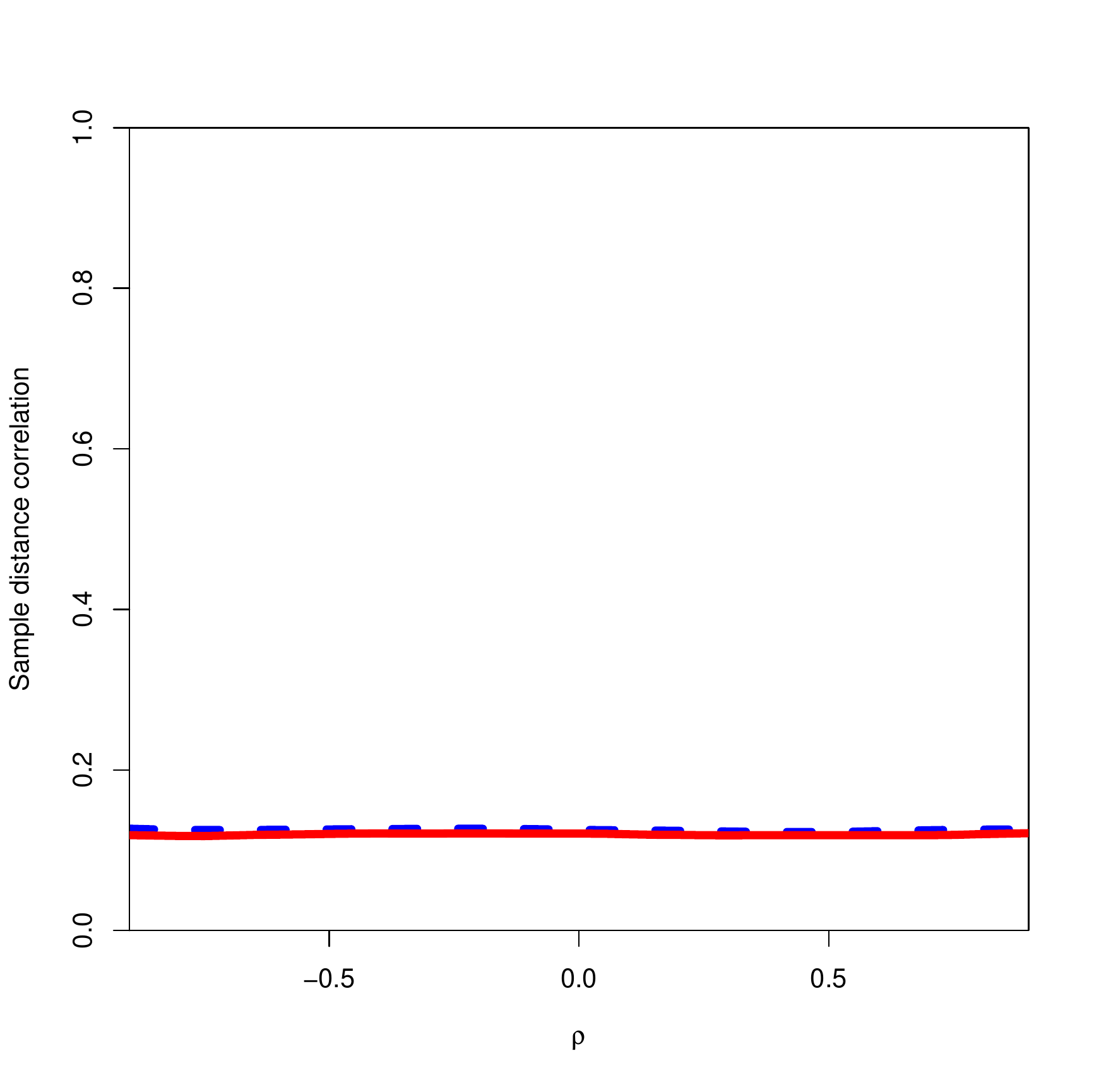}%
\vspace{-0.2in}
 \caption{Plots of sample distance correlations as a function of correlation level $\rho$ based on model (\ref{eq: example-1}). Top left: $\widehat{\dcorr}(X_1^2, Y^2)$ (solid) and $\widehat{\dcorr}(X_1, Y)$  (dashed); top right: $\widehat{\dcorr}(X_3^2, Y^2)$ (solid) and $\widehat{\dcorr}(X_3, Y)$  (dashed); bottom left: $\widehat{\dcorr}(X_{10}^2, Y^2)$ (solid) and $\widehat{\dcorr}(X_{10}, Y)$  (dashed); bottom right: $\widehat{\dcorr}(X_{1000}^2, Y^2)$ (solid) and $\widehat{\dcorr}(X_{1000}, Y)$  (dashed). 
}
\label{fig1}
\end{figure}

It is worth mentioning that \cite{li2012feature} introduced a model-free feature screening procedure based on the distance correlations of the original response and covariates. Their method is applicable to the cases of multiple responses and grouped covariates. Yet we found that the use of distance correlations for the transformed response vector and covariates, $\by \circ \by$ and $X_j^2$, can result in improved performance in interaction variable screening. The specific example considered in Section \ref{Sec2.1} provides some intuitive explanation of this phenomenon. To further illustrate this point, we generated 200 
data sets from the following
simple interaction model
  \begin{eqnarray}\label{eq: example-1}
       Y = X_1X_2 +  W,
   \end{eqnarray}
where the covariate vector $\mathbf{x}=(X_1, \cdots, X_p)^T \sim N(\bzero, \bSig)$ with $p=1000$ and $\bSig = (\rho^{|j-k|})_{1\leq j,k \leq p}$, $\rho$ ranging in $(-1, 1)$ 
measures the correlation level among covariates, and the random error $W \sim N(0,1)$ is independent of $\mathbf{x}$.  As shown in Figure \ref{fig1}, 
the sample distance correlation between $X_1^2$ and $Y^2$ 
is much larger than that between $X_1$ and $Y$. 
For covariates having weak correlation with active interaction variables $X_1$ and $X_2$, such as $X_{10}$ and $X_{1000}$, the square transformation does not increase their distance correlations with the response. The numerical studies in Sections \ref{sec: simulation} and \ref{sec: RealData} also confirm the advantages of our method over the procedure in \cite{li2012feature}.


\subsection{Interaction screening}

Suppose we have a sample $(\by_i, \bx_i)_{i=1}^n$ of $n$ independent and identically distributed (i.i.d.) observations from $(\by, \bx)$ in the multi-response interaction model (\ref{eq: model}).  For each $1\leq k\leq p$, denote by $\dcorr(X_k^2, \by\circ\by)$ the distance correlation between the squared covariate $X_k^2$ and squared response vector $\by\circ\by$. 
The idea of the screening step of our IPDC procedure is to rank the importance of the interaction variables $X_k$ using the sample version of distance correlations $\dcorr(X_k^2, \by\circ\by)$. Similarly, we conduct screening of main effects based on the sample version of distance correlations $\dcorr(X_j, \by)$ between covariates $X_j$ and response vector $\by$.


For notational simplicity, we write $X_{k}^{\ast}=X_{k}^2$,  $\widetilde{\by}=\by/\sqrt{q}$, and $\by^{\ast}=\widetilde{\by}\circ\widetilde{\by}=\by\circ\by/q$.  Define two population quantities
 \begin{align}
      \omega_k^{\ast}=\frac{\dcov^2(X_k^{\ast}, \by^{\ast})}{\sqrt{\dcov^2(X_k^{\ast}, X_k^{\ast})}}
     \quad\mbox{and}\quad
       \omega_j=\frac{\dcov^2(X_j, \widetilde{\by})}{\sqrt{\dcov^2(X_j, X_j)}}
 \end{align}
with $1\leq k, j\leq p$ for interaction variables and main effects, respectively.  Denote by $\widehat{\omega}_k^{\ast}$ and $\widehat{\omega}_j$ the empirical versions of $\omega_k^{\ast}$ and $\omega_j$, respectively, constructed by plugging in the corresponding sample distance covariances based on the sample $(\by_i, \bx_i)_{i=1}^n$. 
According to \cite{szekely:Rizzo:Bakirov:2007}, the sample distance covariance between any two random vectors $\bu$ and $\bv$ based on a sample $(\bu_i, \bv_i)_{i=1}^n$ is given by
\begin{equation*}
\widehat{\dcov}^2(\bu, \bv)=\widehat{S}_1+\widehat{S}_2-2\widehat{S}_3,
\end{equation*}
where the three quantities are defined as $\widehat{S}_1=n^{-2}\sum\nolimits_{i, j=1}^n\|\bu_i-\bu_j\|\|\bv_i-\bv_j\|$, $\widehat{S}_2=[n^{-2}\sum\nolimits_{i, j=1}^n\|\bu_i-\bu_j\|][n^{-2}\sum\nolimits_{i, j=1}^n\|\bv_i-\bv_j\|]$, and $\widehat{S}_3=n^{-3}\sum\nolimits_{i, j,k=1}^n\|\bu_i-\bu_k\|\|\bv_j-\bv_k\|$.
In view of
 \begin{align*}
    \dcorr^2(X_k^2, \by\circ\by)=  \dcorr^2(X_k^{\ast}, \by^{\ast})=\omega_k^{\ast}/\{\dcov^2(\by^{\ast}, \by^{\ast})\}^{1/2}
 \end{align*}
 and
  \begin{align*}
    \dcorr^2(X_j, \by)=\dcorr^2(X_j, \widetilde{\by})= \omega_j/\{\dcov^2(\widetilde{\by}, \widetilde{\by})\}^{1/2},
 \end{align*}
the procedure of screening the interaction variables and main effects via distance correlations $\dcorr(X_k^2, \by\circ\by)$ and $\dcorr(X_j, \by)$ suggested above is equivalent to that of thresholding the quantities $\omega_k^{\ast}$'s and $\omega_j$'s, respectively.

More specifically, in the screening step of IPDC we estimate the sets of important main effects $\mathcal{M}$ and active interaction variables $\mathcal{A}$ as
 \begin{equation}\label{eq: set}
 \widehat{\mathcal{M}}=\{1\leq j\leq p: \widehat{\omega}_j\geq \tau_1\}
       \quad \text{ and } \quad
      \widehat{\mathcal{A}}=\{1\leq k\leq p: \widehat{\omega}_k^{\ast}\geq \tau_2\},
 \end{equation}
where $\tau_1$ and $\tau_2$ are some positive thresholds. With the set $\widehat{\mathcal{A}}$ of retained interaction variables, we construct a set of pairwise interactions
 \begin{eqnarray}\label{eq: set-I}
      \widehat{\mathcal{I}}=\{(k,\,l): 1\leq k<l\leq p \mbox{ and } k,l\in \widehat{\mathcal{A}}\}.
 \end{eqnarray}
This gives a new interaction screening procedure. It is worth mentioning that the set of constructed interactions $\widehat{\mathcal{I}}$ tends to overestimate the set of all important interactions $\mathcal{I}$ since the goal of the first step of IPDC is screening interaction variables. Such an issue can be addressed in the selection step of IPDC investigated in Section \ref{sec: selection}.

\subsection{Sure screening property}


We now study the sampling properties of the newly proposed
interaction screening procedure. Some mild regularity conditions are needed for our analysis. 

\begin{condition}\label{con: lower-bound}
Both $\dcov(X_k, X_k)$ and $\dcov(X_k^2, X_k^2)$ are bounded away from zero uniformly in $k$. 
\end{condition}

\begin{condition}\label{con: tail}
There exists some constant $c_0 > 0$ such that $E\{\exp(c_0X_{k}^2)\}$ and $E\{\exp(c_0\|\by\|/\sqrt{q})\}$ are uniformly bounded.
\end{condition}

\begin{condition}\label{con: min-signal}
There exist some constants $c_1,c_2>0$ and $0\leq \kappa_1, \kappa_2<1/2$
such that $\min_{j\in\mathcal{M}}\omega_j \ge 3c_1n^{-\kappa_1}$ and $\min_{k\in\mathcal{A}}\omega_k^{\ast} \ge 3c_2n^{-\kappa_2}$.
\end{condition}

Condition \ref{con: lower-bound} is a basic assumption requiring that the distance variances of covariates $X_k$ and squared covariates $X_k^2$ are at least of a constant order. Conditions \ref{con: tail} and  \ref{con: min-signal} are analogous to the regularity conditions in \cite{li2012feature}. In particular, Condition \ref{con: tail} controls the tail behavior of the covariates and responses, which facilitates the derivation of deviation probability bounds.  
Condition \ref{con: min-signal} also shares the same spirit as Condition 3 in \cite{Fan:Lv:2008}, and can be understood as an assumption on the minimum signal strength in the feature screening setting.
Smaller constants $\kappa_1$ and $\kappa_2$ correspond to stronger marginal signal strength for active interaction variables and important main effects, respectively. With these regularity conditions, we establish the sure screening property of IPDC in the following theorem.


\begin{theorem} \label{thm:Screening}
Under Conditions \ref{con: lower-bound}--\ref{con: tail} with $\log p = o(n^{\eta_0})$ for $\eta_0 = \min\{(1-2\kappa_1)/3,\, (1-2\kappa_2)/5\}$,
there exists some positive constant $C$ such that
\begin{align}
\label{eq:omega-main-bound}
P \left(\max_{1\leq j\leq p}|\widehat\omega_j - \omega_j| \ge c_1n^{-\kappa_1}\right)
   & \le O\left(
           \exp\big\{-Cn^{(1-2\kappa_1)/6}\big\}\right),  \\
           \label{eq:omega-inter-bound}
P \left(\max_{1\leq k\leq p}|\widehat\omega_k^{\ast} - \omega_k^{\ast}| \ge c_2n^{-\kappa_2}\right)
   & \le O\left(\exp\big\{-Cn^{(1-2\kappa_2)/10}\big\}\right).
\end{align}
Assume in addition that Condition \ref{con: min-signal} holds and set $\tau_1=2c_1n^{-\kappa_1}$ and $\tau_2=2c_2n^{-\kappa_2}$. 
Then we have
\begin{equation} \label{neweq001}
    P \left(\mathcal{M} \subset \widehat{\mathcal{M}} \ \mbox{ and } \ \mathcal{I} \subset \widehat{\mathcal{I}} \right)
   = 1 - O\left\{\exp\big(-C n^{\eta_0/2}\big)\right\}.
\end{equation}
\end{theorem}

Theorem \ref{thm:Screening} reveals that the IPDC enjoys the sure screening property that all active interaction variables and all important main effects can be retained in the reduced model with high probability. In particular, we see that it can handle ultra-high dimensionality with $\log p = o(n^{\eta_0})$.
A comparison of the deviation probability bounds in \eqref{eq:omega-main-bound} and \eqref{eq:omega-inter-bound} shows that interaction screening is generally more challenging and thus needs more restrictive constraint on dimensionality $p$ than main effect screening; see the probability bound (\ref{eq: omgea-star}) and its main effect counterpart for more details. It is also seen that when the marginal signal strength for interactions and
main effects becomes stronger, the sure screening property of IPDC holds for higher dimensionality $p$.


For any feature screening procedure, it is of practical importance to control the dimensionality of the reduced feature space, since feature selection usually follows the screening for further selection of important features in such a space. 
We next investigate such an aspect for IPDC. Let $s_1$ and $s_2$ be the cardinalities of sets of all important main effects $\mathcal{M}$ and all active interaction variables $\mathcal{A}$,
respectively. With the thresholds $\tau_1=2c_1n^{-\kappa_1}$ and $\tau_2=2c_2n^{-\kappa_2}$ specified in Theorem \ref{thm:Screening}, we introduce two sets of unimportant main effects and inactive interaction variables
 \begin{equation}\label{neweq002}
 \mathcal{M}_1=\{j\in {\cal{M}}^c: \omega_j \ge c_1n^{-\kappa_1}\}
       \ \text{ and } \
      \mathcal{A}_1=\{k\in {\cal{A}}^c: \omega_k^{\ast} \ge c_2n^{-\kappa_2}\}
 \end{equation}
that are of significant marginal effects. Denote by $s_3$ and $s_4$ the cardinalities of these two sets $\mathcal{M}_1$ and $\mathcal{A}_1$, respectively. Larger values of $s_3$ and $s_4$ indicate more difficulty in the problem of interaction and main effect screening in the high-dimensional multi-response interaction model (\ref{eq: model}).

\begin{theorem} \label{thm:Size}
Assume that all the conditions of Theorem \ref{thm:Screening} hold and set $\tau_1=2c_1n^{-\kappa_1}$ and $\tau_2=2c_2n^{-\kappa_2}$. Then we have
  \begin{align} \label{neweq003}
      P & \left\{|\widehat{\cal{M}}|\leq s_1+s_3 \ \mbox{ and } \ |\widehat{\cal{I}}|\leq (s_2+s_4)(s_2+s_4-1)/2\right\} \\
      \nonumber
      & \quad = 1 - O\left\{\exp\big(-C n^{\eta_0/2}\big)\right\}
  \end{align}
 for some positive constant $C$.
\end{theorem}

Theorem \ref{thm:Size} quantifies how the size of the reduced model for interactions and main effects is related to the
thresholding parameters $\tau_1$ and $\tau_2$, and the cardinalities of the two sets $\mathcal{M}_1$ and $\mathcal{A}_1$. In particular, we see that when $s_i=O(n^{\delta_i})$ with some constants $\delta_i \geq 0$ for $1 \leq i \leq 4$, the total number of retained interactions and main effects in the reduced feature space can be controlled as $O(n^{\delta})$ with $\delta = \max\{\delta_1 \vee \delta_3, 2(\delta_2 \vee \delta_4)\}$, where $\vee$ denotes the maximum of two values. In contrast, the dimensionality $p$ is allowed to grow nonpolynomially with sample size $n$ in the rate of $\log p = o(n^{\eta_0})$ with $\eta_0 = \min\{(1-2\kappa_1)/3,\, (1-2\kappa_2)/5\}$. The reduced model size can fall below the sample size and be a smaller order of $n$ when both $\max\{\delta_1, \delta_3\} < 1$ and $\max\{\delta_2, \delta_4\} < 1/2$ are satisfied. The post-screening interaction selection and its sampling properties are further investigated in Section \ref{sec: selection} of the Supplementary Material.

\section{Simulation studies}\label{sec: simulation}
We illustrate the finite-sample performance of our method using several simulation examples. Two sets of models are considered for the single-response case and the multi-response case, respectively. 
This section evaluates the screening performance, while the post-screening selection performance is investigated in Section \ref{Sec.Selection} of the Supplementary Material.


\subsection{Screening in single-response models} \label{Sec4.1.1}

We begin with the following four high-dimensional single-response interaction models:
   \begin{eqnarray*}
        && \mbox{Model 1:} \ Y =2X_1+2X_2 + X_1X_2+W,  \\
        && \mbox{Model 2:} \ Y=2X_1 + 3X_1X_2 + 3X_1X_3 +W,  \\
        && \mbox{Model 3:} \ Y=3X_1X_2 + 3X_1X_3 +W,  \\
        && \mbox{Model 4:} \ Y = 3 \mathbb{I}(X_{12}\geq 0) + 2 X_{22} + 3 X_1 X_2 + W,
   \end{eqnarray*}
where all the notation is the same as in (\ref{eq: model}) and $\mathbb{I}(\cdot)$ denotes the indicator function. 
The covariate vector $\bx = (X_1, \cdots, X_p)^T$ is sampled from the distribution $N(\textbf{0}, \bSig)$ with covariance matrix $\bSig = (\rho^{|j-k|})_{1 \leq j,k \leq p}$ for $\rho\in (-1,1)$, and the error term $W \sim N(0, 1)$ is generated independently of $\bx$ to form an i.i.d. sample of size $n = 200$.
For each of the four models, we further consider three different settings with $(p, \rho)=(2000, 0.5)$, $(5000, 0.5)$, and $(2000, 0.1)$, respectively. In particular, Models 2 and 3 are adapted from simulation scenarios 2.2 and 2.3 in 
\citet{jiang2014sliced}, whereas Model 4 is adapted from simulation example 2.b of 
\citet{li2012feature} and accounts for model misspecification since without any prior information, our working model treats $X_{12}$ as a linear predictor instead of $\mathbb{I}(X_{12}\geq 0)$.
We see that Model 1 satisfies the strong heredity assumption 
and Model 2 obeys the weak heredity assumption, 
while Models 3 and 4 violate the heredity assumption since none of the active interaction variables are important main effects.


We compare the interaction and main effect screening performance of the IPDC with the SIS \citep{Fan:Lv:2008}, DCSIS \citep{li2012feature}, SIRI \citep{jiang2014sliced}, IP \citep{fan2014ip}, and iFORT and iFORM \citep{hao2014interaction}. Like IPDC, SIRI and IP were developed for screening interaction variables and main effects separately. In particular, SIRI is an iterative procedure, while all others are non-iterative ones.  For a fair comparison, we
adopt the initial screening step described in Section 2.3 of 
\citet{jiang2014sliced} to implement SIRI in a non-iterative fashion, and keep the top ranked covariates. Since the SIS is originally designed only for main effect screening and the original DCSIS screens variables without the distinction between main effects and interaction variables, 
we construct pairwise interactions based on the covariates recruited by SIS and DCSIS, and refer to the resulting procedures as SIS2 and DCSIS2, respectively, to distinguish them from the original ones. It is worth mentioning that the SIS2 shares a similar spirit to the TS-SIS procedure proposed in \cite{LiZhongLiWu2014}, where the difference is that the latter constructs pairwise interactions between the main effects retained by SIS   and all $p$ covariates.
Following the suggestions of 
\citet{Fan:Lv:2008} and \citet{li2012feature}, we keep the top $[n/(\log n)]$ variables after ranking for each
screening procedure.  We examine the screening performance by the proportions of important main effects, important interactions, and all of them being retained by each screening procedure over 100 replications. 

\begin{table}
\caption{Proportions of important main effects, important interactions, and all of them retained by different screening methods. For SIS2, DCSIS2, and SIRI, interactions are constructed using the top $[n/(\log n)]$ covariates ranked by their marginal utilities with the response. \vspace{0.1in}} \label{tab:Screen}
\centering
\scalebox{0.7}{
\begin{tabular}{ l c c c c c c c c ccccccc }
  \toprule
   Method & \multicolumn{4}{c}{Model 1} & \multicolumn{4}{c}{Model 2} & \multicolumn{3}{c}{Model 3}  & \multicolumn{4}{c}{Model 4} \\
    \cmidrule(lr{.75em}){2-5}  \cmidrule(lr{.75em}){6-9} \cmidrule(lr|{.75em}){10-12}  \cmidrule(lr|{.75em}){13-16}
    & $X_1$ & $X_2$ &  $X_1X_2$  & All & $X_1$ & $X_1X_2$ & $X_1X_3$  & All & $X_1X_2$ & $X_1X_3$ & All  & $X_{12}$ & $X_{22}$ & $X_{1}X_{2}$ & All  \\
  \midrule
    \multicolumn{16}{c}{{Setting 1}: $(p, \rho)=(2000, 0.5)$} \\
  SIS2      & 1.00  & 1.00  & 1.00  & 1.00   & 0.95 & 0.48 & 0.23 & 0.20   & 0.08 & 0.04 & 0.04 &   0.93  & 1.00 & 0.05 & 0.05\\
  iFORT & 1.00  & 1.00  & 1.00  & 1.00   & 0.67 & 0.00 & 0.00 & 0.00   & 0.00 & 0.00 & 0.00 &   0.56  & 1.00 & 0.00 & 0.00\\
  iFORM & 1.00  & 1.00  & 1.00  & 1.00   & 0.85 & 0.07 & 0.02 & 0.00   & 0.00 & 0.00 & 0.00 &   0.53  & 1.00 & 0.00 & 0.00\\
  DCSIS2 & 1.00  & 1.00  & 1.00  & 1.00   & 1.00 & 1.00 & 0.91 & 0.91   & 1.00 & 0.68 & 0.68 &   0.99  & 1.00 & 0.92 & 0.91\\
  SIRI     & 1.00  & 1.00  & 1.00  & 1.00   & 1.00 & 0.99 & 0.88 & 0.87   & 1.00 & 0.73 & 0.73 &   0.89  & 1.00 & 0.86 & 0.78\\
  IP     & 1.00  & 1.00  & 1.00  & 1.00   & 0.95 & 1.00 & 0.87 & 0.83   & 1.00 & 0.90 & 0.90 &   0.93  & 1.00 & 1.00 & 0.93\\
  IPDC  & 1.00  & 1.00  & 1.00  & 1.00   & 1.00 & 1.00 & 0.99 & 0.99   & 1.00 & 0.99 & 0.99 &   0.99  & 1.00 & 1.00 & 0.99\\\vspace{-0.05in}\\
    \multicolumn{16}{c}{{Setting 2}: $(p, \rho)=(5000, 0.5)$} \\
  SIS2      & 1.00  & 1.00  & 1.00  & 1.00   & 0.94 & 0.39 & 0.17 & 0.15   & 0.03 & 0.00 & 0.00 &   0.86  & 1.00 & 0.03 & 0.02\\
  iFORT & 1.00  & 1.00  & 1.00  & 1.00   & 0.62 & 0.00 & 0.00 & 0.00   & 0.00 & 0.00 & 0.00 &   0.40  & 1.00 & 0.00 & 0.00\\
  iFORM & 1.00  & 1.00  & 1.00  & 1.00   & 0.85 & 0.04 & 0.01 & 0.00   & 0.00 & 0.00 & 0.00 &   0.41  & 1.00 & 0.00 & 0.00\\
  DCSIS2 & 1.00  & 1.00  & 1.00  & 1.00   & 1.00 & 0.99 & 0.80 & 0.79   & 1.00 & 0.46 & 0.46 &   0.99  & 1.00 & 0.86 & 0.85\\
  SIRI     & 1.00  & 1.00  & 1.00  & 1.00   & 1.00 & 0.99 & 0.81 & 0.80   & 1.00 & 0.63 & 0.63 &   0.83  & 1.00 & 0.84 & 0.71\\
  IP     & 1.00  & 1.00  & 1.00  & 1.00   & 0.94 & 1.00 & 0.73 & 0.69   & 1.00 & 0.85 & 0.85 &   0.86  & 1.00 & 1.00 & 0.86\\
  IPDC  & 1.00  & 1.00  & 1.00  & 1.00   & 1.00 & 1.00 & 0.96 & 0.96   & 1.00 & 0.98 & 0.98 &   0.99  & 1.00 & 1.00 & 0.99\\\vspace{-0.05in}\\
        \multicolumn{16}{c}{{Setting 3}: $(p, \rho)=(2000, 0.1)$} \\
  SIS2      & 1.00  & 1.00  & 1.00  & 1.00   & 1.00 & 0.08 & 0.04 & 0.00   & 0.02 & 0.00 & 0.00 &   0.97  & 1.00 & 0.00 & 0.00\\
  iFORT & 1.00  & 1.00  & 1.00  & 1.00   & 0.89 & 0.00 & 0.00 & 0.00   & 0.00 & 0.00 & 0.00 &   0.64  & 1.00 & 0.00 & 0.00\\
  iFORM & 1.00  & 1.00  & 1.00  & 1.00   & 0.92 & 0.06 & 0.01 & 0.01   & 0.01 & 0.00 & 0.00 &   0.62  & 1.00 & 0.00 & 0.00\\
  DCSIS2 & 1.00  & 1.00  & 1.00  & 1.00   & 1.00 & 0.71 & 0.72 & 0.55   & 0.19 & 0.11 & 0.00 &   1.00  & 1.00 & 0.06 & 0.06\\
  SIRI     & 1.00  & 1.00  & 1.00  & 1.00   & 1.00 & 0.58 & 0.58 & 0.34   & 0.35 & 0.37 & 0.16 &   0.86  & 1.00 & 0.19 & 0.15\\
  IP     & 1.00  & 1.00  & 0.99  & 0.99   & 1.00 & 0.64 & 0.64 & 0.38   & 0.79 & 0.75 & 0.58 &   0.97  & 1.00 & 0.98 & 0.95\\
  IPDC  & 1.00  & 1.00  & 1.00  & 1.00   & 1.00 & 0.80 & 0.79 & 0.62   & 0.93 & 0.90 & 0.84 &   1.00  & 1.00 & 1.00 & 1.00\\
  \midrule
\end{tabular}}
\end{table}

Table \ref{tab:Screen} reports the screening results of different methods. In Model 1, all screening methods are able to retain almost all important main effects and interactions across all three settings. The IPDC outperforms SIS2, DCSIS2, SIRI, IP, iFORT, and iFORM in Models 2--4 over all three settings. It is seen that SIS2 can barely identify important interactions for those three models. The advantage of IPDC over SIS2, DCSIS2, and SIRI is most pronounced when the heredity assumption is violated as in Models 3 and 4. We also observe significant improvement of IPDC over IP in many of those model settings. When the dimensionality increases from $2000$ to $5000$ (settings 1 and 2), the problem of interaction and main effect screening becomes more challenging as indicated by the drop of the screening probabilities. Compared to others, IPDC consistently performs well.

It is interesting to observe that in view of settings 1 and 3, the interaction screening performance can be improved in the presence of a higher level of correlation among covariates.
One possible explanation is that high correlation among covariates may increase the dependence of the response on the interaction variables and thus benefit interaction screening. For instance, in Model 2 due to the correlation between the interaction variable $X_2$ (or $X_3$) and main effect $X_1$, the response $Y$ depends on $X_2$ (or $X_3$) not only directly through the interaction $X_1X_2$ (or $X_1 X_3$) but also indirectly through the main effect $X_1$. 
Therefore, in this case high correlation among covariates can boost the performance of interaction screening. Similar phenomenon has been documented for DCSIS in the literature; see, for example, Models 1.b and 1.c in Table 2 of 
\citet{li2012feature}.

\subsection{Screening in multi-response model}\label{sec: uni-screen}

We next consider the setting of interaction model with multiple responses and specifically Model 5 with $q = 10$ responses:
{\small\begin{align*}
Y_1 & = \beta_1 X_1 + \beta_2 X_2 +  \beta_3 X_1X_2 +W_1, \ Y_2 = \beta_4 X_1 + \beta_5 X_2 +  \beta_6 X_1X_3 +W_2, \\
Y_3 & = \beta_7 X_1 + \beta_8 X_2 +  \beta_9 X_6X_7 +W_3, \  Y_4 = \beta_{10} X_1 + \beta_{11} X_2 +  \beta_{12} X_8X_9 +W_4, \\
Y_5 & = \beta_{13} X_6X_7  + \beta_{14} X_8X_9 + W_5, \ Y_6 = \beta_{15} X_1 + \beta_{16} X_2 +  \beta_{17} X_1X_2 +W_6, \\
Y_7 & = \beta_{18} X_1 + \beta_{19} X_2 +  \beta_{20} X_1X_3 +W_7, \ Y_8 = \beta_{21} X_1 + \beta_{22} X_2 +  \beta_{23} X_6X_7 +W_8, \\
Y_9 & = \beta_{24} X_1 + \beta_{25} X_2 +  \beta_{26} X_8X_9 +W_9, \ Y_{10} = \beta_{27} X_6X_7  + \beta_{28} X_8X_9 + W_{10},
\end{align*}}
\hspace{-0.13in} where all the notation and setup are the same as in Section \ref{Sec4.1.1},
{the covariate vector $\bx = (X_1, \cdots, X_p)\t$ is sampled from distribution $N(\bzero, \bSig)$ with covariance matrix $\bSig=(0.5^{|j-k|})_{1\leq j, k\leq p}$, and the error vector $\bw = (W_1, \cdots, W_q)\t \sim N(\bzero, I_q)$ is independent of $\bx$}. The nonzero regression coefficients $\beta_k$ with $1 \leq k \leq 28$ for all important main effects and interactions are generated independently as $\beta_{k} = (-1)^{U}\text{Uniform}(1, 2)$, where $U$ is a Bernoulli random variable with success probability $0.5$ and $\text{Uniform}(1, 2)$ is the uniform distribution on $[1, 2]$. 
For simplicity, we consider only the setting of $(n, p, \rho)=(100, 1000, 0.5)$. In Model 5,
covariates $X_1$ and $X_2$ are both active interaction variables and important main effects, whereas covariates $X_3$ and $X_j$ with $6 \leq j \leq 9$ are active interaction variables only.

To simplify the presentation, we examine only the proportions of active interaction variables and important main effects retained by different screening procedures.
A direct application of SIS to each response $Y_k$ with $1 \leq k \leq q$ results in $q$ marginal correlations for each covariate $X_j$ with $1 \leq j \leq p$. We thus consider two modifications of SIS to deal with multi-response data. Specifically, we exploit two new marginal measures, $\max_{1 \leq k \leq q}|\widehat{\corr}(Y_k, X_j)|$ and $\sum_{k = 1}^q|\widehat{\corr}(Y_k, X_j)|$, to quantify the importance of covariates $X_j$, where $\widehat{\corr}$ denotes the sample correlation. We refer to these two methods as SIS.max and SIS.sum, respectively. The SIRI and IP are not included for comparison in this model since both methods were not designed for multi-response models, while the DCSIS is still applicable since the distance correlation is well defined in such a scenario.


Since feature screening is more challenging in multi-response models, we implement IPDC in a slightly different fashion than in single-response models. Recall that in Section \ref{Sec4.1.1}, IPDC screens interaction variables and main effects separately, and keeps the top $[n/(\log n)]$ of each type of variables. For Model 5, we take a union of these two sets of variables and regard an active interaction variable or important main effect as being retained if such a variable belongs to the union, which can contain up to $2[n/(\log n)]$ variables. Consequently we construct pairwise interactions of all variables in the union.
To ensure fair comparison, we keep the top $2[n/(\log n)]$ variables for the other screening methods SIS.max, SIS.sum, and DCSIS.

\begin{table}
\caption{\label{tab2} Proportions of important main effects and active interaction variables retained by different screening
methods.\vspace{0.1in}}
\centering
\scalebox{0.82}{
\begin{tabular}{ l c c c c c c c }
  \toprule
Method & $X_1$ & $X_2$ & $X_3$ & $X_6$ & $X_7$ & $X_8$ &  $X_9$ \\
\midrule
SIS.max	& 1.00 (0.00)	& 1.00 (0.00)	& 0.98 (0.01)	& 0.12 (0.03)	& 0.18 (0.04)	& 0.13 (0.03)	& 0.08 (0.03) \\
SIS.sum	& 1.00 (0.00)	& 1.00 (0.00)	& 0.99 (0.01)	& 0.17 (0.04)	& 0.17 (0.04)	& 0.17 (0.04)	& 0.17 (0.04) \\
DCSIS	& 1.00 (0.00)	& 1.00 (0.00)	& 1.00 (0.00)	& 0.61 (0.05)	& 0.57 (0.05)	& 0.72 (0.05)	& 0.68 (0.05) \\
IPDC	& 1.00 (0.00)	& 1.00 (0.00)	& 0.99 (0.01)	& 0.91 (0.03)	& 0.90 (0.03)	& 0.95 (0.02)	& 0.90 (0.03) \\
  \toprule
\end{tabular}}
\end{table}

Table \ref{tab2} summarizes the screening results under Model 5. We see that all methods perform well in recovering variables $X_1$, $X_2$, and $X_3$. Yet only IPDC is able to retain active interaction variables $X_6, \cdots, X_9$ with large probability. 


\subsection{Screening in multi-response model with discrete covariates} \label{newsimu}
We now turn to the scenario of multi-response interaction model with mixed covariate types and specifically Model 6 with $q = 50$ responses and $(n, p) = (100, 1000)$:
{\small\begin{align*}
Y_1 & = \beta_1 X_1 + \beta_2 X_2 +  \beta_3 X_3 + \beta_4 X_4 + \beta_5 X_1X_2 + \beta_6 X_3X_4 +W_1, \\
Y_2 & = \beta_7 X_1 + \beta_8 X_2 +  \beta_9 X_3 + \beta_{10} X_4 + \beta_{11} X_1X_3 + \beta_{12} X_4X_5 +W_2, \\
Y_3 & = \beta_{13} X_1 + \beta_{14} X_2 +  \beta_{15} X_3 + \beta_{16} X_4 + \beta_{17} X_4X_5 + \beta_{18} X_9X_{13} +W_3, \\
Y_4 & = \beta_{19} X_1 + \beta_{20} X_2 +  \beta_{21} X_3 + \beta_{22} X_4 + \beta_{23} X_9X_{12} + \beta_{24} X_{12}X_{13} +W_4, \\
Y_5 & = \beta_{25} X_9X_{12} + \beta_{26} X_{9}X_{13} + \beta_{27} X_{12}X_{13} +W_5,
\end{align*}}
\hspace{-0.09in} and the remaining nine groups of five responses are defined in a similar way to how $Y_6, \cdots, Y_{10}$ were defined in Model 5 in Section \ref{sec: uni-screen}, that is, repeating the support of each response but with regression coefficients $\beta_{k}$ generated independently from the same distribution as in Model 5. There are several key differences with Model 5. We consider higher response dimensionality $q = 50$, higher population collinearity level $\rho = 0.8$, and larger true model sizes for the responses. The covariates $X_1, \cdots, X_p$ are sampled similarly as in Model 5, but the even numbered covariates are further discretized. More specifically, each even numbered covariate is assigned values $0, 1$, or $2$ if the original continuous covariate takes values below $0$, between $0$ and $1.5$, or above $1.5$, respectively, and then centered with mean zero.
These discrete covariates are included in the model because in real applications some covariates can also be discrete. For instance, the covariates in the single nucleotide polymorphism (SNP) data are typically coded to take values $0, 1$, and $2$.
In addition, the random errors $W_1, \cdots, W_q$ are sampled independently from the $t$-distribution with $5$ degrees of freedom. Thus Model 6 involves both non-Gaussian design matrix with mixed covariate types and non-Gaussian error vector.

We list in Table \ref{tab:newsimu} the screening performance of all the methods as in Section \ref{sec: uni-screen}. Note that the standard errors are omitted in this table to save space. Comparing these results to those in Table \ref{tab2}, we see that the problem of interaction screening becomes more difficult in this model. This result is reasonable since the scenario of Model 6 is more challenging than that of Model 5. Nevertheless IPDC still improves over other methods in retaining active interaction variables $X_9$, $X_{12}$, and $X_{13}$.

\begin{table}
\centering
\caption{Proportions of important main effects and active interaction variables retained by different screening
methods.\vspace{0.1in}} \label{tab:newsimu}
\scalebox{1}{
\begin{tabular}{ l c c c c c c c c }
  \toprule
Method & $X_1$ & $X_2$ & $X_3$ & $X_4$ & $X_5$ & $X_9$ & $X_{12}$ & $X_{13}$ \\
\midrule
SIS.max	& 1.00 & 1.00	& 1.00 & 1.00	& 1.00 & 0.65 & 0.44 & 0.24 \\
SIS.sum	& 1.00 & 1.00	& 1.00 & 1.00	& 1.00 & 0.76 & 0.45 & 0.25 \\
DCSIS	& 1.00 & 1.00	& 1.00 & 1.00	& 1.00 & 0.93 & 0.28 & 0.80 \\
IPDC	& 1.00 & 1.00	& 1.00 & 1.00	& 1.00 & 0.92 & 0.67 & 0.85 \\
  \toprule
\end{tabular}}
\end{table}

\section{Real data analysis}\label{sec: RealData}
We further evaluate the performance of our proposed procedure on 
a multivariate yeast cell-cycle data set from 
\citet{Spellmanetal:1998}, which can be accessed in the R package ``spls."
This data set has been studied in 
\citet{ChunKeles:2010} and \citet{ChenHuang:2012}. Our goal is to predict how much mRNA is produced by 542 genes related to the yeast cell's replication process. For each gene, the binding levels of 106 transcription factors (TFs) are recorded. The binding levels of the TFs play a role in determining which genes are expressed and help detail the process behind eukaryotic cell-cycles. Messenger RNA is collected for two cell-cycles for a total of eighteen time points. 
Thus this data set has sample size $n=542$, number of covariates $p=106$, and number of response $q=18$, with all variables being continuous.

Considering the relatively large sample size, we use $30\%$ of the data as training and the rest as testing, and repeat such random splitting for $100$ times. 
We follow the same screening and selection procedures as in the simulation study for the setting of multiple responses in Section \ref{sec: simulation}. 
Similarly we take a union of the set of retained interaction variables and the set of retained main effects.
For fair comparison, we keep $2[n/(\log n)] = 62$ variables in the screening procedures of SIS.max, SIS.sum, and DCSIS, and use those variables to construct pairwise interactions for the selection step.

\begin{table}
\caption{Means and standard errors (in parentheses) of prediction error as well as numbers of selected main effects and interactions for each method in yeast cell-cycle data.\vspace{0.1in}\label{tab:yeast}}
\centering
\scalebox{1}{
\begin{tabular}{ l c c c}
  \toprule
     & &  \multicolumn{2}{c}{Model Size} \\
     \cline {3-4}
  Method  & PE ($\times 10^{-3}$) & Main & Interaction  \\
   \midrule
  SIS.max-GLasso &  	224.05 (1.20)	&  73.73 (7.96) & 755.35 (61.38) \\
  SIS.sum-GLasso 	&  	223.42 (1.17)	& 50.76 (3.97)  & 764.68 (63.52) \\
  DCSIS-GLasso    & 	223.93 (1.16) &  63.67 (7.47) & 705.11 (61.46) \\
  IPDC-GLasso  	& 	220.44 (1.14)  & 113.78 (9.74)  & 801.70 (54.86) \\
    SIS.max-GLasso-Lasso &  	226.66 (1.45)	&  47.56 (3.66) & 327.32 (19.38) \\
  SIS.sum-GLasso-Lasso 	&  	225.07 (1.48)	& 50.76 (3.97)  & 319.25 (19.95) \\
  DCSIS-GLasso-Lasso    & 	226.40 (1.43) &  47.12 (3.92) & 306.18 (20.75) \\
  IPDC-GLasso-Lasso  	& 	222.43 (1.39)  & 56.08 (3.33)  & 300.93 (15.17) \\
  \midrule
\end{tabular}
}
\end{table}

Table \ref{tab:yeast} presents the results on the prediction error and selected model size. Paired $t$-tests of prediction errors on the $100$ splits of IPDC-GLasso against SIS.max-GLasso, SIS.sum-GLasso, and DCSIS-GLasso result in $p$-values $2.86\times10^{-11}$, $1.70\times10^{-13}$, and $1.15\times10^{-14}$, respectively. Moreover, paired $t$-tests of prediction errors on the $100$ splits of IPDC-GLasso-Lasso against SIS.max-GLasso-Lasso, SIS.sum-GLasso-Lasso, and DCSIS-GLasso-Lasso give $p$-values $2.92\times10^{-4}$, $2.30\times10^{-2}$, and $9.73\times10^{-5}$, respectively. These results show significant improvement of our method over existing ones.

\section{Discussions}\label{sec: discussion}


We have investigated the problem of interaction identification in the setting where the numbers of responses and covariates can both be large. Our suggested two-stage procedure IPDC provides a scalable approach with the idea of interaction screening and selection. It exploits the joint information among all the responses by using the distance correlation in the screening step and the regularized multi-response regression in the selection step. One key ingredient is the use of the square transformation to responses and covariates for effective interaction screening. The established sure screening and model selection properties enable its broad applicability beyond the heredity assumption. 

Although we have focused our attention on the square transformation of the responses and covariates due to its simplicity and the motivation discussed in Section \ref{Sec2.1}, it is possible that other functions can also work for the idea of IPDC. It would be interesting to investigate and characterize what class of functions is optimal for the purpose of interaction screening.

Like all independence screening methods using the marginal utilities including the SIS and DCSIS, our feature screening approach may fail to identify some important interactions or main effects that are marginally weakly related to the responses. One possible remedy is to exploit the idea of the iterative SIS proposed in \cite{Fan:Lv:2008} which has been shown to be capable of ameliorating the SIS. Recently, 
\citet{zhong2014iterative} also introduced an iterative DCSIS procedure and demonstrated that it can improve the finite-sample performance of the DCSIS.
The theoretical properties of these iterative feature screening approaches are, however, less well understood.
It would be interesting to develop an effective iterative IPDC procedure for further improving on the IPDC and investigate its sampling properties. For more flexible modeling, it is also of practical importance to extend the idea of IPDC to high-dimensional multi-response interaction models in the more general settings of the generalized linear models, nonparametric models, and survival models, as well as other single-index models and multi-index models. These possible extensions are beyond the scope of the current paper and will
be interesting topics for future research.


\appendix

\section{Proofs of main results} \label{AppA}
We provide the main steps of the proof of Theorem \ref{thm:Screening} and the proof of Theorem \ref{thm:Size} in this appendix. Some intermediate steps of the proof of Theorem \ref{thm:Screening} and additional technical details are included in the Supplementary Material. Hereafter we denote by $\widetilde{C}_i$ with $i \geq 0$ some generic positive constants whose values may vary from line to line.

\subsection{Proof of Theorem \ref{thm:Screening}}

The proof of Theorem \ref{thm:Screening} consists of two parts. The first part establishes the exponential probability bounds for $\widehat{\omega}_j-\omega_j$ and $\widehat{\omega}_k^{\ast}-\omega_k^{\ast}$, and the second part proves the sure screening property.

\smallskip

{\bf Part 1.} We first prove inequalities \eqref{eq:omega-main-bound} and \eqref{eq:omega-inter-bound}, which give the exponential probability bounds for $\widehat{\omega}_j-\omega_j$ and $\widehat{\omega}_k^{\ast}-\omega_k^{\ast}$, respectively.
Since the proofs of \eqref{eq:omega-main-bound} and \eqref{eq:omega-inter-bound} are similar, here we focus on  \eqref{eq:omega-inter-bound} to save space.
Recall that
\begin{eqnarray*}
    \omega^{\ast}_k
    =\frac{\dcov^2(X_{k}^{\ast}, \by^{\ast})}{\sqrt{\dcov^2(X_{k}^{\ast}, X_{k}^{\ast})}}
    \quad\mbox{and}\quad
    \widehat{\omega}^{\ast}_k
    =\frac{\widehat{\dcov}^2(X_{k}^{\ast}, \by^{\ast})}{\sqrt{\widehat{\dcov}^2(X_{k}^{\ast}, X_{k}^{\ast})}}.
 \end{eqnarray*}
The key idea of the proof is to show that for any positive constant 
$\widetilde{C}$, there exist some positive constants $\widetilde{C}_1, \cdots, \widetilde{C}_4$ such that
\begin{align}\label{eq:dcov-Xksq-Ysq}
     P&\left(\max_{1\leq k\leq p}|\widehat{\dcov}^2(X_{k}^{\ast}, \by^{\ast}) - \dcov^2(X_{k}^{\ast}, \by^{\ast})|\geq {\widetilde{C}n^{-\kappa_2}}\right) \\
     \nonumber
     &\quad\ \leq  p\widetilde{C}_1\exp\{-\widetilde{C}_2n^{(1-2\kappa_2)/5}\}
           +\widetilde{C}_3\exp\{-\widetilde{C}_4n^{(1-2\kappa_2)/10}\}, \\
           \label{eq:dcov-Xksq-Xksq}
     P&\left(\max_{1\leq k\leq p}|\widehat{\dcov}^2(X_{k}^{\ast}, X_{k}^{\ast}) - \dcov^2(X_{k}^{\ast}, X_{k}^{\ast})|  \geq {\widetilde{C}n^{-\kappa_2}}\right)\\
     \nonumber
     & \quad
 \leq  p\widetilde{C}_1\exp\{-\widetilde{C}_2n^{(1-2\kappa_2)/5}\}
\end{align}
for all $n$ sufficiently large.
Once these two probability bounds are obtained,
it follows from Conditions \ref{con: lower-bound}--\ref{con: tail} and Lemmas \ref{lem: Bj-root}--\ref{lem: AjBj-ratio} and  \ref{lem: distance-covariance-bound} that
\begin{align}\label{eq: omgea-star}
    P&\left(\max_{1\leq k\leq p}|\widehat\omega_k^{\ast} - \omega_k^{\ast}| \ge c_2n^{-\kappa_2}\right)
   \le O\Big(p\exp\{-C_1n^{(1-2\kappa_2)/5}\} \\
   \nonumber
   & \quad\quad
           +\exp\{-C_2n^{(1-2\kappa_2)/10}\}\Big)
      {\leq O\left(\exp\big\{-Cn^{(1-2\kappa_2)/10}\big\}\right),}
\end{align}
{where $C_1$, $C_2$, and $C$ are some positive constants, and the last inequality follows from the condition that $\log p=o(n^{\eta_0})$ with $\eta_0=\min\{(1-2\kappa_1)/3, (1-2\kappa_2)/5\}$.}

It thus remains to prove \eqref{eq:dcov-Xksq-Ysq} and \eqref{eq:dcov-Xksq-Xksq}.  Again we concentrate on \eqref{eq:dcov-Xksq-Ysq} since \eqref{eq:dcov-Xksq-Xksq} can be shown using similar arguments. Define  $\phi(X_{1k}^{\ast},X_{2k}^{\ast})=|X_{1k}^{\ast}-X_{2k}^{\ast}|$ and $\psi(\by_1^{\ast}, \by_2^{\ast})=\|\by_1^{\ast}-\by_2^{\ast}\|$.  According to  \cite{szekely:Rizzo:Bakirov:2007}, we have
\begin{equation*}
  \dcov^2(X_{k}^{\ast}, \by^{\ast}) = T_{k1}+T_{k2}-2T_{k3}\quad\mbox{and}\quad
  \widehat{\dcov}^2(X_{k}^{\ast},\by^{\ast}) = \widehat{T}_{k1}+\widehat{T}_{k2}-2\widehat{T}_{k3},
\end{equation*}
where $T_{k1}= E\left[\phi(X_{1k}^{\ast},X_{2k}^{\ast})\psi(\by_1^{\ast}, \by_2^{\ast})\right]$,
$T_{k2}=E\left[\phi(X_{1k}^{\ast},X_{2k}^{\ast})\right]E\left[\psi(\by_1^{\ast}, \by_2^{\ast})\right]$,
$T_{k3}= E[\phi(X_{1k}^{\ast},X_{2k}^{\ast}) \psi(\by_1^{\ast}, \by_3^{\ast})]$, and
\begin{align*}
    \widehat{T}_{k1} &=n^{-2}\sum_{i,j=1}^n\phi(X_{ik}^{\ast},X_{jk}^{\ast})\psi(\by_i^{\ast}, \by_j^{\ast}), \\
    \widehat{T}_{k2} &=\big[n^{-2}\sum_{i,j=1}^n\phi(X_{ik}^{\ast},X_{jk}^{\ast})\big]\big[n^{-2}\sum_{i,j=1}^n\psi(\by_i^{\ast}, \by_j^{\ast})\big], \\
    \widehat{T}_{k3} &=n^{-3}\sum_{i=1}^n\sum_{j,\,l=1}^n\phi(X_{ik}^{\ast},X_{jk}^{\ast})\psi(\by_i^{\ast}, \by_l^{\ast}).
\end{align*}
It follows from the triangle inequality that
\begin{align}\label{eq: dcov-Xsq-Ysq-max}
    \max_{1\leq k\leq p} & | \widehat{\dcov}^2(X_{k}^{\ast}, \by^{\ast})-\dcov^2(X_{k}^{\ast}, \by^{\ast}) | \leq  \max_{1\leq k\leq p}|\widehat{T}_{k1}-T_{k1}| \\
    \nonumber
   & \quad + \max_{1\leq k\leq p} |\widehat{T}_{k2}-T_{k2}| + 2 \max_{1\leq k\leq p}|\widehat{T}_{k3}-T_{k3}|.
\end{align}
To establish the probability bound for the term $\max_{1\leq k\leq p}| \widehat{\dcov}^2(X_{k}^{\ast}, \by^{\ast})-\dcov^2(X_{k}^{\ast}, \by^{\ast}) |$, it is sufficient to bound each term on the right hand side above. To enhance the readability, we proceed with three main steps.

\smallskip

{\bf Step 1}. We start with the first term $\max_{1\leq k\leq p}|\widehat{T}_{k1}-T_{k1}|$.  An application of the Cauchy-Schwarz inequality gives
\begin{equation*}
T_{k1}
\leq  \left\{E[\phi^2(X_{1k}^{\ast},X_{2k}^{\ast})]E[\psi^2(\by_1^{\ast}, \by_2^{\ast})]\right\}^{1/2}.
\end{equation*}
It follows from the triangle inequality that
\begin{equation}\label{eq: X1k-X2k}
   \phi(X_{1k}^{\ast},X_{2k}^{\ast})=|X_{1k}^{\ast}-X_{2k}^{\ast}|\leq |X_{1k}^{\ast}|+|X_{2k}^{\ast}|=X_{1k}^2+X_{2k}^2
\end{equation}
and
\begin{equation}\label{eq: y1-y2}
   \psi(\by_1^{\ast}, \by_2^{\ast})=\|\by_1^{\ast}-\by_2^{\ast}\|\leq \|\by_1^{\ast}\|+\|\by_2^{\ast}\|\leq \|\widetilde{\by}_1\|^2+\|\widetilde{\by}_2\|^2, 
\end{equation}
in view of $\by_1^{\ast}=\widetilde{\by}_1\circ\widetilde{\by}_1$ and the fact that $\|\ba\circ\ba\|\leq \|\ba\|^2$ for any $\ba\in\mathbb{R}^q$. By \eqref{eq: X1k-X2k}, 
we have
$E[\phi^2(X_{1k}^{\ast},X_{2k}^{\ast})]\leq E\{2(X_{1k}^4+X_{2k}^4)\}=4E(X_{1k}^4)$. Similarly, it holds that  $E[\psi^2(\by_1^{\ast}, \by_2^{\ast})]\leq 4E(\|\widetilde{\by}_1\|^4)$.
Combining these results leads to $0\leq T_{k1}
\leq  4\left\{E(X_{1k}^4)E(\|\widetilde{\by}_1\|^4)\right\}^{1/2}$.
By Condition \ref{con: tail},  $E(X_{1k}^4)$ and $E(\|\widetilde{\by}_1\|^4)$ are uniformly bounded by some positive constant for all $1\leq k\leq p$.  Thus for any positive constant 
{$\widetilde{C}$, $|T_{k1}/n|<\widetilde{C}n^{-\kappa_2}/8$} holds uniformly for all $1\leq k\leq p$ when $n$ is sufficiently large.

Let $\widehat T_{k1}^{\ast}=n(n-1)^{-1}\widehat{T}_{k1}=\{n(n-1)\}^{-1}\sum_{i\ne j}\phi(X_{ik}^{\ast},X_{jk}^{\ast})\psi(\by_i^{\ast}, \by_j^{\ast})$.  Then we have
\begin{align*}
   |\widehat T_{k1} - T_{k1}| \leq n^{-1}(n-1)|\widehat T_{k1}^{\ast}- T_{k1}|+|T_{k1}/n|\leq |\widehat T_{k1}^{\ast}- T_{k1}|+{\widetilde{C}n^{-\kappa_2}/8}
\end{align*}
for all $1\leq k\leq p$, which entails
\begin{align}\label{eq:Tk1-bound-1}
  P \big(\max_{1\leq k\leq p}|\widehat T_{k1} - T_{k1}|\ge {\widetilde{C}}n^{-\kappa_2}/4\big)
  \leq  P\big(\max_{1\leq k\leq p}|\widehat T_{k1}^\ast - T_{k1}|\ge {\widetilde{C}}n^{-\kappa_2}/8\big)
\end{align}
for sufficiently large $n$. Thus it is sufficient to bound $\widehat T_{k1}^\ast - T_{k1}$.

Since $X_{ik}^{\ast}$ and $\by_i^{\ast}$ are generally unbounded, we apply the technique of truncation in the technical analysis.  Define \begin{align*}
   \widehat T_{k1,\,1}^\ast
   &=\{n(n-1)\}^{-1}\sum_{i\ne j}\phi(X_{ik}^{\ast}, X_{jk}^{\ast})\psi(\by_i^{\ast}, \by_j^{\ast})\mathbb{I}\{\phi(X_{ik}^{\ast}, X_{jk}^{\ast})\leq M_1\} \\
   & \quad \cdot \mathbb{I}\{\psi(\by_i^{\ast}, \by_j^{\ast})\leq M_2\},\\
   \widehat T_{k1,\,2}^\ast
   &=\{n(n-1)\}^{-1}\sum_{i\ne j}\phi(X_{ik}^{\ast}, X_{jk}^{\ast})\psi(\by_i^{\ast}, \by_j^{\ast})\mathbb{I}\{\phi(X_{ik}^{\ast}, X_{jk}^{\ast})\leq M_1\} \\
   &\quad\cdot \mathbb{I}\{\psi(\by_i^{\ast}, \by_j^{\ast})> M_2\},\\
   \widehat T_{k1,\,3}^\ast
   &=\{n(n-1)\}^{-1}\sum_{i\ne j}\phi(X_{ik}^{\ast}, X_{jk}^{\ast})\psi(\by_i^{\ast}, \by_j^{\ast})\mathbb{I}\{\phi(X_{ik}^{\ast}, X_{jk}^{\ast})> M_1\},
\end{align*}
where $\mathbb{I}\{\cdot\}$ denotes the indicator function and the thresholds $M_1, M_2>0$ will be specified later.  Then we have $\widehat T_{k1}^\ast=\widehat T_{k1,\,1}^\ast +\widehat T_{k1,\,2}^\ast +\widehat T_{k1,\,3}^\ast$. Consequently, we can rewrite $T_{k1}$ as
$T_{k1}=T_{k1,\,1}+T_{k1,\,2}+T_{k1,\,3}$ with
\begin{align*}
 T_{k1,\,1} & =E\left[\phi(X_{1k}^{\ast}, X_{2k}^{\ast})\psi(\by_1^{\ast}, \by_2^{\ast})\mathbb{I}\{\phi(X_{1k}^{\ast}, X_{2k}^{\ast})\leq M_1\}\mathbb{I}\{\psi(\by_1^{\ast}, \by_2^{\ast})\leq M_2\}\right],\\
 T_{k1,\,2} & =E\left[\phi(X_{1k}^{\ast}, X_{2k}^{\ast})\psi(\by_1^{\ast}, \by_2^{\ast})\mathbb{I}\{\phi(X_{1k}^{\ast}, X_{2k}^{\ast})\leq M_1\}\mathbb{I}\{\psi(\by_1^{\ast}, \by_2^{\ast})> M_2\}\right],\\
 T_{k1,\,3} &=E\left[\phi(X_{1k}^{\ast}, X_{2k}^{\ast})\psi(\by_1^{\ast}, \by_2^{\ast})\mathbb{I}\{\phi(X_{1k}^{\ast}, X_{2k}^{\ast})> M_1\}\right].
\end{align*}
Clearly,  $\widehat T_{k1,\,1}^\ast$, $\widehat T_{k1,\,2}^\ast$, and $\widehat T_{k1,\,3}^\ast$ are unbiased estimators of $T_{k1,\,1}$, $T_{k1,\,2}$, and $T_{k1,\,3}$, respectively.  Therefore, it follows from Bonferroni's inequality that
\begin{align}\label{eq:Tk1-bound-2}
     P & \big(\max_{1\leq k\leq p}|\widehat T_{k1}^\ast - T_{k1}|\ge {\widetilde{C}}n^{-\kappa_2}/8\big) \\
     \nonumber
     & \quad
   \leq \sum_{j=1}^3P\big(\max_{1\leq k\leq p}|\widehat T_{k1, \,j}^\ast - T_{k1, \,j}|\ge {\widetilde{C}}n^{-\kappa_2}/24 \big).
\end{align}
In what follows, we will provide details on deriving an exponential tail probability bound for each term on the right hand side above.

\smallskip

{\bf Step 1.1}. We first consider $\widehat T_{k1,\,1}^\ast-T_{k1,\,1}$. For any $\delta>0$, by Markov's inequality we have
\begin{align}\label{eq: TK1-1}
   P(\widehat T_{k1,\,1}^\ast-T_{k1,\,1}\geq \delta)
  \leq \exp(-t\delta)\exp(-tT_{k1,\,1})E[\exp(t\widehat T_{k1,\,1}^\ast)]
\end{align}
for $t>0$.
Let $h(X_{1k}^{\ast}, \by_1^{\ast}; X_{1k}^{\ast}, \by_2^{\ast})=\phi(X_{1k}^{\ast}, X_{2k}^{\ast})\psi(\by_1^{\ast}, \by_2^{\ast})\mathbb{I}\{\phi(X_{1k}^{\ast}, X_{2k}^{\ast})\leq M_1\}\mathbb{I}\{\psi(\by_1^{\ast}, \by_2^{\ast})\leq M_2\}$ be the kernel of the $U$-statistic $\widehat T_{k1,\,1}^\ast$ and define
\begin{align}\label{eq: W}
    W & (X_{1k}^{\ast}, \by_1^{\ast}; \cdots; X_{nk}^{\ast}, \by_n^{\ast}) =  m^{-1} \Big\{h(X_{1k}^{\ast}, \by_1^{\ast}; X_{1k}^{\ast}, \by_2^{\ast}) \\
    \nonumber
    &\quad + h(X_{3k}^{\ast}, \by_3^{\ast}; X_{4k}^{\ast}, \by_4^{\ast})+\cdots+h(X_{2m-1,\,k}^{\ast}, \by_{2m-1}^{\ast}; X_{2m,\,k}^{\ast}, \by_{2m}^{\ast})\Big\},
\end{align}
where $m=\lfloor n/2\rfloor$ is the integer part of $n/2$. According to the theory of $U$-statistics \citep[Section 5.1.6]{Serfling1980Approximation}, any $U$-statistic can be expressed as an average of averages of i.i.d. random variables. This representation gives
\[ \widehat T_{k1,\,1}^\ast=(n!)^{-1}\sum\nolimits_{n!}W(X_{i_1k}^{\ast}, \by_{i_1}^{\ast}; \cdots; X_{i_nk}^{\ast}, \by_{i_n}^{\ast}), \]
where  $\sum_{n!}$ represents the summation over all possible permutations $(i_1, \cdots, \\i_n)$ of $(1, \cdots, n)$. An application of Jensen's inequality yields that for any $t>0$,
\begin{align*}
  E[\exp(t\widehat T_{k1,\,1}^\ast)]
  &= E\left\{\exp\left[(n!)^{-1}\sum_{n!}tW(X_{i_1k}^{\ast}, \by_{i_1}^{\ast}; \cdots; X_{i_nk}^{\ast}, \by_{i_n}^{\ast})\right]\right\}\\
  &\leq E\left\{(n!)^{-1}\sum_{n!}\exp\left[tW(X_{i_1k}^{\ast}, \by_{i_1}^{\ast}; \cdots; X_{i_nk}^{\ast}, \by_{i_n}^{\ast})\right]\right\}\\
  &=E\left\{\exp\left[tW(X_{1k}^{\ast}, \by_{1}^{\ast}; \cdots; X_{nk}^{\ast}, \by_{n}^{\ast})\right]\right\}\\
  &=E^m\left\{\exp\left[tm^{-1}h(X_{1k}^{\ast}, \by_1^{\ast}; X_{2k}^{\ast}, \by_2^{\ast}) \right]\right\},
\end{align*}
where the last equality follows from \eqref{eq: W}. The above inequality together with \eqref{eq: TK1-1} leads to
\begin{align*}
   P(\widehat T_{k1,\,1}^\ast-T_{k1,\,1}\geq \delta)
  \leq \exp(-t\delta)E^m\left\{e^{tm^{-1}\left[h(X_{1k}^{\ast}, \by_1^{\ast}; X_{2k}^{\ast}, \by_2^{\ast})-T_{k1,\,1}\right]}\right\}.
\end{align*}

Note that $E[h(X_{1k}^{\ast}, \by_1^{\ast}; X_{2k}^{\ast}, \by_2^{\ast})-T_{k1,\,1}]=0$ and \[ -T_{k1,\,1}\leq h(X_{1k}^{\ast}, \by_1^{\ast}; X_{2k}^{\ast}, \by_2^{\ast})\\-T_{k1,\,1}\leq M_1M_2-T_{k1,\,1}. \]
Hence it follows from Lemma \ref{lem: Hoeffding-Lemma} that
\begin{align*}
   P(\widehat T_{k1,\,1}^\ast-T_{k1,\,1}\geq \delta)
   \leq \exp[-t\delta+t^2M_1^2M_2^2/(8m)]
\end{align*}
for any $t>0$.  Minimizing the right hand side above with respect to $t$ gives $P(\widehat T_{k1,\,1}^\ast-T_{k1,\,1}\geq \delta)\leq \exp(-2m\delta^2/M_1^2M_2^2)$
for any $\delta>0$. Similarly, we can show that $P(\widehat T_{k1,\,1}^\ast-T_{k1,\,1}\leq -\delta)\leq \exp(-2m\delta^2/M_1^2M_2^2)$. Therefore, it holds that
\begin{equation*}
   P(|\widehat T_{k1,\,1}^\ast-T_{k1,\,1}|\geq \delta)
   \leq 2\exp(-2m\delta^2/M_1^2M_2^2).
\end{equation*}
Recall that $m=\lfloor n/2\rfloor$. If we set $M_1=n^{\xi_1}$ and $M_2=n^{\xi_2}$ with some positive constants $\xi_1$ and $\xi_2$, then for 
{$\delta=\widetilde{C}n^{-\kappa_2}/24$ with any positive constant $\widetilde{C}$},
there exists some positive constant $\widetilde{C}_1$ such that when $n$ is sufficiently large,
\begin{equation*}
   P(|\widehat T_{k1,\,1}^\ast-T_{k1,\,1}|\geq {\widetilde{C}}n^{-\kappa_2}/24)
   \leq 2\exp(-{\widetilde{C}}^2\widetilde{C}_1n^{1-2\kappa_2-2\xi_1-2\xi_2})
\end{equation*}
for all $1\leq k\leq p$.  This along with Bonferroni's inequality entails
\begin{align}\label{eq: max-Tk1-1}
   P&(\max_{1\leq k\leq p}|\widehat T_{k1,\,1}^\ast-T_{k1,\,1}|\geq {\widetilde{C}}n^{-\kappa_2}/24)\\
   \nonumber
   &\quad\leq 2p\exp(-{\widetilde{C}}^2\widetilde{C}_1n^{1-2\kappa_2-2\xi_1-2\xi_2}).
\end{align}


{\bf Step 1.2}. We next deal with $\widehat T_{k1,\,2}^\ast-T_{k1,\,2}$.  Note that
\[ 0\leq T_{k1,\,2}\leq M_1E\left[\psi(\by_1^{\ast}, \by_2^{\ast})\mathbb{I}\{\psi(\by_1^{\ast}, \by_2^{\ast})> M_2\}\right] \] for all $1\leq k\leq p$. It follows from the Cauchy-Schwarz inequality that
\begin{align}\label{eq: y1y2-tail}
   E&\left[\psi(\by_1^{\ast}, \by_2^{\ast})\mathbb{I}\{\psi(\by_1^{\ast}, \by_2^{\ast})> M_2\}\right]\\
   \nonumber
   &\quad
   \leq   \left[E[\psi^2(\by_1^{\ast}, \by_2^{\ast})]P\{\psi(\by_1^{\ast}, \by_2^{\ast})> M_2\}\right]^{1/2}.
\end{align}
In view of \eqref{eq: y1-y2}, we see that
\begin{align}\label{eq: y1y2-moment}
    E&[\psi^2(\by_1^{\ast}, \by_2^{\ast})]
  \leq E[ (\|\widetilde{\by}_1\|^2+\|\widetilde{\by}_2\|^2)^2]\\
  \nonumber
  &\quad
 \leq E[2(\|\widetilde{\by}_1\|^4+\|\widetilde{\by}_2\|^4)]=4E(\|\widetilde{\by}_1\|^4)
\end{align}
and the probability term in \eqref{eq: y1y2-tail} is bounded from above by
\begin{align}\label{eq: y1y2-prob}
   P&(\|\widetilde{\by}_1\|^2+\|\widetilde{\by}_2\|^2> M_2)
   \leq   P(\|\widetilde{\by}_1\|^2> M_2/2) + P(\|\widetilde{\by}_2\|^2> M_2/2) \\
   \nonumber
   = & 2P(\|\widetilde{\by}_1\|> \sqrt{M_2/2})
    \leq 2 \exp(-c_0\sqrt{M_2/2})E[\exp(c_0\|\widetilde{\by}_1\|)],
\end{align}
where $c_0$ is a positive constant given in Condition \ref{con: tail} and the last inequality follows from Markov's inequality. Combining inequalities \eqref{eq: y1y2-tail}--\eqref{eq: y1y2-prob} and by Condition \ref{con: tail},  we obtain
\begin{equation}\label{eq: y1y2-tail-final}
     E\left[\psi(\by_1^{\ast}, \by_2^{\ast})\mathbb{I}\{\psi(\by_1^{\ast}, \by_2^{\ast})> M_2\}\right]
   \leq   \widetilde{C}_2 \exp(-2^{-1}c_0\sqrt{M_2/2})
\end{equation}
and thus $0\leq T_{k1,\,2}\leq \widetilde{C}_2M_1 \exp(-2^{-1}c_0\sqrt{M_2/2})$, where $\widetilde{C}_2$ is some positive constant.  Recall that
$M_1=n^{\xi_1}$ and $M_2=n^{\xi_2}$. Then for any positive constant ${\widetilde{C}}$, it holds that
\begin{equation*}
   0\leq T_{k1,\,2}
   \leq \widetilde{C}_2n^{\xi_1}\exp(-2^{-3/2}c_0n^{\xi_2/2})
   \leq {\widetilde{C}}n^{-\kappa_2}/48
\end{equation*}
for all $1\leq k\leq p$ when $n$ is sufficiently large.  This inequality gives
\begin{equation}\label{eq: Tk1-2-part1}
  P(\max_{1\leq k\leq p}|\widehat T_{k1,\,2}^\ast-T_{k1,\,2}|\geq {\widetilde{C}}n^{-\kappa_2}/24)
  \leq  P(\max_{1\leq k\leq p}|\widehat T_{k1,\,2}^\ast|\geq {\widetilde{C}}n^{-\kappa_2}/48)
\end{equation}
for all $n$ sufficiently large.

Note that for all $1\leq k\leq p$, $|\widehat T_{k1,\,2}^\ast|$ is uniformly bounded from above by $M_1[n(n-1)]^{-1}\sum_{i\neq j}\psi(\by_i^{\ast}, \by_j^{\ast})
      \mathbb{I}\{\psi(\by_i^{\ast}, \by_j^{\ast})>M_2\}$. Thus in view of \eqref{eq: y1y2-tail-final}, applying Markov's inequality yields that for any $\delta>0$,
\begin{align*}
   P&(\max_{1\leq k\leq p}|\widehat T_{k1,\,2}^\ast|\geq \delta/2)
   \leq    P\Big\{M_1[n(n-1)]^{-1}\sum_{i\neq j}\psi(\by_i^{\ast}, \by_j^{\ast})
      \mathbb{I}\{\psi(\by_i^{\ast}, \by_j^{\ast})>M_2\}\\
      & \quad \geq \delta/2\Big\} \leq   (\delta/2)^{-1} E\Big\{M_1[n(n-1)]^{-1}\sum_{i\neq j}\psi(\by_i^{\ast}, \by_j^{\ast})
      \mathbb{I}\{\psi(\by_i^{\ast}, \by_j^{\ast})>M_2\}\Big\} \nonumber\\
   &=  (\delta/2)^{-1}M_1E\left[\psi(\by_1^{\ast}, \by_2^{\ast})\mathbb{I}\{\psi(\by_1^{\ast}, \by_2^{\ast})> M_2\}\right]  \nonumber\\
  &\leq   (\delta/2)^{-1}M_1\widetilde{C}_2 \exp(-2^{-1}c_0\sqrt{M_2/2}).
\end{align*}
Since $M_1=n^{\xi_1}$ and $M_2=n^{\xi_2}$, setting $\delta={\widetilde{C}}n^{-\kappa_2}/24$ in the above inequality entails
\[ P(\max_{1\leq k\leq p}|\widehat T_{k1,\,2}^\ast|\geq {\widetilde{C}}n^{-\kappa_2}/48) \leq  48{\widetilde{C}}^{-1}\widetilde{C}_2n^{\kappa_2+\xi_1} \exp(-2^{-3/2}c_0n^{\xi_2/2}). \]
Combining this inequality with \eqref{eq: Tk1-2-part1} gives
\begin{align}\label{eq: max-Tk1-2}
   P&(\max_{1\leq k\leq p}|\widehat T_{k1,\,2}^\ast-T_{k1,\,2}| \geq {\widetilde{C}}n^{-\kappa_2}/24)\\
   \nonumber
   & \quad
  \leq  48{\widetilde{C}}^{-1}\widetilde{C}_2n^{\kappa_2+\xi_1} \exp(-2^{-3/2}c_0n^{\xi_2/2}).
\end{align}

{\bf Step 1.3}. We finally handle the term $\widehat T_{k1,\,3}^{\ast}-T_{k1,\,3}$ and show that it satisfies
\begin{align}\label{eq: max-Tk1-3}
   P&(\max_{1\leq k\leq p}|\widehat T_{k1,\,3}^{\ast}-T_{k1,\,3}|\geq {\widetilde{C}}n^{-\kappa_2}/24)\\\nonumber
  &\quad\leq  48p{\widetilde{C}}^{-1}\widetilde{C}_3 n^{\kappa_2}\exp(-8^{-1}c_0n^{\xi_1})
\end{align}
with $\widetilde{C}_3$ some positive constant in Section \ref{Thm1Part1Step1.3} of the Supplementary Material.

\smallskip

Combining the results in \eqref{eq:Tk1-bound-2}, \eqref{eq: max-Tk1-1}, and \eqref{eq: max-Tk1-2}--\eqref{eq: max-Tk1-3} leads to
\begin{align*}
P&(\max_{1\leq k\leq p}|\widehat T_{k1}^{\ast}-T_{k1}|\geq {\widetilde{C}}n^{-\kappa_2}/8) \leq  2p\exp(-{\widetilde{C}}^2\widetilde{C}_1n^{1-2\kappa_2-2\xi_1-2\xi_2}) \\\nonumber
   &+ 48p{\widetilde{C}}^{-1}\widetilde{C}_3 n^{\kappa_2}\exp(-8^{-1}c_0n^{\xi_1})+48{\widetilde{C}}^{-1}\widetilde{C}_2n^{\kappa_2+\xi_1} \exp(-2^{-3/2}c_0n^{\xi_2/2}).
\end{align*}
Let  $\xi_1=(1-2\kappa_2)/3-2\eta$ and $\xi_2=3\eta$ with some $0<\eta<(1-2\kappa_2)/6$. Then we have
\begin{align}\label{eq: max-Tk1-star}
  P&(\max_{1\leq k\leq p}|\widehat T_{k1}^{\ast}-T_{k1}|\geq {\widetilde{C}}n^{-\kappa_2}/8)
  \leq p\widetilde{C}_1\exp\{-\widetilde{C}_2n^{(1-2\kappa_2)/3-2\eta}\}\\\nonumber
      &\quad + \widetilde{C}_3\exp\{-\widetilde{C}_4n^{3\eta/2}\},
\end{align}
where $\widetilde{C}_1, \cdots, \widetilde{C}_4$ are some positive constants. This inequality along with \eqref{eq:Tk1-bound-1} yields
\begin{align} \label{eq: max-Tk1-final}
   P&(\max_{1\leq k\leq p}|\widehat T_{k1}-T_{k1}|\geq {\widetilde{C}}n^{-\kappa_2}/4)\leq p\widetilde{C}_1\exp\{-\widetilde{C}_2n^{(1-2\kappa_2)/3-2\eta}\}\\\nonumber
  &\quad
       + \widetilde{C}_3\exp\{-\widetilde{C}_4n^{3\eta/2}\}.
\end{align}

{\bf Step 2}. For the second term $\max_{1\leq k\leq p}|\widehat T_{k2}-T_{k2}|$, we show in Section \ref{Thm1Part1Step2} of the Supplementary Material that
\begin{align} \label{eq: max-Tk2-final}
       P&\big(\max_{1\leq k\leq p}|\widehat{T}_{k2} - T_{k2}|\ge {\widetilde{C}}n^{-\kappa_2}/4\big)\leq  \sum_{k=1}^p P\big(|\widehat{T}_{k2} - T_{k2}|\ge {\widetilde{C}}n^{-\kappa_2}/4\big) \\\nonumber
   &
   \leq p\widetilde{C}_5\exp\{-\widetilde{C}_6n^{(1-2\kappa_2)/5}\}
\end{align}
holds, where $\widetilde{C}_5$ and $\widetilde{C}_6$ are some positive constants.

\smallskip

{\bf Step 3}. We further prove that the third term $\widehat T_{k3}-T_{k3}$ satisfies
\begin{align}\label{eq: max-Tk3-final}
   P&(\max_{1\leq k\leq p}|\widehat T_{k3}-T_{k3}|\geq {\widetilde{C}}n^{-\kappa_2}/4)
  \leq p\widetilde{C}_{1}\exp\{-\widetilde{C}_{2}n^{(1-2\kappa_2)/3-2\eta}\}\\\nonumber
  &\quad
       + \widetilde{C}_{3}\exp\{-\widetilde{C}_4n^{3\eta/2}\}
\end{align}
with $\widetilde{C}_1, \cdots, \widetilde{C}_4$ some positive constants in Section \ref{Thm1Part1Step3} of the Supplementary Material.

\smallskip

Combining inequalities \eqref{eq: dcov-Xsq-Ysq-max} and \eqref{eq: max-Tk1-final}--\eqref{eq: max-Tk3-final} and setting $\eta=(1-2\kappa_2)/15$ entail
\begin{align}\label{eq: dcov-Xsq-Ysq-bound}
    P&\left\{\max_{1\leq k\leq p}| \widehat{\dcov}^2(X_{k}^{\ast}, \by^{\ast})-\dcov^2(X_{k}^{\ast}, \by^{\ast}) |\geq {\widetilde{C}}n^{-\kappa_2}\right\}\\\nonumber
  &\quad\leq  p\widetilde{C}_1\exp\{-\widetilde{C}_2n^{(1-2\kappa_2)/5}\}
           +\widetilde{C}_3\exp\{-\widetilde{C}_4n^{(1-2\kappa_2)/10}\}
\end{align}
with $\widetilde{C}_1, \cdots, \widetilde{C}_4$ some positive constants,
which completes the proof for the first part of Theorem \ref{thm:Screening}.

\smallskip

{\bf Part 2.} We now proceed to prove the second part of Theorem 1. The main idea is to build the probability bounds for two events $\{\mathcal{M}\subset \widehat{\mathcal{M}}\}$ and $\{\mathcal{I}\subset \widehat{\mathcal{I}}\}$. We first bound $P(\mathcal{M}\subset \widehat{\mathcal{M}})$. Define an event
$\Omega_1=\{\max_{j\in \mathcal{M}}|\widehat{\omega}_j-\omega_j|<c_1n^{-\kappa_1}\}$. Then by Condition \ref{con: min-signal}, conditional on the event $\Omega_1$ we have $\widehat{\omega}_j\geq 2c_1n^{-\kappa_1}$ for all $j\in \mathcal{M}$, which gives
\begin{align}\label{eq: main-bound}
     P&(\mathcal{M}\subset \widehat{\mathcal{M}})\geq P(\Omega_1) =1-P(\Omega_1^c) \\\nonumber
     &\quad=1-P(\max_{j\in \mathcal{M}}|\widehat{\omega}_j-\omega_j|\geq c_1n^{-\kappa_1}).
\end{align}
Following similar arguments as for proving \eqref{eq: omgea-star}, it can be shown that there exist some positive constants ${\widetilde{C}_5}$ and ${\widetilde{C}_6}$ such that 
\begin{align*}
    P&\left(\max_{j\in \mathcal{M}}|\widehat\omega_j - \omega_j| \ge c_1n^{-\kappa_1}\right)
   = O\Big(s_1\exp\{-{\widetilde{C}_5}n^{(1-2\kappa_1)/3}\}\\
   &\quad
           +\exp\{-{\widetilde{C}_6}n^{(1-2\kappa_1)/6}\}\Big),
\end{align*}
where $s_1$ is the cardinality of $\mathcal{M}$.  This inequality together with \eqref{eq: main-bound} yields
\begin{align}\label{eq: sure-main}
     P&(\mathcal{M}\subset \widehat{\mathcal{M}})\geq 1-O\Big(s_1\exp\{-{\widetilde{C}_5}n^{(1-2\kappa_1)/3}\}\\\nonumber
     &\quad
           +\exp\{-{\widetilde{C}_6}n^{(1-2\kappa_1)/6}\}\Big).
\end{align}

We next bound $P(\mathcal{I}\subset \widehat{\mathcal{I}})$. Note that $P(\mathcal{I}\subset \widehat{\mathcal{I}})\geq P(\mathcal{A}\subset \widehat{\mathcal{A}})$ since conditional on the event $\{\mathcal{A}\subset \widehat{\mathcal{A}}\}$ it holds that $\{\mathcal{I}\subset \widehat{\mathcal{I}}\}$. Define an event
$\Omega_2=\{\max_{k\in \mathcal{A}}|\widehat{\omega}_k^{\ast}-\omega_k^{\ast}|<c_2n^{-\kappa_2}\}$. Then by Condition \ref{con: min-signal}, we have $\widehat{\omega}_k\geq 2c_2n^{-\kappa_2}$ for all $k\in \mathcal{A}$ conditional on the event $\Omega_2$, which leads to $P(\mathcal{A}\subset \widehat{\mathcal{A}})\geq P(\Omega_2)$. Combining these results yields
\begin{align}\label{eq: inter-bound}
     P(\mathcal{I}\subset \widehat{\mathcal{I}})
    \geq P(\Omega_2)=1-P(\Omega_2^c)=1-P(\max_{k\in \mathcal{A}}|\widehat{\omega}_k^{\ast}-\omega_k^{\ast}|\geq c_2n^{-\kappa_2}).
\end{align}
Using similar arguments as for proving \eqref{eq: omgea-star} shows that there exist some positive constants ${\widetilde{C}_7}$ and ${\widetilde{C}_8}$ such that 
\begin{align*}
    P&\left({\max_{k\in \mathcal{A}}}|\widehat\omega_k^{\ast} - \omega_k^{\ast}| \ge c_2n^{-\kappa_2}\right)
   = O\Big(s_2\exp\{-{\widetilde{C}_7}n^{(1-2\kappa_2)/5}\}\\
   &\quad
           +\exp\{-{\widetilde{C}_8}n^{(1-2\kappa_2)/10}\}\Big),
\end{align*}
where $s_2$ is the cardinality of $\mathcal{A}$. This together with \eqref{eq: inter-bound} entails
\begin{align}\label{eq: sure-inter}
     P(\mathcal{I}\subset \widehat{\mathcal{I}})\geq 1-O\Big(s_2\exp\{-{\widetilde{C}_7}n^{(1-2\kappa_2)/5}\}
           +\exp\{-{\widetilde{C}_8}n^{(1-2\kappa_2)/10}\}\Big).
\end{align}

Finally combining \eqref{eq: sure-main} and \eqref{eq: sure-inter}, we obtain
\begin{align*}
            P& (\mathcal{M}\subset \widehat{\mathcal{M}}\,\,\mbox{and}\,\,\mathcal{I}\subset \widehat{\mathcal{I}})
    \geq   P(\mathcal{M}\subset \widehat{\mathcal{M}})+P(\mathcal{I}\subset \widehat{\mathcal{I}})-1\nonumber\\
    &\quad\geq 1-O\left(s_1\exp\{-{\widetilde{C}_5}n^{(1-2\kappa_1)/3}\}
           +\exp\{-{\widetilde{C}_6}n^{(1-2\kappa_1)/6}\}\right)\nonumber\\
         &\quad \quad -O\left(s_2\exp\{-{\widetilde{C}_7}n^{(1-2\kappa_2)/5}\}
           +\exp\{-{\widetilde{C}_8}n^{(1-2\kappa_2)/10}\}\right)\\
    &{\quad\geq 1-O\left(\exp\big\{-C n^{\eta_0/2}\big\}\right),}
\end{align*}
{where $C$ is some positive constant, and the last inequality follows from the facts $s_1, s_2\leq p$ and the condition that $\log p=o(n^{\eta_0})$ with $\eta_0=\min\{(1-2\kappa_1)/3, (1-2\kappa_2)/5\}$. This concludes the proof for the second part of Theorem \ref{thm:Screening}.}

\subsection{Proof of Theorem \ref{thm:Size}} \label{ProofThm2}

Define an event $\Omega_3=\{\max_{1\leq j\leq p}|\widehat{\omega}_j-\omega_j|<c_1n^{-\kappa_1}\}$.
For any $j\in {\cal{M}}^c$, if $\omega_j< c_1n^{-\kappa_1}$ and $|\widehat{\omega}_j-\omega_j|<c_1n^{-\kappa_1}$, then we have  $\widehat{\omega}_j< 2c_1n^{-\kappa_1}$. Thus
conditional on the event $\Omega_3$, the cardinality of  $\{j: \widehat{\omega}_j\geq 2c_1n^{-\kappa_1}\,\,\mbox{and}\,\, j\in {\cal{M}}^c\}$ cannot exceed that of $\{j: \omega_j\geq c_1n^{-\kappa_1}\,\,\mbox{and}\,\, j\in {\cal{M}}^c\}$. This entails that the cardinality of  $\{j: \widehat{\omega}_j\geq 2c_1n^{-\kappa_1}\}$ is no larger than that of $\{j: \omega_j\geq c_1n^{-\kappa_1}\,\,\mbox{and}\,\, j\in {\cal{M}}^c\}\cup \cal{M}$, which is in turn bounded from above by $|{\cal{M}}|+s_3$.  Thus it follows from \eqref{eq:omega-main-bound} in Theorem \ref{thm:Screening} that 
  \begin{align*}
      P&(|\widehat{\cal{M}}|\leq |{\cal{M}}|+s_3)
   \ge   P(\Omega_3)
   =1-P({\max_{1\leq j\leq p}|\widehat{\omega}_j-\omega_j|}\geq c_1n^{-\kappa_1}) \\
   & \quad \ge  1-O\left(\exp\{-Cn^{(1-2\kappa_1)/6}\}\right).
  \end{align*}
Similarly we can show that
  \begin{align*}
      P&\left\{|\widehat{\cal{I}}|\leq (|{\mathcal{A}}|+s_4)(|{\mathcal{A}}|+s_4-1)/2\right\}
   \ge 1-O\Big(\exp\{-Cn^{(1-2\kappa_2)/10}\}\Big).
  \end{align*}
Combining these two probability bounds yields
  \begin{align*}
      P&\left\{|\widehat{\cal{M}}|\leq |{\cal{M}}|+s_3\,\,\mbox{and}\,\,|\widehat{\cal{I}}|\leq (|{\mathcal{A}}|+s_4)(|{\mathcal{A}}|+s_4-1)/2\right\} \\
   &\quad \ge  1-O\left(\exp\{-Cn^{(1-2\kappa_1)/6}\}\right)
                       -O\left(\exp\{-Cn^{(1-2\kappa_2)/10}\}\right)\\
    &\quad\geq 1-O\left(\exp\big\{-C n^{\eta_0/2}\big\}\right),
  \end{align*}
where $C$ is some positive constant, and the last inequality follows from the condition that $\log p=o(n^{\eta_0})$ with $\eta_0=\min\{(1-2\kappa_1)/3, (1-2\kappa_2)/5\}$. This completes the proof of Theorem \ref{thm:Size}.

\begin{supplement}[id=IPDC]
\stitle{Supplementary material to ``Interaction Pursuit in High-Dimensional Multi-Response Regression via Distance Correlation''}
  \slink[doi]{10.1214/00-AOSXXXXSUPP}
  \sdatatype{.pdf}
  \sdescription{Due to space constraints, some intermediate steps of the proof of Theorem \ref{thm:Screening} and additional numerical studies and technical details are provided in the Supplementary Material \citep{KongLiFanLv2015}.}
\end{supplement}

\vspace{4mm}

\begin{minipage}[t]{0.6\textwidth}
{\scriptsize
\hspace{-0.45in}
\begin{tabular}{ll}
& \\
& \textsc{Department of Information Systems and Decision Sciences}\\
& \textsc{Mihaylo College of Business and Economics}\\
& \textsc{California State University at Fullerton}\\
& \textsc{Fullerton, CA 92831}\\
& \textsc{USA}\\
& \textsc{\printead{e1}}
\end{tabular}}
\end{minipage}
\begin{minipage}[t]{0.5\textwidth}
{\scriptsize
\hspace{0.03in}
\begin{tabular}{ll}
& \textsc{Department of Statistics}\\
& \textsc{University of Central Florida}\\
& \textsc{Orlando, FL 32816-2370}\\
& \textsc{USA}\\
& \textsc{\printead{e2}}
\end{tabular}}
\end{minipage}

\vspace{-0.15in}
\begin{center}
\begin{minipage}[t]{0.5\textwidth}
{\scriptsize
\hspace{-0.24in}
\begin{tabular}{ll}
& \\
& \textsc{Data Sciences and Operations Department}\\
& \textsc{Marshall School of Business}\\
& \textsc{University of Southern California}\\
& \textsc{Los Angeles, CA 90089}\\
& \textsc{USA}\\
& \textsc{\printead{e3}}\\
& \textsc{\phantom{E-mail:\ }\printead*{e4}}
\end{tabular}}
\end{minipage}
\end{center}
%
%


\newpage
\title{Supplementary Material to ``Interaction pursuit in high-dimensional multi-response regression via distance correlation''}

\begin{aug}
\author{\fnms{Yinfei}~\snm{Kong}\thanksref{m1}\ead[label=e1]{yinfeiko@usc.edu}},
\author{\fnms{Daoji}~\snm{Li}\thanksref{m2}\ead[label=e2]{daoji.li@ucf.edu}},
\author{\fnms{Yingying}~\snm{Fan}\thanksref{m3}\ead[label=e3]{fanyingy@marshall.usc.edu}}
\and
\author{\fnms{Jinchi}~\snm{Lv}\thanksref{m3}\ead[label=e4]{jinchilv@marshall.usc.edu}
}

\runauthor{Y. Kong, D. Li, Y. Fan and J. Lv}

\affiliation{California State University at Fullerton\thanksmark{m1},  University of Central Florida\thanksmark{m2} and University of Southern California\thanksmark{m3}}

\end{aug}

\bigskip

This Supplementary Material contains some intermediate steps of the proof of Theorem \ref{thm:Screening} and additional numerical studies and technical details, as well as the details about the post-screening interaction selection.

\smallskip

\appendix
\setcounter{page}{1}
\setcounter{section}{1}
\renewcommand{\theequation}{A.\arabic{equation}}
\setcounter{equation}{0}

\section{Post-screening interaction selection}\label{sec: selection}
The screening step of IPDC can reduce the problem of interaction identification from a huge scale to a moderate one as shown in Section \ref{sec: screening}. In particular, the reduced interaction model can be of dimensionality smaller than the sample size. After the screening step, IPDC further selects important interactions and main effects. Thanks to the much reduced scale, the selection step can be conducted in a computationally efficient fashion by exploiting regularization methods for the multi-response regression.
Various regularization methods have been developed for multi-response linear models. See, for example, 
\cite{Bunea2012}, \cite{ChenHuang:2012}, \cite{chen2012reduced}, \cite{chen2013reduced}, \cite{lounici2011oracle}, and references therein. Those methods were usually investigated for the scenario of no interactions. For the selection step of IPDC, we aim at interaction model recovery by employing a two-step variable selection procedure, where we first recover the support union using the idea of group variable selection \citep{yuan2006model} 
and then estimate the individual supports for each column of the regression coefficient matrix via an additional refitting step of 
Lasso \citep{tibshirani1996regression} applied to the recovered support union (see, e.g., \cite{FanLv2013Asym} for connections and differences among  regularization methods). 

To simplify the presentation, hereafter we assume that the response vector $\by$ is centered with mean zero 
and all interactions $X_kX_\ell$ are also centered to have mean zero with a slight abuse of notation, which eliminates the intercept vector $\balpha$ in model \eqref{eq: model}. Thus given an i.i.d. sample $(\by_i, \bx_i)_{i=1}^n$, the multi-response interaction model (\ref{eq: model}) can be rewritten in the matrix form
\begin{equation} \label{neweq004}
      \bY=\widetilde{\bX}\bB + \bW,
\end{equation}
where $\bY=(\by_1, \cdots, \by_n)^T\in \mathbb{R}^{n\times q}$ is the response matrix,
$\widetilde{\bX}=(\bX, \bZ)\in\mathbb{R}^{n\times \widetilde{p}}$ with $\widetilde{p}=p(p+1)/2$ is the full augmented design matrix with $\bX = (\bx_1, \cdots, \bx_n)\t\in\mathbb{R}^{n\times p}$ the main effect matrix 
and $\bZ = (\bz_1, \cdots, \bz_n)\t \in\mathbb{R}^{n\times [p(p-1)/2]}$ the interaction matrix, $\bz_i$'s are defined similarly to $\bz$, 
$\bB=(\bB_{\bx}^T, \bB_{\bz}^T)^T\in\mathbb{R}^{\widetilde{p}\times q}$ is the regression coefficient matrix with $\bB_{\bx}\in \mathbb{ R}^{p\times q}$ and $ \bB_{\bz}\in \mathbb{R}^{[p(p-1)/2]\times q}$, and $\bW=(\bw_1, \cdots, \bw_n)^T\in\mathbb{R}^{n\times q}$ is the error matrix.


\subsection{Interaction and main effect selection}


Let $S$ be the row support of the true regression coefficient matrix $\bB^*$ in model (\ref{neweq004}), which corresponds to the index set of nonzero rows of $\bB^*$; that is, if $k\in S$, then the $k$th row of $\bB^*$ has at least one nonzero component.
Denote by $\mathcal{J} =\{j_1, \cdots, j_{d_1}\}\subset\{1, \cdots, p\}$ and $\mathcal{K}=\{k_1, \cdots, k_{d_2}\}\subset\{1, \cdots, p\}$ the index sets of retained main effects and interaction variables after the screening step of IPDC, respectively. Then the reduced design matrix is $(\widetilde{\bx}_{j_1}, \cdots, \widetilde{\bx}_{j_{d_1}}, \widetilde{\bx}_{k_1}\circ\widetilde{\bx}_{k_2}, \cdots, \widetilde{\bx}_{k_{d_2-1}}\circ\widetilde{\bx}_{k_{d_2}})\in\mathbb{R}^{n\times d}$ with $d=d_1+d_2(d_2-1)/2$, where $\widetilde{\bx}_\ell$ is the $\ell$th column of $\bX$. 
Let $\widetilde{S}\subset \{1, \cdots, \widetilde{p}\}$ be the index set given by the columns of such a reduced matrix in the full matrix $\widetilde{\bX}$.  As guaranteed by Theorem  \ref{thm:Screening}, the true row support $S$ can be contained in the reduced set $\widetilde{S}$ by IPDC with high probability that converges to one at a fast rate as sample size $n$ increases. 

Observe that the true row support $S$ is the union of individual supports of the columns of the true regression coefficient matrix $\bB^*$ corresponding to the $q$ responses. Given any set $J \subset \{1, \cdots, \widetilde{p}\}$, denote by $\bB_{J}$ a submatrix of $\bB$ formed by the rows indexed by $J$. For the support union recovery, we exploit the multivariate group Lasso given by the following
regularization problem
  \begin{align}\label{eq: problem}
      \min_{\bB_{\widetilde{S}^c}=\bzero}\left\{\frac{1}{2nq}\|\bY-\widetilde{\bX}\bB\|_F^2
                 +\lambda\|\bB\|_{2,\, 1}\right\},
  \end{align}
where $\widetilde{S}^c$ is the complement of the set $\widetilde{S}$, $\|\cdot\|_F$ denotes the Frobenius norm of a matrix, $\lambda \geq 0$ is a regularization parameter, and $\|\cdot\|_{2, 1}$ stands for the matrix rowwise ($2, 1$)-norm defined as $\|\bM\|_{2, 1}=\sum_{i}(\sum_{j}m_{ij}^2)^{1/2}$ for any matrix $\bM=(m_{ij})$.
Note that $\widetilde{S}$ and $\widehat{\cal{M}} \cup \widehat{\cal{I}}$ share the same cardinality. We should remark that as ensured by Theorem \ref{thm:Size}, the computational cost of solving the optimization problem \eqref{eq: problem} can be substantially reduced compared to that of solving the same optimization problem without the screening step, that is, with $\widetilde{S} = \{1, \cdots, \widetilde{p}\}$.

The multivariate group Lasso has been widely used in the multi-response linear regression models typically without interaction terms. For example, \cite{lounici2009taking} and \cite{lounici2011oracle} established the oracle inequalities for the case when the design matrix is deterministic and the error matrix has i.i.d. Gaussian entries. \cite{obozinski2011support} investigated the model selection consistency in terms of support union recovery of the multivariate group Lasso under the assumptions that the design matrix is drawn with i.i.d. Gaussian row vectors and all the entries of the error matrix are i.i.d. Gaussian. 
We will relax such Gaussianity assumptions and justify that this group variable selection procedure continues to perform well in the presence of interactions.

Once the row support of the true regression coefficient matrix is recovered, it is straightforward to recover the individual supports of each column of the regression coefficient matrix by an additional refitting step of applying the ordinary Lasso to the recovered support union. Since the sampling properties of Lasso have been extensively studied and are now well understood in the literature, we will provide only theoretical analysis of the multivariate group Lasso problem \eqref{eq: problem}.

\subsection{Support union recovery and oracle inequalities}

To facilitate our technical analysis for the selection step of IPDC, we impose a few additional regularity conditions.

\begin{condition}\label{con: X}
The covariate vector $\bx$ has a sub-Gaussian distribution and $s=|S|=O(n^{\xi})$ 
for some constant $0\leq \xi<1/4$.
\end{condition}

\begin{condition}[RE($s$) assumption]
\label{con:RE-population}
There exists some positive constant $\kappa$ such that
  \begin{align*}
     \kappa(s)=\min_{|J|\leq s,\, \bDel\in \mathbb{R}^{\widetilde{p}\times q}\backslash\{\bzero\},\, \|\bDel_{J^c}\|_{2, 1}\leq 3\|\bDel_J\|_{2, 1}}
    \frac{\|\bSig^{1/2}\bDel\|_F}{\|\bDel_J\|_F} \geq \kappa,
  \end{align*}
where $\bSig$ is the covariance matrix of $\widetilde{\bx} = (\bx\t, \bz\t)\t$. 
\end{condition}


\begin{condition}\label{con: error}
The error vector $\bw$ has a sub-exponential distribution.
\end{condition}

The first part of Condition \ref{con: X} is a mild assumption on the distribution of the covariates. It can be satisfied by many light-tailed distributions such as Gaussian distributions and distributions with bounded support.  
The second part of Condition \ref{con: X} puts a row sparsity constraint on the true regression coefficient matrix. In particular, the requirement of $\xi<1/4$ reflects the difficulty of interaction selection in high dimensions.

Condition \ref{con:RE-population} is a natural extension of the restricted
eigenvalue (RE) assumption introduced in \cite{BRT09} since here we use the rowwise ($2, 1$)-norm in place of the $L_1$-norm.  The RE assumption has been commonly used to establish the oracle inequalities for the Lasso 
and Dantzig selector \citep{candes2007dantzig}. For simplicity we still refer to Condition \ref{con:RE-population} as the RE($s$) assumption.
This condition is also similar to Condition 3.1 in \cite{lounici2009taking} and Condition 4.1 in \cite{lounici2011oracle}, who considered the scenario of deterministic design matrix and no interactions. 
Condition \ref{con: error}  assumes the sub-exponential distribution for the error vector, which is key to establishing
the deviation probability bound for $\|\widetilde{\bX}^T\bW\|_{2, \infty}$. Here $\|\cdot\|_{2, \infty}$ denotes the matrix rowwise ($2, \infty$)-norm defined as $\|\bM\|_{2, \infty}=\max_{i}(\sum_{j}m_{ij}^2)^{1/2}$ for any matrix $\bM=(m_{ij})$. Hereafter $p$ involved in the regularization parameter $\lambda$ and probability bounds is understood implicitly as $\max\{n, p\}$.

\begin{theorem}\label{thm: selection}
Assume that all the conditions of Theorem \ref{thm:Screening} and Conditions \ref{con: X}--\ref{con: error} hold, $q\leq p$, $\log p=o(n^{\eta})$ with 
$\eta=\min\{\eta_0, 1/2-2\xi\}$, and set $\lambda=c_3\sqrt{(\log p) /(nq)}$ with $c_3 > 0$ some constant. Then with probability at least $1-O\{\exp(-C n^{\eta_0/2})\}-O(p^{-c_4})$ for some constants $C, c_4 > 0$, 
the minimizer $\widehat{\bB}$ of (\ref{eq: problem}) satisfies
  \begin{align}
            (nq)^{-1/2}\|\widetilde{\bX}(\widehat{\bB}-\bB^*)\|_F
  & \leq \frac{8c_3}{\kappa}\sqrt{s(\log p)/n},  \label{eq: pred-loss}\\
               \frac{1}{\sqrt{q}}\|\widehat{\bB}-\bB^*\|_{2,\, 1} &\leq \frac{64c_3}{\kappa^2}s\sqrt{(\log p)/n}. \label{eq: est-loss-21}
  \end{align}
If in addition $\min_{j\in S}\|\bB^*_j\|/\sqrt{q}>128c_3\kappa^{-2}s\sqrt{(\log p)/n}$, then 
with the same probability
the row support of $\widetilde{\bB}$ is identical to 
$S$, where the matrix $\widetilde{\bB}$ is obtained by thresholding the $j$th row of $\widehat{\bB}$ to zero for each $j$ if $\|\widehat{\bB}_{j}\|/\sqrt{q}\leq 64c_3\kappa^{-2}s\sqrt{(\log p)/n}$.
Moreover, if the $RE(s)$ assumption in Condition \ref{con:RE-population} is replaced by $RE(2s)$, then it holds with the same probability that
  \begin{align} \label{eq: est-loss-L2}
     \frac{1}{\sqrt{q}}\|\widehat{\bB}-\bB^*\| _F
   \leq \frac{16\sqrt{10}c_3}{\kappa^2(2s)}\sqrt{s(\log p)/n}.
  \end{align}
\end{theorem}


Theorem \ref{thm: selection} establishes the model selection consistency of the IPDC followed by hard thresholding in terms of support union recovery. It also extends the oracle inequalities in Theorem 3.3 of \cite{lounici2009taking} and Corollary 4.1 of \cite{lounici2011oracle} in three important aspects: the inclusion of interaction terms, the analysis of large random design matrix, and the relaxed distributional assumption. Such extensions make the technical analyses more involved and challenging.  
We should remark that the same results as in Theorem \ref{thm: selection} hold with probability at least $1-O(p^{-c_4})$ for the regularized estimator with $d = \widetilde{p}$, that is, without the screening step.
It is also worth mentioning that the value of the regularization parameter  $\lambda$ in our Theorem \ref{thm: selection} is slightly larger than those  
used in \cite{lounici2009taking} and \cite{lounici2011oracle}, due to the more general model setting considered in our paper. In fact, such larger value of $\lambda$ is needed to suppress the additional noise caused by the presence of interactions and the heavier-tailed distribution of model errors. 

\section{Additional numerical studies} \label{SuppNum}

\subsection{Comparison of IPDC with individual components} \label{newsimu.dcsis2}
Recall that the new interaction screening approach of IPDC treats the screening for interactions and the screening for main effects as two separate components. Since the distance correlation can capture nonlinear dependency between variables, a natural question is whether either of these two components might suffice for the purpose of the joint screening for both interactions and main effects. To ease the presentation, the component for interaction screening is referred to as DCSIS-square, and the component for main effect screening is called DCSIS2 as described in Section \ref{Sec4.1.1}. Thus it is of interest to compare IPDC with both DCSIS2 and DCSIS-square. To this end, we revisit the setting 3 of Model 4 investigated in Section \ref{Sec4.1.1}; see Table \ref{tab:Screen} for the screening performance of DCSIS2 and IPDC.

Table \ref{dcsis2new} reports the comparison results of these three methods. We see that DCSIS2, which is designed specifically for main effect screening, fails to retain the important interaction $X_1 X_2$, and DCSIS-square, which is designed specifically for interaction screening, fails to retain the important main effect $X_{12}$. In contrast, the IPDC combines the strengths of its two individual components in screening for both interactions and main effects. Such an observation is in line with a key message spelled out in the Introduction, that is, a separate screening step for interactions can significantly enhance the screening performance if one aims at finding important interactions.

\begin{table}
\caption{Proportions of important main effects, important interaction, and all of them retained by different screening methods.\vspace{0.1in}} \label{dcsis2new}
\centering
\scalebox{1}{
\begin{tabular}{ l c c c c c}
  \toprule
Method & $X_{12} $ & $X_{22} $ & $X_1 X_2$ & All \\
\midrule
DCSIS2 & 1.00 & 1.00 & 0.06 & 0.06 \\
DCSIS-square &0.00	& 0.81 & 1.00 & 0.00 \\
IPDC & 1.00 & 1.00 & 1.00 & 1.00 \\
  \toprule
\end{tabular}}
\end{table}

\subsection{Performance of interaction and main effect selection} \label{Sec.Selection}

\subsubsection{Selection in single-response models}\label{Sec-Selection-univariate}
After the screening step, we further investigate the performance of selection for the interactions and main effects in the reduced feature space.
The selection step in the single-response examples (Models 1--4) is implemented by the Lasso. Thus we refer to each two-stage interaction screening and selection procedure by the SIS2-Lasso, DCSIS2-Lasso, SIRI-Lasso, IP-Lasso, and IPDC-Lasso, respectively. The oracle procedure, which assumes that the true underlying sparse interaction model is known in advance, is used as a benchmark for comparison. In particular, in Model 4 the indicator covariate $\mathbb{I}(X_{12}\geq 0)$ instead of the linear predictor $X_{12}$ is used in the oracle procedure.

Three performance measures, the prediction error (PE), the number of false positives (FP), and the number of false negatives (FN),  are employed to assess the variable selection performance of each method in the single-response examples. The PE is defined as $E(Y-\widehat{Y})^2$ with $\widehat Y$ the predicted response.  
We generate an independent test sample of size $10,000$ to calculate the PE. The FP is defined as the total number of unimportant interactions and main effects included in the final model, while the FN is defined as the total number of important interactions and main effects missed by the final model. 

\begin{table}
\caption{Means and standard errors (in parentheses) of different selection performance measures for Models 3 and 4 over 100 replications. \label{tab:Select}\vspace{0.1in}}
\centering
\scalebox{0.84}{
\begin{tabular}{lccccccc}
\midrule
   Method &  \multicolumn{3}{c}{Model 3}  & \multicolumn{3}{c}{Model 4} \\
    \cmidrule(lr{.75em}){2-4}  \cmidrule(lr{.75em}){5-7}
              & PE & FP & FN & PE & FP & FN \\
\midrule
\multicolumn{7}{c}{{Setting 1}: $(p, \rho)=(2000, 0.5)$} \\
SIS2-Lasso             &  33.79 (0.87)     & 39.25 (3.79) & 1.88 (0.04)	& 15.00 (0.32)	& 22.05 (2.50) &	1.23 (0.06) \\
DCSIS2-Lasso        &  3.94 (0.41)	& 0.51 (0.17)	& 0.32 (0.05)		& 	3.36 (0.38)	& 4.16 (0.72)	& 0.11 (0.04) \\
SIRI-Lasso        &  3.54 (0.40)	& 0.54 (0.36)	& 0.27 (0.05)		& 	4.34 (0.49)	& 1.63 (0.40)	& 0.31 (0.07)	\\
IP-Lasso        &  2.08 (0.28)	& 0.46 (0.11)	& 0.10 (0.03)		& 	2.38 (0.05)	& 4.24 (0.64)	& 0.07 (0.03)	\\
IPDC-Lasso               &  1.27 (0.10)	& 0.63 (0.20)	& 0.01 (0.01)		& 	2.27 (0.02)	& 3.32 (0.47) & 	0.01 (0.01)\\
Oracle                  &  1.017 (0.002)	 & 0 (0) & 	0 (0)		& 	1.022 (0.002)	& 0 (0)	& 0 (0)\\\vspace{-0.05in}\\
\multicolumn{7}{c}{{Setting 2}: $(p, \rho)=(5000, 0.5)$} \\
SIS2-Lasso             &  36.31 (0.51)	& 61.78 (2.85) 	& 1.97 (0.02)		& 	15.27 (0.24) & 	39.29 (2.33)	& 1.15 (0.04) \\
DCSIS2-Lasso        &  5.81 (0.44)	& 1.20 (0.55)	& 0.54 (0.05)		& 	4.17 (0.48)  & 	3.45 (0.53)	& 0.20 (0.05)	\\
SIRI-Lasso        &  4.48 (0.45)	& 0.42 (0.16)	& 0.37 (0.05)		& 	4.70 (0.52)  & 	2.24 (0.44)	& 0.37 (0.07)	\\
IP-Lasso        & 2.52 (0.33)	& 1.83 (0.61)	& 0.15 (0.04)		& 	2.51 (0.06)  & 	6.75 (1.24)	& 0.14 (0.04)	\\
IPDC-Lasso               &  1.38 (0.16)	& 0.91 (0.31)	& 0.02 (0.01)	& 		2.30 (0.02)  & 	4.39 (0.63)	& 0.01 (0.01) \\
Oracle                  &  1.009 (0.002)	 & 0 (0)	& 0 (0)		& 	1.014 (0.002) & 	0 (0)	& 0 (0)\\\vspace{-0.05in}\\
\multicolumn{7}{c}{{Setting 3}: $(p, \rho)=(2000, 0.1)$} \\
SIS2-Lasso             &  21.96 (0.19) & 	22.84 (3.18)	& 1.98 (0.01)		& 	13.15 (0.10)	& 15.68 (2.04)	& 1.24 (0.05) \\
DCSIS2-Lasso        &  18.85 (0.47)	& 9.32 (2.47)	& 1.70 (0.05)		& 	12.58 (0.28)	& 8.51 (1.73)	& 1.15 (0.05)	\\
SIRI-Lasso        &  14.45 (0.72)	& 0.40 (0.17)	& 1.28 (0.07)		& 	11.55 (0.45) & 	1.35 (0.43)	& 1.28 (0.07)	\\
IP-Lasso        &  6.23 (0.63)	& 4.20 (1.37)	& 0.46 (0.06)		& 	2.54 (0.17)  & 	6.28 (1.54)	& 0.05 (0.02)	\\
IPDC-Lasso               &  3.08 (0.44)	& 0.99 (0.21)	& 0.17 (0.04)		& 	2.26 (0.01) & 	4.00 (0.79) & 	0.00 (0.00)\\
Oracle                  &  1.017 (0.002)	 & 0 (0)	& 0 (0)		& 	1.022 (0.002)	& 0 (0)	& 0 (0)\\
\midrule
\end{tabular}
}
\end{table}

\begin{table}
\caption{Means and standard errors (in parentheses) of different selection performance measures for setting 2 of Model 2 over 100 replications.\label{tab: m2case2}\vspace{0.1in}}
\centering
\scalebox{1}{
\begin{tabular}{lccc}
\midrule
Method     & PE & FP & FN \\
\cline{1-4}
SIS2-Lasso             &  25.57 (1.61)     & 30.20 (3.31) & 1.62 (0.10) \\
DCSIS2-Lasso        &  3.20 (0.40)	& 1.85 (0.44) & 0.21 (0.04)\\
SIRI-Lasso        &  3.03 (0.38)	& 1.30 (0.23) & 0.20 (0.04)\\
IP-Lasso        &  4.05 (0.45)	& 4.79 (1.06) & 0.33 (0.05)\\
IPDC-Lasso        &  1.61 (0.20)	& 2.55 (0.49)	& 0.04 (0.02)\\
Oracle  &  1.014 (0.002)  &  0 (0)  &  0 (0) \\
\midrule
\end{tabular}
}
\end{table}

Table \ref{tab:Select} presents the means and standard errors of different performance measures.
Since the screening results by all methods in Model 1 under three different settings are almost identical, one can expect that the corresponding selection results should be very similar, which is indeed the case. Thus we omit the selection results for Model 1 to save space. For Model 2, the performance of all methods is very similar across three settings. As a result, we only present the selection results under setting 2 of this model in Table \ref{tab: m2case2} as a representative. The complete results are available upon request. 
Based on Tables \ref{tab:Select} and \ref{tab: m2case2}, the following observations can be made.

\begin{itemize}
\item In Model 2, we see that IPDC-Lasso performs the best and closest to the oracle procedure across all measures, as shown in Table \ref{tab: m2case2}. For Model 3 under all settings, our method IPDC-Lasso has far lower mean prediction error than all other methods except for the oracle, according to Table \ref{tab:Select}. 
The advantage of IPDC-Lasso over other methods is also evident in Model 4.
\item We remark that the gap between the prediction errors of the IPDC-Lasso and oracle in Model 4 is mainly because as mentioned before, the latter exploits the indicator covariate $\mathbb{I}(X_{12}\geq 0)$ whereas such prior information is unavailable to all other procedures. Even in this scenario of model misspecification, our method still performs well in identifying important interactions and main effects.

\end{itemize}

\subsubsection{Selection in multi-response model}

As mentioned before, interaction and main effect selection in the multi-response model setting (Model 5) is conducted through a two-step procedure in the reduced feature space obtained by screening. Such a method first selects rows of the regression coefficient matrix using the group Lasso, and then applies the Lasso to each individual response for further selection of the rows.
The goal of the individual Lasso is to eliminate the unimportant interactions and main effects that are included in the model recovered by the group Lasso. The resulting interaction screening and selection procedures are referred to as the SIS.max-GLasso-Lasso, SIS.sum-GLasso-Lasso, DCSIS-GLasso-Lasso, and IPDC-GLasso-Lasso, respectively. We also include for comparison the procedures that exploit only the group Lasso in the selection step, which are named by dropping the Lasso component. The oracle procedure is the ordinary least-squares estimation applied to each response separately on the corresponding true support.


The same performance measures as defined in Section \ref{Sec-Selection-univariate} are employed to evaluate different methods, except that the PE is now calculated as the average prediction error across all $q = 10$ responses. To further differentiate the false positives and false negatives for the main effects and interactions, we attach ``.main" and ``.int" to both measures of FP and FN as shown in Table \ref{tab: m5}.

\begin{table}
\caption{\label{tab: m5}Means and standard errors (in parentheses) of different selection performance measures for Model 5 over 100 replications.\vspace{0.1in}}
\centering
\scalebox{0.86}{
\begin{tabular}{lccccc}
\midrule
Method     & PE & FP.main & FP.int & FN.main & FN.int \\
\cline{1-6}
SIS.max-GLasso             &  6.24 (0.06)     & 127.64 (4.72) & 699.08 (27.37) & 1.22 (0.23) & 9.50 (0.12) \\
SIS.sum-GLasso        &  6.12 (0.08)	& 185.28 (3.64) & 810.72 (20.19) & 0.74 (0.14) & 9.06 (0.17) \\
DCSIS-GLasso        &  4.29 (0.14)	& 157.74 (3.98)	& 830.19 (23.00) & 0.50 (0.10) & 5.63 (0.29) \\
IPDC-GLasso             &  2.74 (0.09)     & 124.77 (3.02) & 878.93 (20.60) & 0.04 (0.02) & 2.46 (0.23) \\
SIS.max-GLasso-Lasso             &  5.11 (0.05)     & 11.92 (0.84) & 52.40 (3.17) & 3.44 (0.19) & 9.60 (0.10)\\
SIS.sum-GLasso-Lasso        &  4.99 (0.07)	& 15.75 (0.86) & 63.73 (3.39) & 3.07 (0.20) & 9.35 (0.16)\\
DCSIS-GLasso-Lasso        &  3.40 (0.12)	& 11.87 (0.70) & 62.98 (2.94) & 1.41 (0.17) & 6.36 (0.28) \\
IPDC-GLasso-Lasso             &  2.08 (0.09)     & 7.42 (0.36) & 58.95 (2.32) & 0.37 (0.09) & 3.28 (0.24)\\
Oracle & 1.048 (0.001) & 0 (0) & 0 (0) &  0 (0) & 0 (0) \\
\midrule
\end{tabular}
}
\end{table}%

Table \ref{tab: m5} reports the selection results for Model 5. The FP.int is relatively large for all methods since even after screening, there are still a large number of interactions left, due to the presence of multiple responses. We observe that a further step of individual Lasso implemented on the support of group lasso for each response separately can substantially reduce the FP for both interactions and main effects. Moreover, our method IPDC-GLasso-Lasso outperforms all competitors under all performance measures.

\subsection{Univariate gene expression study} \label{real:univ}
We study the the inbred mouse microarray gene expression data set in 
\citet{Lanetal:2006}. There are 60 mouse arrays, with 31 from female mice and 29 from male mice. The response variable is the gene expression level of stearoyl-CoA desaturase-1 (Scd1), a gene involved in fat storage. Specifically, Scd1 controls lipid metabolism and insulin sensitivity. The covariates are gene expression levels for 22,690 of the mice's other genes. Therefore, 
the sample size $n=60$, the number of covariates $p=22, 690$, and the number of responses $q=1$.
All variables involved in this studey are continuous. 

This data set is publicly available on the Gene Expression Omnibus website (\url{http://www.ncbi.nlm.nih.gov/geo}; accession number GSE3330),
and has been studied in 
\citet{hao2014interaction} and \citet{narisetty2014}. 
Following 
\citet{narisetty2014}, we randomly split the data into training and test sets of sizes $55$ and $5$, respectively. 
Furthermore, to ameliorate the numerical instability caused by the relatively small sample size we perform 200 random splits and calculate the mean prediction errors and the corresponding standard errors to better evaluate the performance of various methods.
We compare the IPDC with the SIS2, DCSIS2, SIRI, and IP. Detailed descriptions of all these methods can be found in Sections \ref{sec: simulation} and \ref{Sec.Selection}.

\begin{table}
\caption{Means and standard errors (in parentheses) of prediction error as well as numbers of selected main effects and interactions for each method in mice data.
\label{tab:diabetesMice}\vspace{0.1in}}
\centering
\scalebox{1}{
\begin{tabular}{ l c c c}
  \toprule
     & &  \multicolumn{2}{c}{Model Size} \\
     \cline {3-4}
  Method  & PE ($\times 10^{-3}$) & Main & Interaction  \\
   \midrule
  SIS2-Lasso      	&  	100.14 (5.57)	&  0.14 (0.03) & 9.56 (0.11) \\
  DCSIS2-Lasso 	&  	100.91 (6.12)	& 0.12 (0.03)  & 9.22 (0.11) \\
  SIRI-Lasso    	& 	247.53 (10.11) &  1.07 (0.03) & 0.46 (0.13) \\
  IP-Lasso     	& 	101.34 (4.99) &  5.00 (0.09) & 8.55 (0.21) \\
  IPDC-Lasso  	& 	96.55 (5.40)  & 3.04 (0.07)  & 7.31 (0.11) \\
  \midrule
\end{tabular}
}
\end{table}

The final selection results on the prediction error and selected model size are summarized in Table \ref{tab:diabetesMice}.
We see that IPDC-Lasso performs noticeably better than its competitors. Further, paired $t$-tests of prediction errors on the $200$ splits of IPDC-Lasso against SIS2-Lasso, DCSIS2-Lasso, SIRI-Lasso, and IP-Lasso lead to $p$-values $3.40\times10^{-2}$, $1.94\times10^{-2}$, $1.90\times10^{-36}$, and $2.64\times10^{-2}$, respectively. These test results demonstrate the significantly improved performance of IPDC over other methods at the $5\%$ level.

\section{Additional proofs of main results} \label{SuppB}

\subsection{Step 1.3 of Part 1 in the proof of Theorem \ref{thm:Screening}} \label{Thm1Part1Step1.3}

We now consider the term $\widehat T_{k1,\,3}^\ast-T_{k1,\,3}$. Applying the Cauchy-Schwarz inequality twice leads to
\begin{align}\label{eq: Tk1-3-bound-1}
    & T_{k1,\,3}
    \leq \left\{E\left[\psi^2(\by_1^{\ast}, \by_2^{\ast})\right]
            E\left[\phi^2(X_{1k}^{\ast}, X_{2k}^{\ast})\mathbb{I}\{\phi(X_{1k}^{\ast}, X_{2k}^{\ast})> M_1\}\right]\right\}^{1/2} \\\nonumber
   &\quad\leq E^{1/2}[\psi^2(\by_1^{\ast}, \by_2^{\ast})]\left\{E[\phi^4(X_{1k}^{\ast}, X_{2k}^{\ast})]P\{\phi(X_{1k}^{\ast}, X_{2k}^{\ast})> M_1\}\right\}^{1/4}.
\end{align}
In view of \eqref{eq: X1k-X2k}, 
we have
\begin{align}\label{eq: Tk1-3-bound-2}
   E[\phi^4(X_{1k}^{\ast}, X_{2k}^{\ast})]
  &\leq E[(X_{1k}^2+X_{2k}^2)^4]
  \leq E\{[2(X_{1k}^4+X_{2k}^4)]^2\} \\\nonumber
  & \leq E[8(X_{1k}^8+X_{2k}^8)]
    =16E(X_{1k}^8)
\end{align}
and by Bonferroni's inequality,
\begin{align}\label{eq: Tk1-3-bound-3}
        P&\{\phi(X_{1k}^{\ast}, X_{2k}^{\ast})> M_1\}
    \leq P(X_{1k}^2+X_{2k}^2> M_1)
   \leq   P(X_{1k}^2> M_1/2)  \\\nonumber
   &\quad+ P(X_{2k}^2> M_1/2)= 2P(X_{1k}^2> M_1/2)\\\nonumber
   &
    \leq 2 \exp(-c_0M_1/2)E[\exp(c_0X_{1k}^2)].
\end{align}

Combining \eqref{eq: y1y2-moment} with \eqref{eq: Tk1-3-bound-1}--\eqref{eq: Tk1-3-bound-3} and by Condition \ref{con: tail},  we obtain  $T_{k1,\,3}\leq\widetilde{C}_3 \exp(-8^{-1}c_0M_1)$, where $\widetilde{C}_3$ is some positive constant. Since
$M_1=n^{\xi_1}$, it holds that for any positive constant $\widetilde{C}$,
\begin{equation}\label{eq: Tk1-3-part0}
    0\leq T_{k1,\,3}
   \leq   \widetilde{C}_3 \exp(-8^{-1}c_0n^{\xi_1})
   \leq \widetilde{C}n^{-\kappa_2}/48
\end{equation}
for all $1\leq k\leq p$ when $n$ is sufficiently large.  This entails that
\begin{align}\label{eq: Tk1-3-part1}
  P&(\max_{1\leq k\leq p}|\widehat T_{k1,\,3}^\ast-T_{k1,\,3}|\geq \widetilde{C}n^{-\kappa_2}/24) \\ \nonumber
  & \leq  P(\max_{1\leq k\leq p}|\widehat T_{k1,\,3}^\ast|\geq \widetilde{C}n^{-\kappa_2}/48)
\end{align}
for all $n$ sufficiently large.  Thus  applying Markov's inequality and noting that $\widehat T_{k1,\,3}^\ast\geq 0$ and $E(\widehat T_{k1,\,3}^\ast)=T_{k1,\,3}$, we have
\begin{align*}
   P(|\widehat T_{k1,\,3}^\ast|\geq \delta/2)
  & \leq (\delta/2)^{-1}E(|\widehat T_{k1,\,3}^\ast|)
  \leq (\delta/2)^{-1}E(\widehat T_{k1,\,3}^\ast)\\
  &
  = (\delta/2)^{-1}T_{k1,\,3}
\end{align*}
for any $\delta>0$. Choosing $\delta=\widetilde{C}n^{-\kappa_2}/24$ in the above inequality and in view of \eqref{eq: Tk1-3-part0}, it follows that
$P(|\widehat T_{k1,\,3}^\ast|\geq \widetilde{C}n^{-\kappa_2}/48)\leq  48\widetilde{C}^{-1}\widetilde{C}_3 n^{\kappa_2}\exp(-8^{-1}c_0n^{\xi_1})$.
This inequality together with \eqref{eq: Tk1-3-part1} and Bonferroni's inequality yields
\begin{align} 
   P&(\max_{1\leq k\leq p}|\widehat T_{k1,\,3}^{\ast}-T_{k1,\,3}|\geq \widetilde{C}n^{-\kappa_2}/24)\\\nonumber
  &\quad\leq  48p\widetilde{C}^{-1}\widetilde{C}_3 n^{\kappa_2}\exp(-8^{-1}c_0n^{\xi_1}).
\end{align}

\subsection{Step 2 of Part 1 in the proof of Theorem \ref{thm:Screening}} \label{Thm1Part1Step2}

In this step, we handle the term $\max_{1\leq k\leq p}|\widehat T_{k2}-T_{k2}|$.  Define $T_{k2,\,1}=E\left[\phi(X_{1k}^{\ast}, X_{2k}^{\ast})\right]$ and $T_{k2,\,2}=E\left[\psi(\by_1^{\ast}, \by_2^{\ast})\right]$. Then 
$T_{k2}=T_{k2,\,1}T_{k2,\,2}$.  Similarly, $\widehat{T}_{k2}$ can be rewritten as $\widehat{T}_{k2}=\widehat{T}_{k2,\,1}\widehat{T}_{k2,\,2}$ by letting $\widehat{T}_{k2,\,1}=n^{-2}\sum_{i,j=1}^n\phi(X_{ik}^{\ast}, X_{jk}^{\ast})$ and $\widehat{T}_{k2,\,2}=n^{-2}\sum_{i,j=1}^n\psi(\by_i^{\ast}, \by_j^{\ast})$.  An application of similar arguments as in Step 1 results in that for any positive constant $\widetilde{C}$, there exist some positive constants $\widetilde{C}_1, \cdots, \widetilde{C}_4$ such that
\begin{align*}
      P\big(|\widehat{T}_{k2,\,1} - T_{k2,\,1}|\ge \widetilde{C}n^{-\kappa_2}/4\big)
   \leq \widetilde{C}_1\exp\{-\widetilde{C}_2n^{(1-2\kappa_2)/3}\},\nonumber\\
      P\big(|\widehat{T}_{k2,\,2} - T_{k2,\,2}|\ge \widetilde{C}n^{-\kappa_2}/4\big)
   \leq \widetilde{C}_3\exp\{-\widetilde{C}_4n^{(1-2\kappa_2)/5}\}.
\end{align*}

In view of \eqref{eq: X1k-X2k} and \eqref{eq: y1-y2}, we have $T_{k2,\,1}\leq 2E(X_{1k}^2)$ and $T_{k2,\,2}\leq 2E(\|\widetilde{\by}_{1}\|^2)$. By Condition \ref{con: tail}, $T_{k2,\,1}$ and $T_{k2,\,2}$ are uniformly bounded from above by some positive constant for all $1\leq k\leq p$.  Thus it follows from Lemma \ref{lem: AjBj-prod} that for any positive constant $\widetilde{C}$, there exist some positive constants $\widetilde{C}_5$ and $\widetilde{C}_6$ such that
\begin{align*}
P &\big(|\widehat{T}_{k2} - T_{k2}|\ge \widetilde{C}n^{-\kappa_2}/4\big)
   =  P\big(|\widehat{T}_{k2,\,1}\widehat{T}_{k2,\,2} - T_{k2,\,1}T_{k2,\,2}|\ge \widetilde{C}n^{-\kappa_2}/4\big)\\
    &\quad\leq \widetilde{C}_5\exp\{-\widetilde{C}_6n^{(1-2\kappa_2)/5}\}
\end{align*}
for all $1\leq k\leq p$. By Bonferroni's inequality, we obtain
\begin{align} 
       P&\big(\max_{1\leq k\leq p}|\widehat{T}_{k2} - T_{k2}|\ge \widetilde{C}n^{-\kappa_2}/4\big)\leq  \sum_{k=1}^p P\big(|\widehat{T}_{k2} - T_{k2}|\ge \widetilde{C}n^{-\kappa_2}/4\big) \\\nonumber
   &
   \leq p\widetilde{C}_5\exp\{-\widetilde{C}_6n^{(1-2\kappa_2)/5}\}.
\end{align}

\subsection{Step 3 of Part 1 in the proof of Theorem \ref{thm:Screening}} \label{Thm1Part1Step3}


We now consider the term $\widehat T_{k3}-T_{k3}$. Define a $U$-statistic
\begin{align*}
    \widehat T_{k3}^{\ast}=6[n(n-1)(n-2)]^{-1}\sum_{i<j<l}g(X_{ik}^{\ast}, \by_i^{\ast}; X_{jk}^{\ast}, \by_j^{\ast}; X_{lk}^{\ast}, \by_l^{\ast})
\end{align*}
with the kernel $g(X_{ik}^{\ast}, \by_i^{\ast}; X_{jk}^{\ast}, \by_j^{\ast}; X_{lk}^{\ast}, \by_l^{\ast})$ given by
\begin{align*}
g & (X_{ik}^{\ast}, \by_i^{\ast}; X_{jk}^{\ast}, \by_j^{\ast}; X_{lk}^{\ast}, \by_l^{\ast})
  = \phi(X_{ik}^{\ast}, X_{jk}^{\ast})\psi(\by_i^{\ast}, \by_l^{\ast})
         + \phi(X_{ik}^{\ast}, X_{lk}^{\ast})\psi(\by_i^{\ast}, \by_j^{\ast})\\
         &\quad\quad
         + \phi(X_{jk}^{\ast}, X_{ik}^{\ast})\psi(\by_j^{\ast}, \by_l^{\ast}) + \phi(X_{jk}^{\ast}, X_{lk}^{\ast})\psi(\by_j^{\ast}, \by_i^{\ast}) \nonumber \\
     &
         \quad\quad+ \phi(X_{lk}^{\ast}, X_{ik}^{\ast})\psi(\by_l^{\ast}, \by_j^{\ast})
         + \phi(X_{lk}^{\ast}, X_{jk}^{\ast})\psi(\by_l^{\ast}, \by_i^{\ast}).
\end{align*}
Then $\widehat T_{k3}=n^{-2}(n-1)(n-2)[\widehat T_{k3}^{\ast}+(n-2)^{-1}\widehat T_{k1}^{\ast}]$. By the triangle inequality, we deduce
\begin{align}\label{eq: Tk3-bound}
   |\widehat T_{k3} - T_{k3}|
   & =\Big|\frac{(n-1)(n-2)}{n^2}(\widehat T_{k3}^{\ast}-T_{k3})
      - \frac{3n-2}{n^2}T_{k3}
      \\\nonumber
   & \quad +\frac{n-1}{n^2}(\widehat T_{k1}^{\ast}-T_{k1})+ \frac{n-1}{n^2} T_{k1}\Big| \leq |\widehat T_{k3}^{\ast}-T_{k3}|\\\nonumber
   &\quad
       + |\frac{3}{n}T_{k3}|+|\widehat T_{k1}^{\ast}-T_{k1}|
       +|\frac{1}{n}T_{k1}|.
\end{align}
It follows from the Cauchy-Schwarz inequality and \eqref{eq: X1k-X2k}--\eqref{eq: y1-y2} that
\begin{align*}
         T_{k3} &
   \leq  \left\{E[\phi^2(X_{1k}^{\ast}, X_{2k}^{\ast})]
                 E[\psi^2(\by_1^{\ast}, \by_3^{\ast})]\right\}^{1/2}\\
                 &
   \leq \left\{E\left[(X_{1k}^2+X_{2k}^2)^2\right]
                 E\left[(\|\widetilde{\by}_1\|^2+\|\widetilde{\by}_3\|^2)^2\right]\right\}^{1/2}\nonumber\\
   &\leq   \left\{E[2(X_{1k}^4+X_{2k}^4)]
                 E[2(\|\widetilde{\by}_1\|^4+\|\widetilde{\by}_3\|^4)]\right\}^{1/2}\\
                 &
    = 4\left\{E(X_{1k}^4)
                 E(\|\widetilde{\by}_1\|^4)\right\}^{1/2}
\end{align*}
for all $1\leq k\leq p$.

In Step 1, we have shown that $T_{k1}\leq  4\left\{E(X_{1k}^4)E(\|\widetilde{\by}_1\|^4)\right\}^{1/2}$ for all $1\leq k\leq p$. By Condition \ref{con: tail},  $E(X_{1k}^4)$ and $E(\|\by_1\|^4)$ are uniformly bounded from above by some positive constant for all $1\leq k\leq p$. Note that $T_{k1}\geq 0$ and $T_{k3}\geq 0$. Thus for any positive constant $\widetilde{C}$, we have $\max_{1\leq k\leq p}|3n^{-1}T_{k3}|<\widetilde{C}n^{-\kappa_2}/16$ and $\max_{1\leq k\leq p}|n^{-1}T_{k1}|<\widetilde{C}n^{-\kappa_2}/16$ for all $n$ sufficiently large.  These two inequalities along with \eqref{eq: Tk3-bound} entail
\begin{align}\label{eq: Tk3-bound-1}
       P & (\max_{1\leq k\leq p}|\widehat T_{k3} - T_{k3}|\geq \widetilde{C}n^{-\kappa_2}/4)
 \leq   P(\max_{1\leq k\leq p}|\widehat T_{k3}^{\ast} - T_{k3}| \\ \nonumber
 & \quad \geq \widetilde{C}n^{-\kappa_2}/16)
        +P(\max_{1\leq k\leq p}|\widehat T_{k1}^{\ast} - T_{k1}|\geq \widetilde{C}n^{-\kappa_2}/16).
\end{align}
Replacing $\widetilde{C}$ with $\widetilde{C}/2$ in \eqref{eq: max-Tk1-star} gives
\begin{align}\label{eq: Tk3-bound-part2}
  P & (\max_{1\leq k\leq p}|\widehat T_{k1}^{\ast}-T_{k1}|\geq \widetilde{C}n^{-\kappa_2}/16)
  \leq  p\widetilde{C}_1\exp\{-\widetilde{C}_2n^{(1-2\kappa_2)/3-2\eta}\}\\\nonumber
  &\quad
      + \widetilde{C}_3\exp\{-\widetilde{C}_4n^{3\eta/2}\},
\end{align}
where $\widetilde{C}_1, \cdots, \widetilde{C}_4$ are some positive constants.

It remains to bound $P(\max_{1\leq k\leq p}|\widehat T_{k3}^{\ast} - T_{k3}|\geq \widetilde{C}n^{-\kappa_2}/16)$.   Let $T_{k3}=T_{k3,\,1}+T_{k3,\,2}+T_{k3,\,3}$ with
\begin{align*}
    T_{k3, \,1} &= E\left[\phi(X_{1k}^{\ast}, X_{2k}^{\ast})\psi(\by_1^{\ast}, \by_3^{\ast})
                          \mathbb{I}\{\phi(X_{1k}^{\ast}, X_{2k}^{\ast})\leq M_3\}\mathbb{I}\{\psi(\by_1^{\ast}, \by_3^{\ast})\leq M_4\}\right], \nonumber\\
    T_{k3, \,2} &= E\left[\phi(X_{1k}^{\ast}, X_{2k}^{\ast})\psi(\by_1^{\ast}, \by_3^{\ast})
                          \mathbb{I}\{\phi(X_{1k}^{\ast}, X_{2k}^{\ast})\leq M_3\}\mathbb{I}\{\psi(\by_1^{\ast}, \by_3^{\ast})> M_4\}\right], \nonumber\\
    T_{k3, \,3} &= E\left[\phi(X_{1k}^{\ast}, X_{2k}^{\ast})\psi(\by_1^{\ast}, \by_3^{\ast})
                          \mathbb{I}\{\phi(X_{1k}^{\ast}, X_{2k}^{\ast})> M_3\}\right].
\end{align*}
Similarly, write $\widehat T_{k3}^{\ast}$ as $\widehat T_{k3}^{\ast}=\widehat T_{k3,\,1}^{\ast}+\widehat T_{k3,\,2}^{\ast}+\widehat T_{k3,\,3}^{\ast}$, where
\begin{align*}
    \widehat T_{k3, \,1}^{\ast}
  = & \frac{1}{n(n-1)(n-2)}\sum_{i<j<l}\left[\phi(X_{ik}^{\ast}, X_{jk}^{\ast})\psi(\by_i^{\ast}, \by_l^{\ast})\mathbb{I}\{\phi(X_{ik}^{\ast}, X_{jk}^{\ast})\leq M_3\}\right.\\
  &\quad\cdot\mathbb{I}\{\psi(\by_i^{\ast}, \by_l^{\ast})\leq M_4\} \\
   &+ \phi(X_{ik}^{\ast}, X_{lk}^{\ast})\psi(\by_i^{\ast}, \by_j^{\ast})\mathbb{I}\{\phi(X_{ik}^{\ast}, X_{lk}^{\ast})\leq M_3\}\mathbb{I}\{\psi(\by_i^{\ast}, \by_j^{\ast})\leq M_4\} \nonumber\\
     & + \phi(X_{jk}^{\ast}, X_{ik}^{\ast})\psi(\by_j^{\ast}, \by_l^{\ast})\mathbb{I}\{\phi(X_{jk}^{\ast}, X_{ik}^{\ast})\leq M_3\}\mathbb{I}\{\psi(\by_j^{\ast}, \by_l^{\ast})\leq M_4\} \nonumber\\
     & + \phi(X_{jk}^{\ast}, X_{lk}^{\ast})\psi(\by_j^{\ast}, \by_i^{\ast})\mathbb{I}\{\phi(X_{jk}^{\ast}, X_{lk}^{\ast})\leq M_3\}\mathbb{I}\{\psi(\by_j^{\ast}, \by_i^{\ast})\leq M_4\} \nonumber\\
     & + \phi(X_{lk}^{\ast}, X_{ik}^{\ast})\psi(\by_l^{\ast}, \by_j^{\ast})\mathbb{I}\{\phi(X_{lk}^{\ast}, X_{ik}^{\ast})\leq M_3\}\mathbb{I}\{\psi(\by_l^{\ast}, \by_j^{\ast})\leq M_4\} \nonumber\\
     & +\left. \phi(X_{lk}^{\ast}, X_{jk}^{\ast})\psi(\by_l^{\ast}, \by_i^{\ast})\mathbb{I}\{\phi(X_{lk}^{\ast}, X_{jk}^{\ast})\leq M_3\}\mathbb{I}\{\psi(\by_l^{\ast}, \by_i^{\ast})\leq M_4\}\right]\nonumber\\
  =: & \frac{6}{n(n-1)(n-2)}\sum_{i<j<l}\widetilde{g}(X_{ik}^{\ast}, \by_i^{\ast}; X_{jk}^{\ast}, \by_j^{\ast}; X_{lk}^{\ast}, \by_l^{\ast}),
\end{align*}
\vspace{-0.2in}
\begin{align*}
    \widehat T_{k3, \,2}^{\ast}
  = & \frac{1}{n(n-1)(n-2)}\sum_{i<j<l}\left[\phi(X_{ik}^{\ast}, X_{jk}^{\ast})\psi(\by_i^{\ast}, \by_l^{\ast})\mathbb{I}\{\phi(X_{ik}^{\ast}, X_{jk}^{\ast})\leq M_3\}\right.\\
  &\quad \cdot\mathbb{I}\{\psi(\by_i^{\ast}, \by_l^{\ast})> M_4\} \nonumber\\
     & + \phi(X_{ik}^{\ast}, X_{lk}^{\ast})\psi(\by_i^{\ast}, \by_j^{\ast})\mathbb{I}\{\phi(X_{ik}^{\ast}, X_{lk}^{\ast})\leq M_3\}\mathbb{I}\{\psi(\by_i^{\ast}, \by_j^{\ast})> M_4\} \nonumber\\
     & + \phi(X_{jk}^{\ast}, X_{ik}^{\ast})\psi(\by_j^{\ast}, \by_l^{\ast})\mathbb{I}\{\phi(X_{jk}^{\ast}, X_{ik}^{\ast})\leq M_3\}\mathbb{I}\{\psi(\by_j^{\ast}, \by_l^{\ast})> M_4\} \nonumber\\
     & + \phi(X_{jk}^{\ast}, X_{lk}^{\ast})\psi(\by_j^{\ast}, \by_i^{\ast})\mathbb{I}\{\phi(X_{jk}^{\ast}, X_{lk}^{\ast})\leq M_3\}\mathbb{I}\{\psi(\by_j^{\ast}, \by_i^{\ast})> M_4\} \nonumber\\
     & + \phi(X_{lk}^{\ast}, X_{ik}^{\ast})\psi(\by_l^{\ast}, \by_j^{\ast})\mathbb{I}\{\phi(X_{lk}^{\ast}, X_{ik}^{\ast})\leq M_3\}\mathbb{I}\{\psi(\by_l^{\ast}, \by_j^{\ast})> M_4\} \nonumber\\
     & + \left.\phi(X_{lk}^{\ast}, X_{jk}^{\ast})\psi(\by_l^{\ast}, \by_i^{\ast})\mathbb{I}\{\phi(X_{lk}^{\ast}, X_{jk}^{\ast})\leq M_3\}\mathbb{I}\{\psi(\by_l^{\ast}, \by_i^{\ast})> M_4\}\right],
\end{align*}
\vspace{-0.2in}
\begin{align*}
    \widehat T_{k3, \,3}^{\ast}
  = & \frac{1}{n(n-1)(n-2)}\sum_{i<j<l}\left[\phi(X_{ik}^{\ast}, X_{jk}^{\ast})\psi(\by_i^{\ast}, \by_l^{\ast})\mathbb{I}\{\phi(X_{ik}^{\ast}, X_{jk}^{\ast})> M_3\}\right. \nonumber\\
     & + \phi(X_{ik}^{\ast}, X_{lk}^{\ast})\psi(\by_i^{\ast}, \by_j^{\ast})\mathbb{I}\{\phi(X_{ik}^{\ast}, X_{lk}^{\ast})> M_3\} \nonumber\\
     & + \phi(X_{jk}^{\ast}, X_{ik}^{\ast})\psi(\by_j^{\ast}, \by_l^{\ast})\mathbb{I}\{\phi(X_{jk}^{\ast}, X_{ik}^{\ast})> M_3\} \nonumber\\
     & + \phi(X_{jk}^{\ast}, X_{lk}^{\ast})\psi(\by_j^{\ast}, \by_i^{\ast})\mathbb{I}\{\phi(X_{jk}^{\ast}, X_{lk}^{\ast})> M_3\} \nonumber\\
     & + \phi(X_{lk}^{\ast}, X_{ik}^{\ast})\psi(\by_l^{\ast}, \by_j^{\ast})\mathbb{I}\{\phi(X_{lk}^{\ast}, X_{ik}^{\ast})> M_3\} \nonumber\\
     & + \left.\phi(X_{lk}^{\ast}, X_{jk}^{\ast})\psi(\by_l^{\ast}, \by_i^{\ast})\mathbb{I}\{\phi(X_{lk}^{\ast}, X_{jk}^{\ast})> M_3\}\right].
\end{align*}
Clearly, $\widehat T_{k3,\,1}^\ast$, $\widehat T_{k3,\,2}^\ast$, and $\widehat T_{k3,\,3}^\ast$ are unbiased estimators of $T_{k3,\,1}$, $T_{k3,\,2}$, and $T_{k3,\,3}$, respectively. By the
triangle inequality, we have
\begin{align}\label{eq: Tk3-1-tail}
   P&(\max_{1\leq k\leq p}|\widehat T_{k3}^{\ast} - T_{k3}|\geq \widetilde{C}n^{-\kappa_2}/16)\\\nonumber
  &\quad\leq \sum_{j=1}^3P(\max_{1\leq k\leq p}|\widehat T_{k3,\,j}^{\ast} - T_{k3,\,j}|\geq \widetilde{C}n^{-\kappa_2}/48).
\end{align}

Note that $\widetilde{g}$ defined in the expression for $\widehat T_{k3, \,1}^{\ast}$ is the kernel of the $U$-statistic $ \widehat T_{k3, \,1}^{\ast}$ of order 3. Applying similar arguments to those for dealing with $\widehat T_{k1,\,1}^{\ast}$  in Step 1 yields
\begin{align*}
   P(|\widehat T_{k3,\,1}^{\ast}-T_{k3,\,1}|\geq \widetilde{C}n^{-\kappa_2}/48)
   & \leq 2\exp\{-m_3 \widetilde{C}^2n^{-2\kappa_2}/(1152M_3^2M_4^2)\} \nonumber\\
   & \leq 2\exp\{-\widetilde{C}_5n^{1-2\kappa_2-2\xi_3-2\xi_4}\}
\end{align*}
with $\widetilde{C}_5$ some positive constant, by setting $M_3=n^{\xi_3}$ and $M_4=n^{\xi_4}$ with $\xi_3, \xi_4>0$ and noting that $m_3=\lfloor n/3\rfloor$. Thus it follows from Bonferroni's inequality that
\begin{align}\label{eq: Tk3-1-max}
   &P(\max_{1\leq k\leq p}|\widehat T_{k3,\,1}^{\ast}-T_{k3,\,1}|\geq \widetilde{C}n^{-\kappa_2}/48)\\\nonumber
   &\leq \sum_{1\leq k\leq p} P(|\widehat T_{k3,\,1}^{\ast}-T_{k3,\,1}|\geq \widetilde{C}n^{-\kappa_2}/48) \\\nonumber
   &\leq 2p\exp\{-\widetilde{C}_5n^{1-2\kappa_2-2\xi_3-2\xi_4}\}.
\end{align}
Using similar arguments to those for \eqref{eq: max-Tk1-2}--\eqref{eq: max-Tk1-3}, we can show that
\begin{align}
\label{eq: Tk3-2-max}
   P&(\max_{1\leq k\leq p}|\widehat T_{k3,\,2}^{\ast}-T_{k3,\,2}|\geq \widetilde{C}n^{-\kappa_2}/48)\\\nonumber
   &\quad\quad
   \leq \widetilde{C}_6n^{\kappa_2+\xi_3}\exp\{-2^{-3/2}c_0n^{\xi_4/2}\},  \ \\\label{eq: Tk3-3-max}
   P&(\max_{1\leq k\leq p}|\widehat T_{k3,\,3}^{\ast}-T_{k3,\,3}|\geq \widetilde{C}n^{-\kappa_2}/48)\\\nonumber
   &\quad\quad \leq p\widetilde{C}_7 n^{\kappa_2}\exp(-8^{-1}c_0n^{\xi_3}),
\end{align}
where $\widetilde{C}_6$ and $\widetilde{C}_7$ are some positive constants.

Combining the results in \eqref{eq: Tk3-1-tail}--\eqref{eq: Tk3-3-max} leads to
\begin{align*}
   P&(\max_{1\leq k\leq p}|\widehat T_{k3}^{\ast} - T_{k3}|\geq \widetilde{C}n^{-\kappa_2}/16)
  \leq  2p\exp\{-\widetilde{C}_5n^{1-2\kappa_2-2\xi_3-2\xi_4}\}
           \nonumber\\
       &\quad+ p\widetilde{C}_7 n^{\kappa_2}\exp(-8^{-1}c_0n^{\xi_3})+ \widetilde{C}_6n^{\kappa_2+\xi_3}\exp\{-2^{-3/2}c_0n^{\xi_4/2}\}.
\end{align*}
Let $\xi_3=(1-2\kappa_2)/3-2\eta$ and $\xi_4=3\eta$ with some $0<\eta<(1-2\kappa_2)/6$. Then we have
\begin{align*}
  P&(\max_{1\leq k\leq p}|\widehat T_{k3}^{\ast}-T_{k3}|\geq \widetilde{C}n^{-\kappa_2}/16)
  \leq p\widetilde{C}_8\exp\{-\widetilde{C}_9n^{(1-2\kappa_2)/3-2\eta}\}\\
  &\quad
       + \widetilde{C}_{10}\exp\{-\widetilde{C}_{11}n^{3\eta/2}\},
\end{align*}
where $\widetilde{C}_8, \cdots, \widetilde{C}_{11}$ are some positive constants. This inequality together with \eqref{eq: Tk3-bound-1}--\eqref{eq: Tk3-bound-part2} entails
\begin{align} 
   P&(\max_{1\leq k\leq p}|\widehat T_{k3}-T_{k3}|\geq \widetilde{C}n^{-\kappa_2}/4)
  \leq p\widetilde{C}_{1}\exp\{-\widetilde{C}_{2}n^{(1-2\kappa_2)/3-2\eta}\}\\\nonumber
  &\quad
       + \widetilde{C}_{3}\exp\{-\widetilde{C}_4n^{3\eta/2}\}
\end{align}
for some positive constants $\widetilde{C}_1, \cdots, \widetilde{C}_4$.

\subsection{Proof of Theorem \ref{thm: selection}}
For simplicity, we provide here only the proof for the case without variable screening, that is, $d_1=d_2=p$.  The case with variable screening can be proved using similar arguments, in view of the sure screening property established in Theorem \ref{thm:Screening}.  
By the definition of $\widehat{\bB}$, we have
  \begin{align*}
      \frac{1}{2nq}\|\bY-\widetilde{\bX}\widehat{\bB}\|_F^2
                 +\lambda\|\widehat{\bB}\|_{2,\, 1}
   \leq   \frac{1}{2nq}\|\bY-\widetilde{\bX}\bB^*\|_F^2
                 +\lambda\|\bB^*\|_{2,\, 1}.
  \end{align*}
Substituting $Y=\widetilde{\bX}\bB^*+\bW$ and rearranging terms yield
  \begin{align}\label{eq: opt}
      \frac{1}{2nq}\|\widetilde{\bX}\widehat{\bDel}\|_F^2
   \leq   \frac{1}{nq}\tr(\bW^T\widetilde{\bX}\widehat{\bDel})
                 +\lambda(\|\bB^*\|_{2,\, 1}-\|\widehat{\bB}\|_{2,\, 1}),
  \end{align}
where $\widehat{\bDel}=\widehat{\bB}-\bB^*$.
An application of the Cauchy-Schwarz inequality gives
  \begin{align}\label{eq: trace}
     \tr(\bW^T\widetilde{\bX}\widehat{\bDel})
    &=\tr(\widehat{\bDel}\bW^T\widetilde{\bX})
    =\sum_{k} \widehat{\bDel}_k\left[(\widetilde{\bX}^T\bW)_k\right]^T \\\nonumber
    &\leq \sum_{k} \|\widehat{\bDel}_k\|_2\cdot\| (\widetilde{\bX}^T\bW)_k\|_2
    \leq \|\widetilde{\bX}^T\bW\|_{2, \infty}\|\widehat{\bDel}\|_{2, 1},
  \end{align}
where $\widehat{\bDel}_{k}$ and $ (\widetilde{\bX}^T\bW)_{k}$ are the $k$th rows of $\widehat{\bDel}$ and $ \widetilde{\bX}^T\bW$, respectively.

{Note that $q \leq p$ and $\log p=o(n^{\eta})$ with $\eta=\min\{\eta_0, 1/2-2\xi\}$.}
By Lemmas \ref{lem: RE-sample}--\ref{lem: concentration}, with probability at least $1-O\{\exp(-\widetilde{C}_1n^{1/2-2\xi})\}-O(p^{-c})=1-O(p^{-c_4})$ for some constants $\widetilde{C}_1, c, c_4 > 0$ it holds that
  \begin{align}\label{eq: XE}
     \frac{1}{nq}\|\widetilde{\bX}^T\bW\|_{2, \infty}\leq \frac{\lambda}{2}
  \end{align}
and
  \begin{align}\label{eq: RE-sample}
     \min_{|J|\leq s,\, \bDel\in \mathbb{R}^{\widetilde{p}\times q}\backslash\{\bzero\},\, \|\bDel_{J^c}\|_{2, 1}\leq 3\|\bDel_J\|_{2, 1}}
    \frac{\|\widetilde{\bX}\bDel\|_F}{\sqrt{n}\|\bDel_J\|_F}\geq \frac{\kappa}{2}.
  \end{align}
From now on, we condition on the event that these two inequalities hold.  In view of \eqref{eq: opt} and \eqref{eq: trace}, we have the following basic inequality
  \begin{align}\label{eq: basic}
      \frac{1}{2nq}\|\widetilde{\bX}\widehat{\bDel}\|_F^2
   & +\frac{\lambda}{2}\|\widehat{\bB}-\bB^*\|_{2, 1}
   \leq  \lambda(\|\bB^*\|_{2,\, 1}-\|\widehat{\bB}\|_{2,\, 1}\\ \nonumber & \quad +\|\widehat{\bB}-\bB^*\|_{2, 1})
   \leq 2\lambda \|(\widehat{\bB}-\bB^*)_S\|_{2, 1},
  \end{align}
where we have used the fact that $\|(\bB^*)_{S^c}\|_{2,\, 1}-\|(\widehat{\bB})_{S^c}\|_{2,\, 1}+\|(\widehat{\bB}-\bB^*)_{S^c}\|_{2, 1}=0$ and the triangle inequality.

The basic inequality (\ref{eq: basic}) implies
  \begin{align}\label{eq: pred}
      \frac{1}{2nq}\|\widetilde{\bX}\widehat{\bDel}\|_F^2
   \leq 2\lambda \|\widehat{\bDel}_S\|_{2, 1}
    \leq 2\lambda \sqrt{s}\|\widehat{\bDel}_S\|_{F},
  \end{align}
where the last inequality holds since
  \begin{align}\label{eq: norm-inequ}
      \|\widehat{\bDel}_S\|_{2, 1}
    = \sum_{k\in S} \|\widehat{\bDel}_k\|_{2}
   \leq \sqrt{s\sum_{k\in S} \|\widehat{\bDel}_k\|_{2}^2}
     =\sqrt{s}\|\widehat{\bDel}_S\|_{F}.
  \end{align}
Moreover, it follows from \eqref{eq: basic} that $\|\widehat{\bDel}_{S^c}\|_{2, 1}\leq 3\|\widehat{\bDel}_{S}\|_{2, 1}$ and thus by \eqref{eq: RE-sample}, we have
$\|\widehat{\bDel}_S\|_F\leq 2\|\widetilde{\bX}\widehat{\bDel}\|_F/(\kappa\sqrt{n})$.
This inequality along with \eqref{eq: basic}--\eqref{eq: norm-inequ} yields
  \begin{align*} 
      \frac{1}{2nq}\|\widetilde{\bX}\widehat{\bDel}\|_F^2
   +\frac{\lambda}{2}\|\widehat{\bB}-\bB^*\|_{2, 1}
   \leq \frac{4\lambda \sqrt{s}\|\widetilde{\bX}\widehat{\bDel}\|_F}{\kappa\sqrt{n}},
  \end{align*}
which gives $n^{-1/2}\|\widetilde{\bX}\widehat{\bDel}\|_F
   \leq 8q\lambda \sqrt{s}/\kappa$.  Therefore, we obtain
  \begin{align*}
      \frac{1}{2nq}\|\widetilde{\bX}(\widehat{\bB}-\bB^*)\|_F^2
   +\frac{\lambda}{2}\|\widehat{\bB}-\bB^*\|_{2, 1}
   \leq \frac{32qs\lambda^2}{\kappa^2},
  \end{align*}
which completes the proof for the first part of Theorem \ref{thm: selection}.


We next proceed to prove the second part of Theorem \ref{thm: selection}.  We condition on the event that \eqref{eq: est-loss-21} holds.  Denote by $S(\bB)$ the row support of any matrix $\bB$. We need to show that with the same probability $S(\widetilde{\bB})=S(\bB^*)$ holds. To this end, we first prove $S(\bB^*)\subset S(\widetilde{\bB})$. For any $j_0\in S(\bB^*)$, if $j_0\not\in S(\widetilde{\bB})$ then the $j_0$th row of $\widetilde{\bB}$ is zero, which means $\|\widehat{\bB}_{j_0}\|\leq 64c_3\kappa^{-2}s\sqrt{q(\log p)/n}$. It then follows from the condition of
$\min_{j\in S}\|\bB^*_j\|>128{c_3}\kappa^{-2}s\sqrt{q(\log p)/n}$ that
  \begin{align*}
      \frac{1}{\sqrt{q}}\|\widehat{\bB}-\bB^*\|_{2, 1}
    \geq &       \frac{1}{\sqrt{q}}\|\widehat{\bB}_{j_0}-\bB^*_{j_0}\|
    \geq       \frac{1}{\sqrt{q}}(\|\bB^*_{j_0}\|-\|\widehat{\bB}_{j_0}\|) \\
   > & 
    64c_3\kappa^{-2}s\sqrt{(\log p)/n},
  \end{align*}
which leads to a contradiction to the estimation bound (\ref{eq: est-loss-21}). Thus it holds that $S(\bB^*)\subset S(\widetilde{\bB})$.
We can also show that $S(\widetilde{\bB})\subset S(\bB^*)$. In fact, if there exists some $j_0$ such that $j_0\in S(\widetilde{\bB})$ and $j_0\not\in S(\bB^*)$, then we have $\|\widehat{\bB}_{j_0}\|>64c_3\kappa^{-2}s\sqrt{q(\log p)/n}$ and $\bB^*_{j_0}=\bzero$, and thus
  \begin{align*}
      \frac{1}{\sqrt{q}}\|\widehat{\bB}-\bB^*\|_{2, 1}
    \geq        \frac{1}{\sqrt{q}}\|\widehat{\bB}_{j_0}-\bB^*_{j_0}\|
   >   64c_3\kappa^{-2}s\sqrt{(\log p)/n},
  \end{align*}
which contradicts again the bound (\ref{eq: est-loss-21}).  Combining
these results yields that with the same probability, the row support of $\widetilde{\bB}$ is identical to the true row support $S$. 

We finally prove \eqref{eq: est-loss-L2}. By assumption, the RE($2s$) condition holds. Using similar arguments as for proving \eqref{eq: XE}--\eqref{eq: RE-sample}, we can show that with probability at least $1-O\{\exp(-\widetilde{C}_1\\ \cdot n^{1/2-2\xi})\}-O(p^{-c})=1-O(p^{-c_4})$ for some constants $\widetilde{C}_1, c, c_4 > 0$, it holds that
  \begin{align}\label{eq: XE-2}
     \frac{1}{nq}\|\widetilde{\bX}^T\bW\|_{2, \infty}\leq \frac{\lambda}{2}
  \end{align}
and
  \begin{align}\label{eq: RE-sample-2}
     \min_{|J|\leq 2s,\, \bDel\in \mathbb{R}^{\widetilde{p}\times q}\backslash\{\bzero\},\, \|\bDel_{J^c}\|_{2, 1}\leq 3\|\bDel_J\|_{2, 1}}
    \frac{\|\widetilde{\bX}\bDel\|_F}{\sqrt{n}\|\bDel_J\|_F}\geq \frac{\kappa(2s)}{2}.
  \end{align}
Recall that $\widehat{\bDel}=\widehat{\bB}-\bB^*$.  Let $S'$ be a subset of $S^c$ corresponding to the $s$ 
largest values of $\|\widehat{\bDel}_k\|$.  Then we have $|S\cup S'|=2s$.
From now on, we condition on the event that inequalities \eqref{eq: XE-2} and \eqref{eq: RE-sample-2} hold.  Conditional on such an event, the basic inequality \eqref{eq: basic} still holds.  Thus we have  $\|\widehat{\bDel}_{S^c}\|_{2, 1}\leq 3\|\widehat{\bDel}_{S}\|_{2, 1}$, which entails
  \begin{align*}
      \|\widehat{\bDel}_{(S\cup S')^c}\|_{2,\, 1}
      \leq \|\widehat{\bDel}_{{S}^c}\|_{2,\, 1}
      \leq 3\|\widehat{\bDel}_{S}\|_{2,\, 1}
      \leq 3\|\widehat{\bDel}_{S\cup S'}\|_{2,\, 1}.
  \end{align*}
This together with \eqref{eq: RE-sample-2} yields  $\|\widehat{\bDel}_{S\cup S'}\|_F\leq 2\|\widetilde{\bX}\widehat{\bDel}\|_F/(\kappa(2s)\sqrt{n})$. From \eqref{eq: pred}, we have
  \begin{align*}
 \frac{1}{2nq}\|\widetilde{\bX}\widehat{\bDel}\|_F^2 \leq 2\lambda \sqrt{s}\|\widehat{\bDel}_S\|_{F}\leq 2\lambda \sqrt{s}\|\widehat{\bDel}_{S\cup S'}\|_{F}.
  \end{align*}
Combining these two results gives
  \begin{align}\label{eq: bound-2s}
     \|\widehat{\bDel}_{S\cup S'}\|_F\leq 16q\lambda \sqrt{s}/\kappa^2(2s).
  \end{align}

Since the $j$th largest norm in the set $\{\|\widehat{\bDel}_k\|: k\in S^c\}$ is bounded from above by $\|\widehat{\bDel}_{S^c}\|_{2,\, 1}/j$, it holds that
  \begin{align*}
   \sum_{k\in {(S\cup S')^c} } \|\widehat{\bDel}_k\|^2
    &\leq \sum_{k=s+1}^{\widetilde{p}-s}\frac{\|\widehat{\bDel}_{S^c}\|_{2,\, 1}^2}{k^2}
    \leq  \frac{\|\widehat{\bDel}_{S^c}\|_{2,\, 1}^2}{s}
    \leq \frac{9\|\widehat{\bDel}_{S}\|_{2,\, 1}^2}{s}\\
    &
     \leq 9\sum_{k\in S}\|\widehat{\bDel}_{k}\|^2
    \leq  9\sum_{k\in {S\cup S'}}\|\widehat{\bDel}_{k}\|^2,
  \end{align*}
which results in $\|\widehat{\bDel}\|_F^2\leq 10 \|\widehat{\bDel}_{S\cup S'}\|_{F}^2$. This inequality along with \eqref{eq: bound-2s} yields
  \begin{align*}
     \frac{1}{\sqrt{q}}\|\widehat{\bDel}\|_F\leq \frac{16\sqrt{10}c_3}{\kappa^2(2s)}\sqrt{s(\log p)/n},
  \end{align*}
which concludes the proof for the third part of Theorem \ref{thm: selection}.

\medskip
\section{Additional technical details and lemmas} \label{AppB}

\subsection{Terms $E(Y|X_j)$ and $E(Y^2|X_j)$ under model \eqref{eq: simple-model0}} \label{AppD}
Since the covariates $X_1, \cdots, X_p$  are all independent with mean zero and the random error $W$ is of mean zero and independent of all $X_j$'s, it is immediate that  $E(Y|X_j) = \alpha + \beta_j X_j $.
We now calculate $E(Y^2|X_j)$.  Define
\begin{align*}
J_1&=\sum_{j=1}^p\beta_j X_j, \quad J_2=\sum_{k=1}^{p-1}\sum_{\ell = k+1}^p \gamma_{k\ell}X_kX_{\ell},\quad J_3=\sum_{k\neq j}\beta_kX_k,\\
 J_4& =\sum_{k=1}^{j-1}\gamma_{kj}X_k+\sum_{\ell=j+1}^p\gamma_{j\ell}X_{\ell},\quad J_5=\sum_{k=1, k\neq j}^{p-1}\sum_{\ell = k+1, \ell\neq j}^p \gamma_{k\ell}X_kX_{\ell}.
\end{align*}
Then we have
$Y=\alpha+J_1+J_2+W$ with $J_1=\beta_jX_j+J_3$ and $J_2=J_4X_j+J_5$, and $J_3$, $J_4$, and $J_5$ are independent of $X_j$. Applying the properties of conditional expectation yields
\begin{align*}
E[(\alpha+J_1+J_2)W|X_j]
&=  E\{E[(\alpha+J_1+J_2)W|X_1, \cdots, X_p]|X_j\} \\
&= E[(\alpha+J_1+J_2) E(W|X_1, \cdots, X_p)|X_j]
=0
  \end{align*}
and
\begin{align*}
E&[(\alpha+J_1+J_2)^2|X_j]
=  E\{[(\beta_j+J_4)X_j+(\alpha+J_3+J_5)]^2|X_j\}  \\
=& X_j^2E[(\beta_j+J_4)^2] + 2X_jE[(\beta_j+J_4)(\alpha+J_3+J_5)] + E[(\alpha+J_3+J_5)^2]\\
= & \left[\beta_j^2 + \sum_{k=1}^{j-1}\gamma_{kj}^2E(X_k^2) + \sum_{\ell=j+1}^p\gamma_{j\ell}^2E(X_{\ell}^2)\right]X_j^2  \\
&\quad + 2\left[\beta_j\alpha+\sum_{k=1}^{j-1}\beta_k\gamma_{kj}E(X_k^2) +\sum_{\ell=j+1}^{p}\beta_{\ell}\gamma_{j\ell}E(X_{\ell}^2)\right]X_j \\
  &\quad  +\alpha^2+\sum_{k\neq j}\beta_k^2E(X_k^2)+\sum_{k=1, k\neq j}^{p-1}\sum_{\ell = k+1, \ell\neq j}^p \gamma_{k\ell}^2 E(X_k^2)E(X_{\ell}^2).
  \end{align*}
Therefore, it holds that
\begin{align*}
E(Y^2|X_j)
& = E[(\alpha+J_1+J_2)^2|X_j] + 2E[(\alpha+J_1+J_2)W|X_j]  + E(W^2|X_j)  \\
&= \left[\beta_j^2 + \sum_{k=1}^{j-1}\gamma_{kj}^2E(X_k^2)  + \sum_{\ell=j+1}^p\gamma_{j\ell}^2E(X_{\ell}^2)\right]X_j^2\\
 &\quad+ 2\left[\beta_j\alpha+\sum_{k=1}^{j-1}\beta_k\gamma_{kj}E(X_k^2)+\sum_{\ell=j+1}^{p}\beta_{\ell}\gamma_{j\ell}E(X_{\ell}^2)\right]X_j+C_j,
  \end{align*}
where $C_j=\alpha^2+\sum_{k\neq j}\beta_k^2E(X_k^2)+\sum_{k=1, k\neq j}^{p-1}\sum_{\ell = k+1, \ell\neq j}^p \gamma_{k\ell}^2 E(X_k^2)E(X_{\ell}^2)+\sigma^2$ is a constant that is free of $X_j$, and $\sigma^2$ is the variance of $W$.


\subsection{Lemma \ref{lem: AjBj-prod} and its proof}

\begin{lemma}\label{lem: AjBj-prod}
Let $\widehat{A}$ and $\widehat{B}$ be estimates of $A$ and $B$, respectively, based on a sample of size $n$. Assume that both $A$ and $B$ are bounded and for any constant $\widetilde{C} > 0$, there exist positive constants  $\widetilde{C}_1, \cdots, \widetilde{C}_4$ such that
  \begin{align*}
       & P\left( |\widehat{A}-A|\geq \widetilde{C}n^{-\kappa}\right)
            \leq \widetilde{C}_1\exp\left\{-\widetilde{C}_2n^{f(\kappa)}\right\}\\
       & P\left(|\widehat{B}-B|\geq \widetilde{C}n^{-\kappa}\right)
            \leq \widetilde{C}_3\exp\left\{-\widetilde{C}_4n^{f(\kappa)}\right\}
  \end{align*}
with $f(\kappa)$ some function of $\kappa$.
Then for any constant $\widetilde{C}> 0$, there exist positive constants $\widetilde{C}_5$ and $\widetilde{C}_{6}$ such that
   \begin{align*}
      P(|\widehat{A}\widehat{B}-AB|\geq \widetilde{C}n^{-\kappa}) \leq \widetilde{C}_5\exp\left\{-\widetilde{C}_{6}n^{f(\kappa)}\right\}.
  \end{align*}
\end{lemma}

\textit{Proof}.
Note that $|\widehat{A}\widehat{B}-AB|\leq  |\widehat{A}(\widehat{B}-B)| +  |(\widehat{A}-A)B| $. Thus for any positive constant $\widetilde{C}$, we have
  \begin{align} \label{eq: AjB}
       P&(|\widehat{A}\widehat{B}-AB|\geq \widetilde{C}n^{-\kappa})
     \leq P( |\widehat{A}(\widehat{B}-B)|\geq \widetilde{C}n^{-\kappa}/2)\\\nonumber
     &\quad
    + P( |(\widehat{A}-A)B|\geq \widetilde{C}n^{-\kappa}/2).
  \end{align}
We first deal with the second term on the right hand side of \eqref{eq: AjB}.  Since both $A$ and $B$ are bounded, there exists some positive constant $L$ such that $|A|\leq L$ and $|B|\leq L$. It follows that
  \begin{align} \label{eq: AjB-part2}
    P&( |(\widehat{A}-A)B|\geq \widetilde{C}n^{-\kappa}/2)
      \leq  P( |\widehat{A}-A|L\geq \widetilde{C}n^{-\kappa}/2)  \\\nonumber
    &= P\{ |\widehat{A}-A|\geq (2L)^{-1}\widetilde{C}n^{-\kappa_1}\}
    \leq \widetilde{C}_1\exp\left\{-\widetilde{C}_2n^{f(\kappa_1)}\right\},
  \end{align}
where $\widetilde{C}_1$ and $\widetilde{C}_2$ are some positive constants.

We next consider the first term on the right hand side of \eqref{eq: AjB}. Note that
  \begin{align} \label{eq: AjB-part1}
  &P( |\widehat{A}(\widehat{B}-B)|\geq \widetilde{C}n^{-\kappa}/2)
   \leq   P\Big\{ |\widehat{A}(\widehat{B}-B)|\geq \widetilde{C}n^{-\kappa}/2, \\\nonumber
   & \quad |\widehat{A}|\geq L+\frac{\widetilde{C}}{2}n^{-\kappa}\Big\}   + P\Big( |\widehat{A}(\widehat{B}-B)|\geq \frac{\widetilde{C}}{2}n^{-\kappa}, |\widehat{A}|< L+\frac{\widetilde{C}}{2}n^{-\kappa}\Big)  \\\nonumber
   &\leq   P(|\widehat{A}|\geq L+\frac{\widetilde{C}}{2}n^{-\kappa})
   + P( |\widehat{A}(\widehat{B}-B)|\geq \frac{\widetilde{C}}{2}n^{-\kappa}, |\widehat{A}|< L+\widetilde{C})  \\\nonumber
   &\leq   P(|\widehat{A}|\geq L+\widetilde{C}n^{-\kappa}/2)
   + P\{ (L+\widetilde{C})|\widehat{B}-B|\geq \widetilde{C}n^{-\kappa}/2\}.
  \end{align}
We will bound the two terms on the right hand side of \eqref{eq: AjB-part1} separately.  It follows from $|A|\leq L$ that
  \begin{align}\label{eq: AjB-part1-1}
          P &(|\widehat{A}|\geq L+\widetilde{C}n^{-\kappa}/2)
    \leq  P(|\widehat{A}-A| +|A| \geq L+\widetilde{C}n^{-\kappa}/2) \\
    \nonumber
     & \leq  P\{|\widehat{A}-A| \geq 2^{-1}\widetilde{C}n^{-\kappa}\}
    \leq  \widetilde{C}_3\exp\left\{-\widetilde{C}_4n^{f(\kappa_1)}\right\}, \nonumber
  \end{align}
where $\widetilde{C}_3$ and $\widetilde{C}_4$ are some positive constants.  It also holds that
  \begin{align*}
       P( (L+\widetilde{C})|\widehat{B}-B|\geq \widetilde{C}n^{-\kappa}/2)
     & =  P\{ |\widehat{B}-B|\geq (2L+2\widetilde{C})^{-1}\widetilde{C}n^{-\kappa}\} \\
     & \leq  \widetilde{C}_7\exp\left\{-\widetilde{C}_8n^{f(\kappa)}\right\},
  \end{align*}
where $\widetilde{C}_7$ and $\widetilde{C}_8$ are some positive constants.  This inequality together with \eqref{eq: AjB}--\eqref{eq: AjB-part1-1} entails
  \begin{align*}
      P&( |\widehat{A}\widehat{B}-AB|\geq \widetilde{C}n^{-\kappa})
  \leq \widetilde{C}_1\exp\left\{-\widetilde{C}_2n^{f(\kappa)}\right\}
       +\widetilde{C}_3\exp\left\{-\widetilde{C}_4n^{f(\kappa)}\right\}\\
       & \quad
       +\widetilde{C}_7\exp\left\{-\widetilde{C}_8n^{f(\kappa)}\right\}
  \leq     \widetilde{C}_5\exp\left\{-\widetilde{C}_6n^{f(\kappa)}\right\},
  \end{align*}
where $\widetilde{C}_5=\widetilde{C}_1+\widetilde{C}_3+\widetilde{C}_7$ and $\widetilde{C}_6=\min\{\widetilde{C}_2, \widetilde{C}_4, \widetilde{C}_8\}$.

\subsection{Lemma \ref{lem: Bj-root} and its proof}

For any set $\mathcal{D}$, we denote by $|\mathcal{D}|$ its cardinality throughout the paper.

\begin{lemma}\label{lem: Bj-root}
Let $\widehat{B}_j \geq 0$ be an estimate of $B_j$ based on a sample of size $n$ for each $j\in \mathcal{D}\subset\{1,\cdots, p\}$.  Assume that $\min_{j\in\mathcal{D}}B_j\geq L$ for some positive constant $L$, and for any constant $\widetilde{C} > 0$, there exist positive constants $\widetilde{C}_1$ and $ \widetilde{C}_2$ such that
  \begin{eqnarray*}
      P\left(\max_{j\in\mathcal{D}}|\widehat{B}_j-B_j|\geq \widetilde{C}n^{-\kappa}\right)\leq |\mathcal{D}|\widetilde{C}_1\exp\left\{-\widetilde{C}_2n^{f(\kappa)}\right\}
  \end{eqnarray*}
with $f(\kappa)$ some function of $\kappa$. Then for any constant $\widetilde{C} > 0$, there exist positve constants $\widetilde{C}_3$ and $\widetilde{C}_4$ such that
  \begin{eqnarray*}
      P\left( \max_{j\in\mathcal{D}}|\sqrt{\widehat{B}_j}-\sqrt{B_j}|\geq \widetilde{C}n^{-\kappa}\right)
         \leq |\mathcal{D}| \widetilde{C}_3\exp\left\{-\widetilde{C}_4n^{f(\kappa)}\right\}.
  \end{eqnarray*}
\end{lemma}

\textit{Proof}.
Since $\min_{j\in\mathcal{D}}B_j\geq L$ for some positive constant $L$, there exists a constant $L_0$ such that $0<L_0<L$. Note that for any positive constant $\widetilde{C}$,
  \begin{align} \label{eq: Bj-root}
      P&(\max_{j\in\mathcal{D}} |\sqrt{\widehat{B}_j}-\sqrt{B_j}|\geq \widetilde{C}n^{-\kappa})
    \leq  P\Big\{\max_{j\in\mathcal{D}} |\sqrt{\widehat{B}_j}-\sqrt{B_j}|\geq \widetilde{C}n^{-\kappa}, \\\nonumber
    &\quad\min_{j\in\mathcal{D}}|\widehat{B}_j|\leq L-L_0n^{-\kappa}\Big\} +  P\Big\{\max_{j\in\mathcal{D}} |\sqrt{\widehat{B}_j}-\sqrt{B_j}|\geq \widetilde{C}n^{-\kappa}, \\\nonumber
    & \quad \min_{j\in\mathcal{D}}|\widehat{B}_j|> L-L_0n^{-\kappa}\Big\}
    \leq  P( \min_{j\in\mathcal{D}}|\widehat{B}_j|\leq L-L_0n^{-\kappa})\\\nonumber
       &\quad + P( \max_{j\in\mathcal{D}} \frac{|\widehat{B}_j-B_j|}{|\sqrt{\widehat{B}_j}+\sqrt{B_j}|}\geq \widetilde{C}n^{-\kappa},  \min_{j\in\mathcal{D}}|\widehat{B}_j|>L-L_0).
  \end{align}
We first consider the first term on the right hand side of \eqref{eq: Bj-root}.  It follows from $\min_{j\in\mathcal{D}}B_j\geq L$ that
  \begin{align} \label{eq: Bj-bound}
        P&( \min_{j\in\mathcal{D}}|\widehat{B}_j|\leq L-L_0n^{-\kappa})
        \leq P\Big\{ \min_{j\in\mathcal{D}}|B_j|- \max_{j\in\mathcal{D}}|\widehat{B}_j-B_j|\\\nonumber
        &\quad\quad \leq L-L_0n^{-\kappa}\Big\}\leq   P(\max_{j\in\mathcal{D}}  |\widehat{B}_j-B_j |\geq L_0n^{-\kappa}) \\\nonumber
         &\quad
    \leq |\mathcal{D}|\widetilde{C}_1\exp\left\{-\widetilde{C}_2n^{f(\kappa)}\right\},
  \end{align}
where $\widetilde{C}_1$ and $\widetilde{C}_2$ are some positive constants.

Next we consider the second term on the right hand side of \eqref{eq: Bj-root}. For any positive constant $\widetilde{C}$, we have
  \begin{align} \label{eq: Bj-root-part2}
      P&(\max_{j\in\mathcal{D}}  \frac{|\widehat{B}_j-B_j|}{|\sqrt{\widehat{B}_j}+\sqrt{B_j}|}\geq \widetilde{C}n^{-\kappa},  \min_{j\in\mathcal{D}}|\widehat{B}_j|>L-L_0)\\\nonumber
    &\leq    P\{\max_{j\in\mathcal{D}}|\widehat{B}_j-B_j| \geq \widetilde{C}(\sqrt{L-L_0}+\sqrt{L})n^{-\kappa}\}\\\nonumber
    &\leq   |\mathcal{D}|\widetilde{C}_5\exp\left\{-\widetilde{C}_6n^{f(\kappa)}\right\},
  \end{align}
where $\widetilde{C}_5$ and $\widetilde{C}_6$ are some positive constants.
Combining \eqref{eq: Bj-root}--\eqref{eq: Bj-root-part2} yields
  \begin{equation}
           P( \max_{j\in\mathcal{D}}|\sqrt{\widehat{B}_j}-\sqrt{B_j}|\geq \widetilde{C}n^{-\kappa})
         \leq |\mathcal{D}|\widetilde{C}_3\exp\left\{-\widetilde{C}_4n^{f(\kappa)}\right\},
  \end{equation}
where $\widetilde{C}_3=\widetilde{C}_1+\widetilde{C}_5$ and $\widetilde{C}_4=\min\{\widetilde{C}_2, \widetilde{C}_6\}$.

\subsection{Lemma \ref{lem: AjBj-ratio} and its proof}

\begin{lemma}\label{lem: AjBj-ratio}
Let $\widehat{A}_j$ and $\widehat{B}_j$ be estimates of $A_j$ and $B_j$, respectively, based on a sample of size $n$ for each $j\in \mathcal{D}\subset\{1,\cdots, p\}$. Assume that $A_j$ and $B_j$ satisfy $\max_{j\in\mathcal{D}}|A_j|\leq L_1$ and $\min_{j\in\mathcal{D}}|B_j|\geq L_2$ for some constants $L_1, L_2 > 0$, and for any constant $\widetilde{C} > 0$, there exist constants $\widetilde{C}_1, \cdots, \widetilde{C}_6 > 0$ such that
  \begin{align*}
     & P\big(\max_{j\in\mathcal{D}} |\widehat{A}_j-A_j|\geq \widetilde{C}n^{-\kappa}\big)
        \leq  |\mathcal{D}|\widetilde{C}_1\exp\big\{-\widetilde{C}_2n^{f(\kappa)}\big\}+ \widetilde{C}_3\exp\big\{-\widetilde{C}_4n^{g(\kappa)}\big\},\\
     & P\big(\max_{j\in\mathcal{D}} |\widehat{B}_j-B_j|\geq \widetilde{C}n^{-\kappa}\big)
      \leq |\mathcal{D}|\widetilde{C}_5\exp\big\{-\widetilde{C}_6n^{f(\kappa)}\big\}
  \end{align*}
with $f(\kappa)$ and $g(\kappa)$ some functions of $\kappa$.
Then for any constant $\widetilde{C} > 0$, there exist positive constants $\widetilde{C}_7, \cdots, \widetilde{C}_{10}$ such that
  \begin{align*}
     P \left( \max_{j\in\mathcal{D}}\left|\frac{\widehat{A}_j}{\widehat{B}_j}-\frac{A_j}{B_j}\right|\geq \widetilde{C}n^{-\kappa}\right)
         &\leq    |\mathcal{D}|\widetilde{C}_7\exp\left\{-\widetilde{C}_8n^{f(\kappa)}\right\}\\
         &\quad+ \widetilde{C}_9\exp\left\{-\widetilde{C}_{10}n^{g(\kappa)}\right\}.
  \end{align*}
\end{lemma}

\textit{Proof}.  Since $\min_{j\in\mathcal{D}}|B_j|\geq L_2>0$, there exists some constant $L_0$ such that $0<L_0<L_2$. Note that for any positive constant $\widetilde{C}$, we have
  \begin{align} \label{eq: AjBj-ratio}
      P&(\max_{j\in\mathcal{D}} |\frac{\widehat{A}_j}{\widehat{B}_j}-\frac{A_j}{B_j}|\geq \widetilde{C}n^{-\kappa})
    \leq  P\Big\{\max_{j\in\mathcal{D}} |\frac{\widehat{A}_j}{\widehat{B}_j}-\frac{A_j}{B_j}|\geq \widetilde{C}n^{-\kappa}, \\\nonumber &\quad \min_{j\in\mathcal{D}} |\widehat{B}_j|\leq L_2-L_0n^{-\kappa}\Big\}
           +  P\Big\{\max_{j\in\mathcal{D}} |\frac{\widehat{A}_j}{\widehat{B}_j}-\frac{A_j}{B_j}|\geq \widetilde{C}n^{-\kappa}, \\\nonumber
           &\quad \min_{j\in\mathcal{D}} |\widehat{B}_j|> L_2-L_0n^{-\kappa}\Big\}
    \leq  P( \min_{j\in\mathcal{D}}|\widehat{B}_j|\leq L_2-L_0n^{-\kappa})\\\nonumber
        &\quad + P(\max_{j\in\mathcal{D}} |\frac{\widehat{A}_j}{\widehat{B}_j}-\frac{A_j}{B_j}|\geq \widetilde{C}n^{-\kappa}, \min_{j\in\mathcal{D}} |\widehat{B}_j|> L_2-L_0).
  \end{align}
We start with the first term on the right hand side of \eqref{eq: AjBj-ratio}. In light of $\min_{j\in\mathcal{D}}|B_j|\geq L_2$, we deduce
  \begin{align} \label{eq: AjBj-ratio-part1}
       P&( \min_{j\in\mathcal{D}}|\widehat{B}_j|\leq L_2-L_0n^{-k})
     \leq   P\Big\{\min_{j\in\mathcal{D}} |B_j| -  \max_{j\in\mathcal{D}}|\widehat{B}_j-B_j |\\\nonumber
     & \quad\quad \leq L_2-L_0n^{-\kappa}\Big\}
      \leq   P(\max_{j\in\mathcal{D}}  |\widehat{B}_j-B_j|\geq L_0n^{-\kappa})\\\nonumber
      &\quad
     \leq |\mathcal{D}|\widetilde{C}_1\exp\left\{-\widetilde{C}_2n^{f(\kappa)}\right\},
  \end{align}
where $\widetilde{C}_1$ and $\widetilde{C}_2$ are some positive constants.

The second term on the right hand side of \eqref{eq: AjBj-ratio} can be bounded as
  \begin{align} \label{eq: AjBj-ratio-part2}
      P&(\max_{j\in\mathcal{D}} |\frac{\widehat{A}_j}{\widehat{B}_j}-\frac{A_j}{B_j}|\geq \widetilde{C}n^{-\kappa}, \, \min_{j\in\mathcal{D}}|\widehat{B}_j|> L_2-L_0)  \\\nonumber
     &\leq  P( \max_{j\in\mathcal{D}}|\frac{\widehat{A}_j}{\widehat{B}_j}-\frac{A_j}{\widehat{B_j}}|\geq \widetilde{C}n^{-\kappa}/2, \, \min_{j\in\mathcal{D}}|\widehat{B}_j|> L_2-L_0)\\\nonumber
     & \quad+ P(\max_{j\in\mathcal{D}} |\frac{A_j}{\widehat{B}_j}-\frac{A_j}{B_j}|\geq \widetilde{C}n^{-\kappa}/2, \,\min_{j\in\mathcal{D}} |\widehat{B}_j|> L_2-L_0)  \\\nonumber
     &\leq  P\{\max_{j\in\mathcal{D}} |\widehat{A}_j-A_j|\geq2^{-1}(L_2-L_0)\widetilde{C}n^{-\kappa}\}\\\nonumber
     &\quad
     + P\{ \max_{j\in\mathcal{D}}|\widehat{B}_j-B_j|\geq(2L_1)^{-1} (L_2-L_0)L_2\widetilde{C}n^{-\kappa}\} \\\nonumber
    &\leq  |\mathcal{D}|\widetilde{C}_3\exp\left\{-\widetilde{C}_4n^{f(\kappa)}\right\}
          +  \widetilde{C}_{9}\exp\left\{-\widetilde{C}_{10}n^{g(\kappa)}\right\}\\\nonumber
          &\quad
         +  |\mathcal{D}|\widetilde{C}_{5}\exp\left\{-\widetilde{C}_{6}n^{f(\kappa)}\right\},
  \end{align}
where $\widetilde{C}_3, \cdots, \widetilde{C}_{6}$, and $\widetilde{C}_{9}, \widetilde{C}_{10}$ are some positive constants.
Combining \eqref{eq: AjBj-ratio}--\eqref{eq: AjBj-ratio-part2} results in
  \begin{align*}
       P( \max_{j\in\mathcal{D}}|\frac{\widehat{A}_j}{\widehat{B}_j}-\frac{A_j}{B_j}|\geq \widetilde{C}n^{-\kappa})
         &\leq   |\mathcal{D}|\widetilde{C}_7\exp\left\{-\widetilde{C}_8n^{f(\kappa)}\right\}\\
         &\quad+  \widetilde{C}_{9}\exp\left\{-\widetilde{C}_{10}n^{g(\kappa)}\right\},
  \end{align*}
where $\widetilde{C}_7=\widetilde{C}_1+\widetilde{C}_3+\widetilde{C}_{5}$ and $\widetilde{C}_8=\min\{\widetilde{C}_2, \widetilde{C}_4, \widetilde{C}_{6}\}$.

\subsection{Lemma \ref{lem: sub-Gaussian} and its proof}

\begin{lemma}\label{lem: sub-Gaussian}
Let $Z$ be a nonnegative random variable satisfying $P(Z>t)\leq \widetilde{C}_1\exp(-\widetilde{C}_2t^2)$ for all $t>0$ with $\widetilde{C}_1, \widetilde{C}_2 > 0$ some constants. Then
  \begin{align*}
     E\left[\exp\left(\frac{\widetilde{C}_2}{2}Z^2\right)\right]
 \leq  1+ \widetilde{C}_1 \quad \mbox{and} \quad E(Z^{2m})\leq (1+\widetilde{C}_1)(2\widetilde{C}_2^{-1})^m m!
  \end{align*}
for any nonnegative integer $m$.
\end{lemma}

\textit{Proof}.
Let $F(t)$ be the cumulative distribution function of $Z$.  Then
  \begin{align*}
     1-F(t)=P(Z>t)\leq \widetilde{C}_1\exp(-\widetilde{C}_2t^2)
  \end{align*}
for all $t>0$.  Using integration by parts, we have
  \begin{align*}
     E\left[\exp\left(\frac{\widetilde{C}_2}{2}Z^2\right)\right]
   &=-\int_0^{\infty} \exp\left(\frac{\widetilde{C}_2}{2}t^2\right)\,d[1-F(t)]\\
   &
   =1+\int_0^{\infty} \widetilde{C}_2t\exp\left(\frac{\widetilde{C}_2}{2}t^2\right)[1-F(t)]\,dt \\
   & \leq 1+\widetilde{C}_1\int_0^{\infty} c_2t\exp\left(-\frac{\widetilde{C}_2}{2}t^2\right)\,dt
   =1 + \widetilde{C}_1.
  \end{align*}
With the Taylor series of
the exponential function, we obtain
  \begin{align*}
     E\left[\exp\left(\frac{\widetilde{C}_2}{2}Z^2\right)\right]
  =\sum_{k=0}^{\infty}\frac{\widetilde{C}_2^kE(Z^{2k})}{2^k k!}
  \geq  \frac{\widetilde{C}_2^mE(Z^{2m})}{2^m m!}
  \end{align*}
for any nonnegative integer $m$.  Thus $E(Z^{2m})\leq (1+\widetilde{C}_1)(2\widetilde{C}_2^{-1})^m m!$.

\subsection{Lemma \ref{lem: sub-exponential} and its proof}

\begin{lemma}\label{lem: sub-exponential}
Let $Z$ be a nonnegative random variable satisfying $P(Z>t)\leq \widetilde{C}_1\exp(-\widetilde{C}_2t)$ for all $t>0$ with $\widetilde{C}_1, \widetilde{C}_2 > 0$ some constants.  Then
  \begin{align*}
     E\left[\exp\left(\frac{\widetilde{C}_2}{2}Z\right)\right]
 \leq  1+ \widetilde{C}_1 \quad \mbox{and} \quad E(Z^{m})\leq (1+\widetilde{C}_1)(2\widetilde{C}_2^{-1})^m m!
  \end{align*}
for any nonnegative integer $m$.
\end{lemma}

\textit{Proof}.
Let $F(t)$ be the cumulative distribution function of $Z$.  Then
  \begin{align*}
     1-F(t)=P(Z>t)\leq \widetilde{C}_1\exp(-\widetilde{C}_2t)
  \end{align*}
for all $t>0$.  It follows from integration by parts that
  \begin{align*}
     E\left[\exp\left(\frac{\widetilde{C}_2}{2}Z\right)\right]
   &=-\int_0^{\infty} \exp\left(\frac{\widetilde{C}_2}{2}t\right)\,d[1-F(t)]\\
   &
   =1+\int_0^{\infty} \exp\left(\frac{\widetilde{C}_2}{2}t\right)[1-F(t)]\,dt \\
   & \leq 1+\widetilde{C}_1\int_0^{\infty} \frac{\widetilde{C}_2}{2}\exp\left(-\frac{\widetilde{C}_2}{2}t\right)\,dt
   =1 + \widetilde{C}_1.
  \end{align*}
Applying the Taylor series of
the exponential function leads to
  \begin{align*}
     E\left[\exp\left(\frac{\widetilde{C}_2}{2}Z\right)\right]
  =\sum_{k=0}^{\infty}\frac{\widetilde{C}_2^kE(Z^{k})}{2^k k!}
  \geq  \frac{\widetilde{C}_2^mE(Z^{m})}{2^m m!}
  \end{align*}
for any nonnegative integer $m$.  Thus $E(Z^{m})\leq (1+\widetilde{C}_1)(2\widetilde{C}_2^{-1})^m m!$.

\subsection{Lemma \ref{lem: distance-covariance-bound} and its proof}

\begin{lemma}\label{lem: distance-covariance-bound}
Under Condition \ref{con: tail}, both $\dcov^2(X_{k}, \by)$ and $\dcov^2(X_{k}^{\ast}, \by^{\ast})$ are uniformly bounded in $k$.
\end{lemma}

\textit{Proof}.  We will show that $\dcov^2(X_{k}^{\ast}, \by^{\ast})$ are uniformly bounded in $1\leq k\leq p$.  Similar arguments apply to prove that $\dcov^2(X_{k}, \by)$ are also uniformly bounded in $k$. Recall that
$\dcov^2(X_{k}^{\ast}, \by^{\ast}) = T_{k1}+T_{k2}-2T_{k3}$ where
$T_{k1}= E\left[\phi(X_{1k}^{\ast},X_{2k}^{\ast})\psi(\by_1^{\ast}, \by_2^{\ast})\right]$,
$T_{k2}=E\left[\phi(X_{1k}^{\ast},X_{2k}^{\ast})\right] E\left[\psi(\by_1^{\ast}, \by_2^{\ast})\right]$, and
$T_{k3}= E\left[\phi(X_{1k}^{\ast},X_{2k}^{\ast}) \psi(\by_1^{\ast}, \by_3^{\ast})\right]$. Here
$\phi(X_{1k}^{\ast},X_{2k}^{\ast})=|X_{1k}^{\ast}-X_{2k}^{\ast}|$ and $\psi(\by_1^{\ast}, \by_2^{\ast})=\|\by_1^{\ast}-\by_2^{\ast}\|$. Thus an application of the triangle inequality gives
  \begin{align}\label{eq:covariance-bound}
       0\leq \dcov^2(X_{k}^{\ast}, \by^{\ast})
   \leq |T_{k1}|+|T_{k2}|+2|T_{k3}|.
  \end{align}
To prove that $\dcov^2(X_{k}^{\ast}, \by^{\ast})$ are uniformly bounded in $k$, it suffices to show that each term on the right hand side above is uniformly bounded in view of (\ref{eq:covariance-bound}).
As shown in Step 1 of Part 1 in the proof of Theorem \ref{thm:Screening}, under Condition \ref{con: tail} the first quantity $T_{k1}$ is uniformly bounded in $1\leq k\leq p$.  Using similar arguments, we can show that $T_{k2}$ and $T_{k3}$ are also uniformly bounded in $k$, which completes the proof.

\subsection{Lemma \ref{lem: RE-sample} and its proof}

\begin{lemma}\label{lem: RE-sample}
Assume that Conditions \ref{con: X}--\ref{con:RE-population} hold and $\log p=o(n^{1/2-2\xi})$.
Then with probability at least $1-O\{\exp(-\widetilde{C}_1n^{1/2-2\xi})\}$ for some constant $\widetilde{C}_1 > 0$, it holds that
  \begin{align*}
     \min_{|J|\leq s,\, \bDel\in \mathbb{R}^{\widetilde{p}\times q}\backslash\{\bzero\},\, \|\bDel_{J^c}\|_{2, 1}\leq 3\|\bDel_J\|_{2, 1}}
    \frac{\|\widetilde{\bX}\bDel\|_F}{\sqrt{n}\|\bDel_J\|_F}\geq \frac{\kappa}{2}.
  \end{align*}
\end{lemma}

\textit{Proof}.
The main idea of the proof is to first introduce an event with a high probability and then derive the desired inequality conditional on that event. Define an event
  \begin{align*}
       \mathcal{E}=\{\|n^{-1}\widetilde{\bX}^T\widetilde{\bX}-\bSig\|_{\infty}<\epsilon\},
  \end{align*}
where $\|\cdot\|_{\infty}$ denotes the entrywise matrix $L_\infty$-norm and $0<\epsilon<1$ will be specified later.  In view of the first part of Condition \ref{con: X}, it follows from Lemmas \ref{lem: sub-Gaussian} and \ref{lem: Hao-Zhang} that $P(\mathcal{E})\geq 1-\widetilde{C}_2\widetilde{p}^2\exp(-\widetilde{C}_3n^{1/2}\epsilon^2)$ for some constants $\widetilde{C}_2, \widetilde{C}_3 > 0$.

From now on, we condition on the event $\mathcal{E}$. By the definition of the Frobenius norm, we have
  \begin{align}\label{eq2}
     n^{-1}\|\widetilde{\bX}\bDel\|_F^2
     =\tr[\bDel^T(n^{-1}\widetilde{\bX}^T\widetilde{\bX}-\bSig)\bDel] + \tr(\bDel^T\bSig\bDel).
  \end{align}
For any matrix $\bM$, denote by $\bM_{ij}$ the ($i, j$)-entry of $\bM$. Then conditional on the event $\mathcal{E}$, the first term on the right hand of \eqref{eq2} can be bounded as
  \begin{align}\label{eq3}
     \Big|&\tr[\bDel^T(n^{-1}\widetilde{\bX}^T\widetilde{\bX}-\bSig)\bDel]\Big|
    = \left|\tr[(n^{-1}\widetilde{\bX}^T\widetilde{\bX}-\bSig)\bDel\bDel^T]\right|\\\nonumber
   &\quad= |\sum_{i=1}^{\widetilde{p}}\sum_{j=1}^{\widetilde{p}}(n^{-1}\widetilde{\bX}^T\widetilde{\bX}-\bSig)_{ij}(\bDel\bDel^T)_{ij}| \\\nonumber
  & \quad\leq \epsilon \sum_{i=1}^{\widetilde{p}}\sum_{j=1}^{\widetilde{p}} \sum_{k=1}^q |\bDel_{ik}||\bDel_{jk}|
      = \epsilon \sum_{k=1}^q \left (\sum_{j\in J}| \bDel_{jk}|+\sum_{j\in J^c}| \bDel_{jk}|\right)^2 \\\nonumber
    & \quad\leq 2\epsilon \sum_{k=1}^q \left (\sum_{j\in J}| \bDel_{jk}|\right)^2+2\epsilon \sum_{k=1}^q\left(\sum_{j\in J^c}| \bDel_{jk}|\right)^2,
  \end{align}
where we have used the fact that $(a+b)^2\leq 2(a^2+ b^2)$ in the last inequality.

By the Cauchy-Schwarz inequality, for any set $J$ satisfying $|J|\leq s$ we have
  \begin{align}\label{eq4}
        \sum_{k=1}^q \left (\sum_{j\in J}| \bDel_{jk}|\right)^2
    \leq \sum_{k=1}^q |J|\sum_{j\in J}\bDel_{jk}^2
     =|J|\cdot \|\bDel_J\|_F^2
    \leq s \|\bDel_J\|_F^2.
  \end{align}
For any $\bDel\in \mathbb{R}^{\widetilde{p}\times q}\backslash\{\bzero\}$ satisfying $\|\bDel_{J^c}\|_{2, 1}\leq 3\|\bDel_J\|_{2, 1}$ with $|J|\leq s$, similar arguments apply to show that
  \begin{align*}
    & \sum_{k=1}^q  \left(\sum_{j\in J^c}| \bDel_{jk}|\right)^2
    = \sum_{j\in J^c}\sum_{j'\in J^c}\sum_{k=1}^q | \bDel_{jk}|| \bDel_{j'k}|
    \leq  \sum_{j\in J^c}\sum_{j'\in J^c}\left(\sum_{k=1}^q \bDel_{jk}^2\right)^{1/2}  \nonumber\\
   &\quad\quad\cdot \left(\sum_{k=1}^q \bDel_{j'k}^2\right)^{1/2} = \sum_{j\in J^c}\sum_{j'\in J^c}\|\bDel_j\|_2\cdot \|\bDel_{j'}\|_2
     = \left(\sum_{j\in J^c}\|\bDel_j\|_2\right)^2
        \nonumber\\
  & \quad=\|\bDel_{J^c}\|_{2, 1}^2\leq 9\|\bDel_J\|_{2, 1}^2
     = 9 \left(\sum_{j\in J}\|\bDel_j\|_{2}\right)^2
     \leq  9 |J|\left(\sum_{j\in J}\|\bDel_j\|_{2}^2\right) \nonumber\\
   & \quad=  9 |J|\|\bDel_J\|_F^2 \leq 9s\|\bDel_J\|_F^2,
  \end{align*}
which along with \eqref{eq3}--\eqref{eq4} entails
  \[ |\tr[\bDel^T(n^{-1}\widetilde{\bX}^T\widetilde{\bX}-\bSig)\bDel]| \leq 20s\epsilon \|\bDel_J\|_F^2. \]

Combining the above inequality with \eqref{eq2} and by Condition \ref{con:RE-population}, we obtain
  \begin{align*}
     \min_{|J|\leq s, \bDel\in \mathbb{R}^{\widetilde{p}\times q}\backslash\{\bzero\}, \|\bDel_{J^c}\|_{2, 1}\leq 3\|\bDel_J\|_{2, 1}}\left(\frac{\|\widetilde{\bX}\bDel\|_F}{\sqrt{n}\|\bDel_J\|_F}\right)^2
    \geq \kappa^2-20s\epsilon.
  \end{align*}
It follows from 
the second part of Condition that $s\leq \widetilde{C}_4n^{\xi}$ for some positive constant $\widetilde{C}_4$.  We choose $\epsilon=3\kappa^2/(80\widetilde{C}_4n^{\xi})$.  Then we have $\kappa^2-20s\epsilon\geq \kappa^2/4$ and for sufficiently large $n$, $0<\epsilon<1$.  Therefore, it holds with probability at least $1-O\{\exp(-\widetilde{C}_1n^{1/2-2\xi})\}$ for some constant $\widetilde{C}_1 > 0$ that
  \begin{align*}
     \min_{|J|\leq s, \bDel\in \mathbb{R}^{\widetilde{p}\times q}\backslash\{\bzero\}, \|\bDel_{J^c}\|_{2, 1}\leq 3\|\bDel_J\|_{2, 1}}
    \frac{\|\widetilde{\bX}\bDel\|_F}{\sqrt{n}\|\bDel_J\|_F}\geq \frac{\kappa}{2},
  \end{align*}
which completes the proof.

\subsection{Lemma \ref{lem: concentration} and its proof}

\begin{lemma}\label{lem: concentration}
Assume that Condition \ref{con: error} and the first part of Condition \ref{con: X} hold, $q\leq p$, $\log p=o(n^{1/3})$,
and $\lambda= c_3\sqrt{(\log p)/(nq)}$ with $c_3 > 0$ some large enough constant. Then with probability at least $1-O(p^{-c})$ for some positive constant $c$, it holds that
  \begin{align*}
      \frac{1}{nq}\|\widetilde{\bX}^T\bW\|_{2, \infty}\leq \frac{\lambda}{2}.
  \end{align*}
\end{lemma}

\textit{Proof}.
An application of the union bound leads to
  \begin{align}\label{eq: XE-bound-1}
      P(\|\widetilde{\bX}^T\bW\|_{2, \infty}\geq nq\lambda/2)
   \leq  \sum_{j=1}^{\widetilde{p}} P\left\{\sum_{k=1}^q (\widetilde{\bX}^T\bW)_{jk}^2\geq (nq\lambda/2)^2\right\}
  \end{align}
for any $\lambda\geq 0$, where
$(\widetilde{\bX}^T\bW)_{jk}$ is the ($j, k$)-entry of $\widetilde{\bX}^T\bW$. The key ingredient of the proof is to bound $P\{\sum_{k=1}^q (\widetilde{\bX}^T\bW)_{jk}^2\geq (nq\lambda/2)^2\}$.
Define $T_{jk, 1} =\sum_{i=1}^n\widetilde{\bX}_{ij}\bW_{ik}I(|\widetilde{\bX}_{ij}|\leq L)$ and $T_{jk, 2} =\sum_{i=1}^n\widetilde{\bX}_{ij}\bW_{ik}I(|\widetilde{\bX}_{ij}|> L)$, where $L>0$ will be specified later.  Since
$(\widetilde{\bX}^T\bW)_{jk}=\sum_{i=1}^n\widetilde{\bX}_{ij}\bW_{ik}=T_{jk, 1} +T_{jk, 2}$, we deduce
  \begin{align}\label{eq: XE-bound-2}
          P&\left\{\sum_{k=1}^q (\widetilde{\bX}^T\bW)_{jk}^2\geq (nq\lambda/2)^2\right\} \leq   P\left\{\sum_{k=1}^q T_{jk, 1}^2\geq (nq\lambda/4)^2\right\}\\\nonumber
   &\quad\quad
         + P\left\{\sum_{k=1}^q T_{jk, 2}^2\geq (nq\lambda/4)^2\right\}  \\\nonumber
   &\quad\leq   \sum_{k=1}^q P\left\{ |T_{jk, 1}|\geq \sqrt{q}n\lambda/4\right\}
         + P\left\{\sum_{k=1}^q T_{jk, 2}^2\geq (nq\lambda/4)^2\right\}.
  \end{align}
We will deal with the two terms on the right hand side above separately.

We first bound $P\left\{ |T_{jk, 1}|\geq \sqrt{q}n\lambda/4\right\} $.  By the first part of Condition \ref{con: X}, there exist some positive constants $a_1$ and $b_1$ such that
\[ P(|\bv^T\bx_i|>t)\leq a_1\exp(-b_1t^2) \]
for any $\|\bv\|_2=1$ and $t>0$, where $\bx_i\t=(\bX_{i1}, \cdots, \bX_{ip})$ is the $i$th row of the main effect design matrix $\bX$.
Then choosing $\bv$ as a unit vector with the $j$th component being $1$ gives \[ P(|\bX_{ij}|>t)\leq a_1\exp(-b_1t^2) \]
for any $1\leq i\leq n$, $1\leq j\leq p$, and $t>0$.  Thus it follows from Lemma \ref{lem: sub-Gaussian} that
  \begin{align*}
      E(\bX_{ij}^2)\leq 2(1+a_1)/b_1 \quad \mbox{and} \quad E(\bX_{ij}^4)\leq 8(1+a_1)/b_1^2
  \end{align*}
for all $i$ and $j$.

Note that $\widetilde{\bX}_{ij}=\bX_{ij}$ for $1\leq j\leq p$ and $\widetilde{\bX}_{ij}=\bX_{i\ell}\bX_{i\ell'}$ with $1\leq \ell<\ell'\leq p$ for $p+1\leq j\leq \widetilde{p}$.  Thus $E(\widetilde{\bX}_{jk}^2)$ are uniformly bounded from above by some positive constant $\widetilde{C}_1$.  Similarly, by Condition \ref{con: error} and Lemma \ref{lem: sub-exponential} there exist some positive constants $a_2$ and $b_2$ such that $E(|\bW_{ik}|^m)\leq a_2b_2^m m!$ for any nonnegative integer $m$ and indices $i$ and $k$.
Since $\widetilde{\bX}_{ij}$ is independent of $\bW_{ik}$, we have
  \begin{align*}
     E\left[|\widetilde{\bX}_{ij}\bW_{ik}I(|\widetilde{\bX}_{ij}|\leq L)|^m\right]
   &\leq L^{m-2}E(\widetilde{\bX}_{ij}^2)E(|\bW_{ik}|^m)\\
   &
   \leq m!(Lb_2)^{m-2}(2a_2b_2^2\widetilde{C}_1)/2
  \end{align*}
for each integer $m\geq 2$.  In view of $T_{jk, 1} =\sum_{i=1}^n\widetilde{\bX}_{ij}\bW_{ik}I(|\widetilde{\bX}_{ij}|\leq L)$,  applying Bernstein's inequality (Lemma 2.2.11 of \citep{van1996weak}) yields
  \begin{align}\label{eq: XE-bound-3}
      P\left\{ |T_{jk, 1}|\geq \sqrt{q}n\lambda/4\right\}
    \leq & 2\exp\left(-\frac{qn\lambda^2}{64a_2b_2^2\widetilde{C}_1+8b_2L\sqrt{q}\lambda}\right). 
  \end{align}

We next bound $P\left\{\sum_{k=1}^q T_{jk, 2}^2\geq (nq\lambda/4)^2\right\}$.   
By the definition of $T_{jk, 2}$, it is seen that for each $j$, such an event satisfies
  \begin{align*}
     \left\{\sum_{k=1}^q T_{jk, 2}^2\geq (nq\lambda/4)^2\right\}
   \subset \left\{|\widetilde{\bX}_{ij}|> L\,\,\mbox{for some}\,\,1\leq i\leq n\right\}.
  \end{align*}
Thus using the union bound, we obtain
  \begin{align*}
    P\left\{\sum_{k=1}^q T_{jk, 2}^2\geq (nq\lambda/4)^2\right\}
     \leq \sum_{i=1}^nP\{|\widetilde{\bX}_{ij}|> L\}.
  \end{align*}
Combining this inequality with \eqref{eq: XE-bound-1}--\eqref{eq: XE-bound-3} gives
  \begin{align}
      P&(\|\widetilde{\bX}^T\bW\|_{2, \infty}\geq nq\lambda/2) \leq  2\widetilde{p} q  \exp\left(-\frac{qn\lambda^2}{64a_2b_2^2\widetilde{C}_1+8b_2L\sqrt{q}\lambda}\right)\\\nonumber
      &\quad\quad + \sum_{i=1}^n\sum_{j=1}^p P\{|\bX_{ij}|> L\}
       +  \sum_{i=1}^n\sum_{1\leq \ell< \ell'\leq p}^p P\{|\bX_{i\ell}\bX_{i\ell'}|> L\} \\\nonumber
      &\quad\leq  qp^2  \exp\left(-\frac{qn\lambda^2}{64a_2b_2^2\widetilde{C}_1+8b_2L\sqrt{q}\lambda}\right)
       + a_1np\exp(-b_1L^2)\\\nonumber
       &\quad\quad
       +  a_1np^2\exp(-b_1L).
  \end{align}
Note that $\lambda=c_3\sqrt{(\log p)/(nq)}$ with some large enough positive constant $c_3$. Therefore, setting $L=\widetilde{C}_2\sqrt{n/(\log p)}$ for some large positive constant $\widetilde{C}_2$ ensures that there exists some positive constant $c_4$ such that
\begin{align}
P(\|\widetilde{\bX}^T\bW\|_{2, \infty}\geq nq\lambda/2) \leq O(p^{-c}),
\end{align}
where we have used the assupmtion that $q\leq p$ and $\log p=o(n^{1/3})$. This concludes the proof.

\subsection{Additional lemmas}

\begin{lemma}[Hoeffding's inequality]\label{lem: Hoeffding-Lemma}
Let $X$ be a real-valued random variable with $E(X)=0$. If $P(a\leq X\leq b)=1$ for some $a, b \in \mathbb{R}$, then $E[\exp(tX)] \leq \exp[t^2(b-a)^2/8]$ for any $t>0$.
\end{lemma}

\begin{lemma}[Lemma B.4 of \citet{hao2014interaction}]\label{lem: Hao-Zhang}
Let $Z_1, \cdots, Z_n$ be independent random variables with zero mean and  $E[\exp(T_0|Z_i|^{\alpha})]\leq A_0$ for constants $T_0, A_0>0$ and $0< \alpha\leq 1$.  Then there exist some constants $\widetilde{C}_3, \widetilde{C}_4 > 0$ such that
  \begin{align*}
     P\Big(\big|n^{-1} \sum_{i=1}^n Z_i\big|>\epsilon\Big) \leq \widetilde{C}_3\exp(-\widetilde{C}_4n^{\alpha}\epsilon^2)
  \end{align*}
for any 
{$0< \epsilon\leq 1$}.
\end{lemma}

\end{document}